\documentclass[a4paper, 12pt, jpgfig]{report}
\usepackage[T1, T2A]{fontenc}
\usepackage[utf8]{inputenc}
\usepackage[english]{babel}
\usepackage{amsmath}
\usepackage{amsfonts, amssymb, longtable}
\usepackage{caption}
\usepackage{graphicx}

\usepackage{float}
\usepackage[clearempty]{titlesec}
\usepackage{titletoc}
\usepackage{lipsum} 

\titlecontents{section}
[5.5em] 
{\bigskip}
{\contentslabel[\chaptername~\thecontentslabel]{5.0em}}
{\hspace*{-6.5em}}
{\hfill\contentspage}[\medskip]

\newcommand{\eqdef}{\stackrel{def}{=}}

\begin{document}
\title{\textbf{Effects Induced by the Inclusion of the Symmetric Derivatives of the Vector Potential Affinors in Modified Electrodynamics.}}
\author{Yu. A. Alebastrov \thanks{jalebastrov@gmail.com} \\
141980, Dubna }
\date{7.03.2016}
\maketitle

\renewcommand{\abstractname}{Abstract}
\begin{abstract}

This work continues the study of the real vector field theory introduced in Refs. ~\cite{Al1}\, --\, ~\cite{Al9}.
It shows how the Liénard--Wiechert potentials can be used to derive the field and dynamical variables of the wave electrojeitonic field generated by an electrically charged point particle.
The derived expressions provide the angular distributions of the instantaneous electrojeitonic radiated power emitted by an  charged particle in arbitrary motion.
The results predict the dominance of longitudinal electrojeitonic radiation from ultrarelativistic charged particles.


\end{abstract}

\clearpage

\begin{center}
\textbf{\Large Contents}
\end{center}

\vspace{1em}

\makeatletter
\@starttoc{toc}
\makeatother


\section{The Electrojeitonic Field and Its Field and Dynamical Variables}

\subsection*{\S \, 1. General Axioms}
\addcontentsline{toc}{subsection}{\S\, 1. General Axioms}
\setcounter{section}{1}
\setcounter{figure}{0}

The theory of the vector field with symmetric affinors, discussed in \cite{Al1}\, --\, \cite{Al9}, is based on the equal mathematical status and independence of the derivative tensors of the vector field $\mathit A(x)$, 
$$
F(x)={\widetilde{rot} }\, A(x) \, ,                    \eqno (1.1)
$$
$$
G(x)={\widetilde{def}}\, A(x)\, ,                            \eqno (1.2)
$$
where the 4\nobreakdash-tensor differential operators $\widetilde{rot}$  and $\widetilde{def} $ are defined by 
$$
\widetilde{rot}\, \eqdef \, \partial\, \overset{a}{\otimes}\, ,              \eqno (1.3)
$$
$$
\widetilde{def}\, \eqdef \, \partial\, \overset{s}{\otimes}\, .                  \eqno (1.4)
$$
Within this theory, the field determined by the symmetric derivative tensor obtained from the 4\nobreakdash-potential of classical electrodynamics is regarded as an independent \emph{physical} field having the same theoretical status as the electromagnetic field. This field is referred to as the \emph{electrojeitonic} field.

According to this theory, the electromagnetic field is represented by the 4\nobreakdash-tensor $ \mathit F(x)$, defined by relation (1.1), where $ \mathit A(x)$ denotes the conventional 4\nobreakdash-potential of classical electordynamics.

The electrojeitonic field is associated with the 4\nobreakdash-tensor $\mathit G(x)$, defined by relation(1.2) accordingly.

The physical reality of both the classical electromagnetic field and the classical electrojeitonic field is established primarily by the direct and indirect forces exerted by their respective field components on charged particles.

The indirect force exerted by the \emph{jeitonic} component of the electrojeitonic field (jeitonic field) on \emph{systems} of charged particles is mediated by the \emph{deforming} electric fields generated by this component and considered in \cite{Al0}. The jeitonic component is represented, according to (1.2), by the 3-tensor
$$
\mathcal{G}(x) = \tilde{def}\, \vec{A}(x)\, ,                       \eqno (1.5)
$$
where $ \tilde{def} $ is the differential operator defined by \cite{Pob}
$$
\tilde{def} \eqdef \vec{\bigtriangledown}\, \overset{s}{\otimes}\, .                \eqno (1.6)
$$

As in \cite{Al0}, in order to compare the properties of the jeitonic component of the electrojeitonic field with those of the magnetic component of the electromagnetic field, we consider both components simultaneously, representing the latter, in accordance with (1.1), by the 3-tensor

$$
\mathcal{F}(x) = \tilde{rot}\, \vec{A}(x)\, ,                       \eqno (1.7)
$$
where $ \tilde{rot} $ is the differential operator defined by
$$
\tilde{rot} \eqdef {\vec\bigtriangledown}\, \overset{a}{\otimes}.               \eqno (1.8)
$$

According to the unique decomposition of any second-rank tensor into a deviator and a spherical tensor (dilatator), the jeitonic field tensor $\mathcal{G}(x) $ generally has two components \cite{Al0}, which are linearly independent both of each other and of the tensor ${\mathcal F}$:

$$
\mathcal{G}_{d}(x) = \tilde{def}_{d}\, \vec{A}(x)\, ,               \eqno (1.9)
$$
$$
\mathcal{G}_{\ast}(x) = \tilde{def}_{\ast}\, \vec{A}(x)\, ,               \eqno (1.10)
$$
$$
\tilde{def}_{d} \eqdef  \vec{\bigtriangledown} \overset{s}{\otimes} - \frac{g}{tr g}\, \vec{\bigtriangledown}\, \cdot\, ,   \eqno (1.11)
$$
$$
\tilde{def}_{\ast} \eqdef  \frac{g}{tr g}\, \vec{\bigtriangledown}\, \cdot ,  \qquad  g = \vec{e}_{\alpha}\otimes\vec{e}\, ^{\alpha}, \qquad tr g = g\cdot\cdot g\, .                     \eqno (1.12)
$$

The fields represented by these components are referred to, respectively, as the \emph{deviatoric jeitonic} field and the \emph{dilatational jeitonic} field.

As in \cite{Al0}, in this context, a tensor is regarded as an object, invariant under the class of admissible passive coordinate transformations, and represented by different sets of components in different reference frames [10, §16.1].

The term ``vector field'' is used throughout the theory exclusively in the mathematical sense.

The derivative second-rank 3-tensor (the derivative affinor of the vector field $\vec {A}(x)$ \cite[\S11.]{Rash}, or the gradient or local affinor of the vector field $\vec {A}(x)$ \cite[\S16.]{Ko}) is defined as

$$
\partial A(x) \eqdef \vec{\bigtriangledown} \otimes \vec{A} (x)\, ,          \eqno (1.13)
$$

The tensors discussed above, $ \mathcal{ F}(x)$, $ \mathcal{G}_{d}(x)$ and $ \mathcal{G}_{\ast}(x)$, are its linearly independent components. This tensor represents a ``measure of inhomogeneity'' \cite{Bor} of the vector field $\vec {A}(x)$ at a fixed time $t$ in an infinitesimal neighborhood $\delta_{ M}$ of a point $ \mathit M$. Thus, in the linear approximation, this tensor determines the corresponding state of the vector field $\vec {A}(x)$ in $ \delta_{  \mathit M}$ \cite{BSh}.

This state, by virtue of relation \cite{Al0}, 
$$
\vec{ \mathit A}_{\mathit \partial A}(N)=\partial A( \mathit M)^T\! \cdot \vec{\rho}\, ( \mathit N),    \quad    \mathit N \in \delta_{ \mathit M}\, ,   \eqno (1.14)
$$

regarded as a mapping $V\otimes V\mapsto V$, is geometrically represented by a vector field whose congruence of vector lines is determined by the phase portrait of the dynamical system

$$
\frac{d\, \vec{ r}}{d\, \tau}=\partial A( \mathit  M)^T\!\cdot\, {\vec r}\, .               \eqno (1.15)
$$

This \emph{congruence} is proposed as a geometric representation of the derivative tensor $\partial A(x)$ and is referred to as the \emph{phase portrait} of this tensor \cite{Al0}.

Using this approach, \cite{Al0} defines the corresponding geometric representations of the second-rank tensors ${\mathcal F}(x)$, $\mathcal G_d(x)$, and  $\mathcal G_*(x)$, as shown in Fig. 1.1.

As in \cite{Al0}, the point $M$ is also taken to be the origin of the coordinate system; accordingly, the relation $\vec{\rho}(N)=\vec{r}(N)$ will be used frequently in the corresponding formulas.
\\

\begin{figure}[H]
\begin{center}
\vspace{-6pt}
\includegraphics[width=127mm]{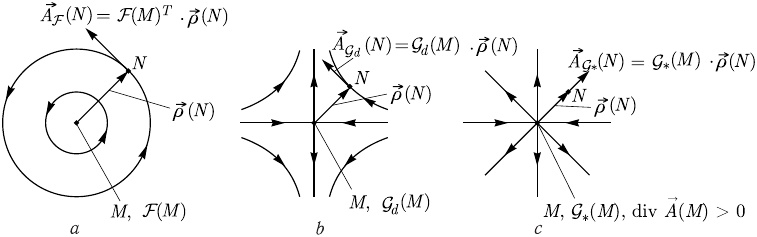}
\vspace{+2pt}
\caption{Phase portraits of the tensors ${\mathcal F}(M)$, ${\mathcal G}_d(M)$ and ${\mathcal G}_*(M)$ in Euclicical space $E^2$, corresponding to a classical center, a saddle, and a dicritical (star) node, respectively (an attractor for $div\, {\vec A}(M)>0$, and a repeller for $div\, {\vec A}(M)<0$).}
\end{center}
\vspace{-5mm}
\end{figure}

Following \cite{Arn}, we use the terms center, saddle, and node for the \emph{phase portraits} of the corresponding dynamical systems and refer to their corresponding singular points as the singular points of the center, saddle, and node, respectively.

These geometric representations \emph{visually}demonstrates both the \emph{fundamental} differences between the tensors ${\mathcal F}(x)$, $\mathcal G_d(x)$, and $\mathcal G_*(x)$, and their linear independence from one another.

On the other hand, \emph{each} of these tensors defines \emph{its own} state of the vector field 
$\vec{\mathit A}(x)$ in $ \delta_{\mathit M}$ at a given time $t$, topologically inequivalent to those defined by the other tensors.

Accordingly, and in accordance with the one-to-one 3-invariant representation,
$$
\partial A(\mathit M)=\mathcal {F}( \mathit M)+ \mathcal G_d (\mathit M)+ \mathcal G_*( \mathit M)\, , 
                                                                                          \eqno (1.16)
$$

we find that each of the tensors $ \mathcal F(x)$, $\mathcal G_d(x)$, $ \mathcal G_*(x)$, makes its own linearly independent contribution to the overall structure (state) of the vector field $\vec{ A}(x)$, in $ \delta_{\mathit M}$ at a given time $t$, as determined by the derivative tensor $\partial \mathit A(x)$ at that time and in that $ \delta_{\mathit M}$. This, in turn, demonstrates the corresponding mathematical independence and equal status of these tensors with respect to one another.

Examples of both separate and combined existence of states of the vector field $\vec {A}(x)$, shown in Fig. 1.1, are discussed in \cite{Al0} for \emph{fundamental physical} field systems.

The electric-field strength of the electromagnetic field represented by the 4\nobreakdash-tensor (1.1) is defined, according to \cite{La}, by the polar 3-vector

$$
\vec{E}_{\mathcal F}(x):= F_{0\alpha}(x)\, \vec{e}\, \, ^{\alpha}\, .                 \eqno (1.17)
$$

Substituting the values

$$
A_{i}(x)=\, (\varphi(x), -\, \vec{A}(x))                \eqno (1.18)
$$

into relation

$$
F_{0\alpha}(x)\, =\, \frac{1}{2}\, (\partial_{0}{A}_{\alpha}(x)-\partial_{\alpha}A_{0}(x))    \eqno (1.19)
$$

yields the following representation of the electric-field strength associated with the magnetic field under consideration:

$$
\vec{E}_{\mathcal F}(x)= -\, \frac{1}{2}\, (\partial_{0}\vec{A}(x)+\, \vec{\bigtriangledown}\varphi(x))
:=\, \frac{1}{2}\, (\vec{E}_{A}(x)+\, \vec{E}_{\varphi}(x))\, .                  \eqno (1.20)
$$

The plus sign ``+'' in the parentheses on the right-hand side of (1.20) results from the antisymmetry of the 4\nobreakdash-tensor (1.1), which is a Lorentz-invariant property of this tensor. Consequently, the magnetic field represented by the 3-tensor$\mathcal F(x)$ is, in general, accompanied by an electric field whose strength is determined by the Lorentz-covariant relation (1.20).

The strength of the electric component of the electrojeitonic field represented by the 4\nobreakdash-tensor (1.2) is defined, by analogy with (1.17), by the polar 3-vector

$$
\vec{E}_{\mathcal G}(x):= G_{0\alpha}(x)\, \vec{e}\, \, ^{\alpha}\, .                 \eqno (1.21)
$$

Substituting (1.18) into relation

$$
G_{0\alpha}(x)\, =\, \frac{1}{2}\, (\partial_{0}{A}_{\alpha}(x)+\partial_{\alpha}A_{0}(x))    \eqno (1.22)
$$

yields the following representation of the electric-field strength accompanying the jeitonic field under consideration \cite{Al0}:

$$
\vec{E}_{\mathcal G}(x)= -\, \frac{1}{2}\, (\partial_{0}\vec{A}(x)-\, \vec{\bigtriangledown}\varphi(x))
=\, \frac{1}{2}\, (\vec{E}_{A}(x)-\, \vec{E}_{\varphi}(x))\, .                  \eqno (1.23)
$$

The minus sign ``$-$'' in the parentheses on the right-hand side of (1.23) results from the symmetry of the 4\nobreakdash-tensor (1.2), which is a Lorentz-invariant property of this tensor. Consequently, the jeitonic field represented by the 3-tensor $\mathcal G(x)$ is, in general, accompanied by an electric field whose strength is determined by the Lorentz-covariant relation (1.23).

As in \cite{La1}, we write the covariant components of the 4\nobreakdash-tensor (1.1) in the form

$$
F_{ik}(x)=\, (\vec{E}_{\mathcal F}(x), -\mathcal F(x))\, .       \eqno (1.24)
$$

Using the well-known relation defining the pseudovector $\vec{\mathcal H}(x)$ accompanying 3-tensor $\mathcal F(x)$ in the three-dimensional space $E^{\,3}$,

$$
\vec{\mathcal H}(x)=\, -\, \frac{1}{2}\, \, \varepsilon\, \cdot\cdot\, \mathcal F(x),     \eqno (1.25)
$$

equation (1.24) can also be written in the form \cite{La1}

$$
F_{ik}(x)=\, (\vec{E}_{\mathcal F}(x), \vec{\mathcal H}(x))\, ,      \eqno (1.26)
$$

showing that the covariant components of the 4\nobreakdash-tensor of the electromagnetic field (1.1) correspond to the components of the electric- and magnetic-field strengths defined by (1.20) and (1.25), respectively.

Similarly, we write the covariant components of the 4\nobreakdash-tensor (1.2) in the form identical to (1.24):

$$
G_{ik}(x)=\, (\vec{E}_{\mathcal G}(x), -\mathcal G(x))\, .       \eqno (1.27)
$$

In \cite{Al0}, relations for the field and dynamical variables of the electrojeitonic field were derived for the class of electrodynamic 4\nobreakdash-potentials $A(x)$ satisfying the well-known system of differential equations
$$
\partial\cdot\partial A(x)=\, j(x)\, ,           \eqno (1.28)
$$
$$
\partial\cdot A(x)\, =\, 0\, ,                       \eqno (1.29)
$$

where the symbol $\partial A(x) := \partial \otimes A(x)$ denotes the 4-gradient of the 4\nobreakdash-potential $A(x)$, whose linearly independent components are the 4\nobreakdash-tensors (1.1) and (1.2), and the symbol $\partial := e^{\,\mu}\partial_{\mu}$ denotes the Hamilton differential operator in pseudo-Euclidean spacetime, also used in (1.3) and (1.4).

To simplify the notation, we use the same symbol $\partial A(x)$ to denote both the 4\nobreakdash-tensor $\partial \otimes A(x)$ and the 3-tensor $\vec{\bigtriangledown} \otimes \vec{A}(x)$.

The following system of field equations \cite{Al0} was used as the basic system for deriving the above relations:

$$
\partial\cdot G_{L}(x)=\, j(x)\, ,           \eqno (1.30)
$$
$$
\partial\cdot A(x)\, =\, 0\, ,                    \eqno (1.31)
$$

This system, as a system of equations for the 4\nobreakdash-potential $A(x)$, is equivalent to (1.28) and (1.29) and thus keeps the analysis within the class of electrodynamic potentials under consideration.

Equation (1.30) is the relativistic form of the second pair of differential equations representing the corresponding field strengths of the electrojeitonic field \cite{Al0}.

It is also worth noting that this equation can also be derived using the conventional Lagrangian formalism, starting from the Lagrangian $\mathfrak{L}_{G}(x)$ \cite[Ch.~V, (14.65).]{Al0}, the corresponding Euler\,--\,Lagrange equations for which directly yield (1.30).

The present work continues the study of electrojeitonic radiation from fundamental field sources initiated in \cite{Al6} and \cite{Al0}.

When deriving the relations for the field variables of the radiation field of a point electric charge moving along a prescribed trajectory in vacuum, the action of the Hamilton operator $\vec{\bigtriangledown}=\vec{e}^{\,\alpha}\partial_{\alpha}$ on the Liénard\,--\,Wiechert potentials reduces to the action of the operator $-\,\vec{n}\,\partial^{\,0}$ on these potentials. Thus, in deriving such relations, we may make the substitution \cite{Jack}

$$
\vec{\bigtriangledown}\, \longmapsto\, -\, \vec{n}\, \partial^{\, 0}\, ,         \eqno (1.32)
$$

where  $\vec{n}$ is a unit vector directed from the position of the charge at the corresponding retarded time toward the observation point.

For example, substituting (1.32) into (1.5) and (1.7) directly yields the following equations relating the field variables $\mathcal{G}(x)$ and $\mathcal{F}(x)$ to $\vec{E}_{A}(x)$ for such radiation:
$$
\mathcal{G}(x)=\vec{n}\overset{s}{\otimes}\vec{E}_{A}(x), \eqno (1.33)
$$
$$
\mathcal{F}(x)=\vec{n}\overset{a}{\otimes}\vec{E}_{A}(x). \eqno (1.34)
$$

When deriving the relations between the field variables in the \emph{wave} zone of the radiation from a bounded \emph{system} of electric currents and charges, the action of the Hamilton operator on the asymptotic expressions for the potentials $\vec{A}(x)$ and $\varphi(x)$ may likewise be replaced by the action of the differential operator $-\,\vec{n}\,\partial^{\,0}$ on these expressions, where $\vec{n}$ is the unit vector defined by $\vec{n}=\vec{r}/r$, and $\vec{r}$ is the position vector of the observation point in a coordinate system whose origin is located at any point ``inside'' this system of charges \cite{Med}, \cite{La}.

Expression (1.23) for the electric-field strength, whose components are defined by the corresponding covariant components of the 4\nobreakdash-tensor (1.2), also follows ``automatically'' by directly passing from the system of equations (1.28), (1.29) to the system (1.30), (1.31) \cite{Al0}.

We show that this important relation can also be obtained by solving the first equation of system (IV) \cite[\S\,14, Ch.~V.]{Al0},

$$
\partial^{\,0}\vec{E}_{\mathcal G}(x)+\vec{\bigtriangledown}\cdot\mathcal{G}(x)=\vec{0} \eqno (1.35)
$$

as a first-order differential equation for $\vec{E}_{\mathcal G}(x)$, with the field variable $\mathcal{G}(x)$ assumed to be known, having been determined by some method.

The 3-tensor of the jeitonic field, $\mathcal{G}(x)$, defined by (1.5), upon using (1.13), can be written in the form

$$
\mathcal{G}(x)=\frac{1}{2}\left(\partial A(x)+\partial A(x)^{T}\right). \eqno (1.36)
$$

In turn, the wave equation for the vector potential $\vec{A}(x)$ can be written, using the relation

$$
\bigtriangleup\,\vec{A}(x)=\vec{\bigtriangledown}\cdot\partial A(x), \eqno (1.37)
$$

as the following first-order differential equation, which serves as a differential equation relating the wave field variables $\partial A(x)=\vec{\bigtriangledown}\otimes\vec{A}(x)$ and $\vec{E}_{A}(x)=-\partial^{\,0}\vec{A}(x)$:

$$
\vec{\bigtriangledown}\cdot\partial A(x)=-\partial^{\,0}\vec{E}_{A}(x). \eqno (1.38)
$$

On the other hand, using the relation

$$
\vec{\bigtriangledown}\cdot\partial A(x)^{T}=\vec{\bigtriangledown}\,div\,\vec{A}(x) \eqno (1.39)
$$

and taking into account that, for the radiation field of a point electric charge considered below, the Liénard\,--\,Wiechert potentials satisfy the Lorenz condition
$div\,\vec{A}(x)=-\partial^{\,0}\varphi(x)$,
we obtain the following relation, which represents a differential equation relating the wave field variables $\partial A(x)^{T}$ and $\vec{E}_{\varphi}(x)=-\vec{\bigtriangledown}\,\varphi(x)$:

$$
\vec{\bigtriangledown}\cdot\partial A(x)^{T}=\partial^{\,0}\vec{E}_{\varphi}(x). \eqno (1.40)
$$

Thus, (1.36), together with (1.38) and (1.40), yields

$$
\vec{\bigtriangledown}\cdot\mathcal{G}(x)=-\frac{1}{2}\partial^{\,0}\left(\vec{E}_{A}(x)-\vec{E}_{\varphi}(x)\right), \eqno (1.41)
$$

Using this relation in (1.35) and subsequently integrating the resulting equation with respect to time yields, up to a constant of integration, relation (1.23).

It is also worth noting that the minus sign in the parentheses on the right-hand side of (1.41) is due to the symmetry of the 3-tensor $\mathcal{G}(x)$.

Thus, the electric-field strength $\vec{E}_{\mathcal{G}}(x)$ can be determined either directly from (1.23), when $\vec{E}_{A}(x)$ and $\vec{E}_{\varphi}(x)$ can be readily calculated, or by using (1.35) as a first-order differential equation for $\vec{E}_{\mathcal{G}}(x)$, when $\vec{\bigtriangledown}\cdot\mathcal{G}(x)$ is easier to calculate.

The expression (1.20) for the electric-field strength, whose components are defined by the corresponding covariant components of the 4\nobreakdash-tensor (1.1), also follows ``automatically'' when passing from the system of equations (1.28), (1.29) to the well-known system \cite{Al0} equivalent with respect to $A(x)$:

$$
\partial\cdot F_{L}(x)=j(x), \eqno (1.42)
$$
$$
\partial\cdot A(x)=0. \eqno (1.43)
$$

Let us show that relation (1.20) can likewise be obtained by solving the first equation of the second pair of Maxwell’s equations in the absence of field sources, represented by the first equation of system (I) \cite[\S\,14, Ch.~V]{Al0},

$$
\partial^{\,0}\vec{E}_{\mathcal F}(x)+\vec{\bigtriangledown}\cdot\mathcal F(x)=\vec{0}, \eqno (1.44)
$$

as a first-order differential equation for $\vec{E}_{\mathcal F}(x)$, with the wave field variable $\mathcal F(x)$ assumed to be known, having been determined by some method.

Using (1.13), the 3-tensor of the magnetic field, $\mathcal F(x)$, defined by relation (1.7), can be written in the form

$$
\mathcal F(x)=\frac{1}{2}\left(\partial A(x)-\partial A(x)^{T}\right). \eqno (1.45)
$$

As a result, (1.45), together with (1.38) and (1.40), now yields

$$
\vec{\bigtriangledown}\cdot\mathcal F(x)=-\frac{1}{2}\partial^{\,0}\left(\vec{E}_{A}(x)+\vec{E}_{\varphi}(x)\right). \eqno (1.46)
$$

Using this relation in (1.44) yields (1.20).

It is also worth noting that the plus sign in the parentheses on the right-hand side of (1.46) is due to the antisymmetry of the 3-tensor $\mathcal F(x)$.

Thus, the electric-field strength $\vec{E}_{\mathcal F}(x)$ can be determined either directly from (1.20), when $\vec{E}_{A}(x)$ and $\vec{E}_{\varphi}(x)$ can be readily calculated (for example, when studying the radiation fields of a point electric charge or analyzing radiation due to the electric dipole moment of an electric system \cite[\S\,63, \S\,72]{La}), or by using (1.44) as a first-order differential equation for $\vec{E}_{\mathcal F}(x)$, when $\vec{\bigtriangledown}\cdot\mathcal F(x)$ is easier to calculate (for example, when calculating the electric-field strengths of radiation from antenna systems \cite{Jack}, \cite{Eis}).

Using the ``rule for differentiating retarded potentials in the wave zone'' \cite[\S\,17.3, Part~II]{Med} in equations (1.44) and (1.35) makes it possible to express the wave-zone field variables $\vec{E}_{\mathcal F}(x)$ and $\vec{E}_{\mathcal G}(x)$ in terms of the electric-field strength $\vec{E}_{A}(x)$.

Indeed, substituting (1.32) into (1.44) yields a relation which, upon subsequent integration with respect to time, gives, up to an ``integration constant'' \cite[\S\,72]{La}, the following algebraic relation between the wave field variables $\vec{E}_{\mathcal F}(x)$ and $\mathcal F(x)$ in the wave zone:

$$
\vec{E}_{\mathcal F}(x)=\vec{n}\cdot\mathcal F(x), \eqno (1.47)
$$

which, upon using (1.34), takes the form

$$
\vec{E}_{\mathcal F}(x)=\vec{n}\cdot\left(\vec{n}\overset{a}{\otimes}\vec{E}_{A}(x)\right), \eqno (1.48)
$$

and, on applying the vector-analysis identity in (1.48),

$$
\vec{a}\cdot\left(\vec{b}\overset{a}{\otimes}\vec{c}\right)=-\frac{1}{2}\vec{a}\times(\vec{b}\times\vec{c}), \eqno (1.49)
$$

yields, without using (1.20), the following expression for the wave field variable $\vec{E}_{\mathcal F}(x)$ in terms of $\vec{E}_{A}(x)$ in the wave zone (see also \cite[(88), \S\,17.3, Part~II]{Med}):

$$
\vec{E}_{\mathcal F}(x)=\frac{1}{2}\vec{n}\times(\vec{E}_{A}(x)\times\vec{n})=\frac{1}{2}\vec{E}_{A}^{\,t}(x), \eqno (1.50)
$$

which at the same time demonstrates the transversality of the wave field variable $\vec{E}_{\mathcal F}(x)$ with respect to $\vec{n}$ in the zone under consideration.

For the axial vector $\vec{\mathcal H}(x)$, dual to the tensor $\mathcal F(x)$,

$$
\vec{\mathcal H}(x)=-\frac{1}{2}\epsilon\cdot\cdot\mathcal F(x)=\frac{1}{2}rot\vec{A}(x), \eqno (1.51)
$$

we accordingly have the relations

$$
\vec{\mathcal H}(x)=\frac{1}{2}\left(\vec{n}\times\vec{E}_{A}^{\,t}(x)\right)=\vec{n}\times\vec{E}_{\mathcal F}(x), \eqno (1.52)
$$

which demonstrate that the field variable $\vec{\mathcal H}(x)$ is transverse both to $\vec{n}$ and to $\vec{E}_{\mathcal F}(x)$, and also demonstrate the equality $|\vec{\mathcal H}(x)|=|\vec{E}_{\mathcal F}(x)|$.

Similarly, substituting (1.32) into (1.35) yields a differential relation which, upon subsequent integration with respect to time, gives, up to an ``integration constant,'' the following algebraic relation between the wave field variables $\vec{E}_{\mathcal G}(x)$ and $\mathcal G(x)$ in the wave zone:

$$
\vec{E}_{\mathcal G}(x)=\vec{n}\cdot\mathcal G(x), \eqno (1.53)
$$

and applying the vector identity

$$
\vec{a}\cdot\left(\vec{b}\overset{s}{\otimes}\vec{c}\right)
=-\frac{1}{2}\vec{a}\times(\vec{b}\times\vec{c})+\vec{c}(\vec{a}\cdot\vec{b}), \eqno (1.54)
$$

to (1.53) yields, without using (1.23), the following expression for the wave field variable $\vec{E}_{\mathcal G}(x)$ in terms of $\vec{E}_{A}(x)$ in the wave zone:

$$
\vec{E}_{\mathcal G}(x)=\frac{1}{2}\vec{n}\times(\vec{E}_{A}(x)\times\vec{n})+\vec{n}(\vec{E}_{A}(x)\cdot\vec{n})
=\frac{1}{2}\vec{E}_{A}^{\,t}(x)+\vec{E}_{A}^{\,l}(x). \eqno (1.55)
$$

Unlike $\vec{E}_{\mathcal F}(x)$, however, $\vec{E}_{\mathcal G}(x)$ contains both the transverse and longitudinal components of the wave field variable $\vec{E}_{A}(x)$ in the zone under consideration, thereby predicting the possible existence of electrodynamic radiation with longitudinal polarization of the electric-field strength \cite[\S\,2]{Al2}, \cite[Part~I, Ch.~I, \S\,2]{Al8}.

Substituting (1.32) into equation (1.35) for the field variables determined by the component of the vector potential longitudinal with respect to $\vec{n}$,

$$
\partial_{0}\vec{E}_{\mathcal{G}^{\,l}}(x)+\vec{\bigtriangledown}\cdot\mathcal{G}^{\,l}(x)=\vec{0}, \eqno (1.56)
$$

where $\mathcal{G}^{\,l}(x)=\widetilde{def}\,\vec{A}^{\,l}(x)$, yields, in the wave zone, the relation

$$
\vec{E}_{\mathcal{G}^{\,l}}(x)=\vec{E}_{A}^{\,l}(x). \eqno (1.57)
$$

On the other hand, equation (1.56), together with the second equation of system (IV) \cite[\S\,14, Ch.~V]{Al0} for the wave field variables determined by the component of the vector potential longitudinal with respect to $\vec{n}$,

$$
\partial_{0}\mathcal{G}^{\,l}(x)=-\widetilde{def}\,\vec{E}_{\mathcal{G}^{\,l}}(x), \eqno (1.58)
$$

yields the differential form of the conservation law for the field energy of longitudinal electrojeitonic radiation represented by the wave field complex $\{\vec{E}_{\mathcal{G}^{l}}(x),\mathcal{G}^{l}(x)\}$, as well as the expressions for the energy density and energy flux density of this radiation determined by this law. These expressions are used below as the starting relations (3.1)--(3.3), \S\,3.

\subsection*{\S \, 2. Field Variables of Radiation Fields Carried by the Liénard--Wiechert Potentials}
\addcontentsline{toc}{subsection}{\S\, 2. Field Variables of Radiation Fields Carried by the Liénard--Wiechert Potentials}
\setcounter{section}{1}
\setcounter{figure}{0}

Following \cite{La}, we shall use the Gaussian system of units below, while continuing to restrict our consideration to the ``electrodynamics of vacuum and point electric charges.''

Carrying out the corresponding differentiations of the Liénard\,--\,Wiechert vector potential, we obtain the following asymptotic expressions \cite[\S17, Part II.]{Med} for the field variables $\partial A(x)$ and $\vec{E}_{A}(x)$, containing terms dependent on the acceleration of the particle under consideration \cite[\S63]{La},

$$
\partial A(x)_{a}=-\frac{e}{cR(1-\vec{n}\cdot\vec{\beta})^{3}}
\left\{(1-\vec{n}\cdot\vec{\beta})(\vec{n}\otimes\dot{\vec{\beta}})
+(\vec{n}\cdot\dot{\vec{\beta}})(\vec{n}\otimes\vec{\beta})\right\},
\eqno (2.1)
$$
$$
\vec{E}_{A}(x)_{a}=-\frac{e}{cR(1-\vec{n}\cdot\vec{\beta})^{3}}
\left\{(1-\vec{n}\cdot\vec{\beta})\dot{\vec{\beta}}
+(\vec{n}\cdot\dot{\vec{\beta}})\vec{\beta}\right\}.
\eqno (2.2)
$$
The subscript “$a$” will henceforth be omitted from the symbols denoting such expressions.

Substituting (1.32) into (1.13) immediately yields the field equation relating the wave field variables $\partial A(x)$ and $\vec{E}_{A}(x)$,
$$
\partial A(x)=\vec{n}\otimes\vec{E}_{A}(x),
\eqno (2.3)
$$
The same equation also follows by comparing (2.1) and (2.2).

Writing (2.2) in a form more convenient for further analysis,
$$
\vec{E}_{A}(x)=-\frac{e}{cR(1-\vec{n}\cdot\vec{\beta})^{3}}
\left\{\vec{n}\times(\dot{\vec{\beta}}\times\vec{n})
+\vec{n}\times(\vec{\beta}\times\dot{\vec{\beta}})
+(\vec{n}\cdot\dot{\vec{\beta}})\vec{n}\right\},
\eqno (2.4)
$$

we obtain the relations defining the transverse and longitudinal components of the electric-field strength,

$$
\vec{E}_{A}^{\,t}(x):=\vec{n}\times(\vec{E}_{A}(x)\times\vec{n})
=\frac{e}{cR(1-\vec{n}\cdot\vec{\beta})^{3}}
\left\{\vec{n}\times(\dot{\vec{\beta}}\times\vec{n})
+\vec{n}\times(\vec{\beta}\times\dot{\vec{\beta}})\right\},
\eqno (2.5)
$$
$$
\vec{E}_{A}^{\,\ell}(x):=\vec{n}\cdot(\vec{E}_{A}(x)\cdot\vec{n})
=-\frac{e}{cR(1-\vec{n}\cdot\vec{\beta})^{3}}
(\vec{n}\cdot\dot{\vec{\beta}})\vec{n}.
\eqno (2.6)
$$

Equation (2.5), in turn, yields the relation
$$
\vec{E}_{A}^{\,t}(x)=\frac{e}{cR(1-\vec{n}\cdot\vec{\beta})^{3}}
\left\{\vec{n}\times[(\vec{n}-\vec{\beta})\times\dot{\vec{\beta}}]\right\},
\eqno (2.7)
$$
whose right-hand side coincides with the right-hand side of the equation for the electric-field strength of the \emph{electromagnetic} radiation of a point electric charge in the framework of the traditional classical theory of electromagnetism, \cite{La}, \cite{Jack}.

Carrying out the corresponding differentiation of the Liénard\,--\,Wiechert scalar potential, we obtain the following expression for the corresponding contribution to the electric-field strength due to this potential,
$$
\vec{E}_{\varphi}(x)=\frac{e}{cR(1-\vec{n}\cdot\vec{\beta})^{3}}
(\vec{n}\cdot\dot{\vec{\beta}})\vec{n}.
\eqno (2.8)
$$
Comparing (2.8) and (2.6), we obtain the relation
$$
\vec{E}_{\varphi}(x)=-\vec{E}_{A}^{\,\ell}(x),
\eqno (2.9)
$$
which demonstrates, first and foremost, that the wave field variables $\vec{E}_{\varphi}(x)$ and $\vec{E}_{A}^{\,\ell}(x)$ are not independent field variables of the field system under consideration.

This fact must be taken into account when calculating the corresponding dynamical variables of this field system.

Using (2.4) and (2.8), we obtain the following expressions for the electric-field strengths defined by (1.20) and (1.23):
$$
\vec{E}_{\mathcal{F}}(x)=-\frac{e}{2cR(1-\vec{n}\cdot\vec{\beta})^{3}}
\left\{\vec{n}\times(\dot{\vec{\beta}}\times\vec{n})
+\vec{n}\times(\vec{\beta}\times\dot{\vec{\beta}})\right\},
\eqno (2.10)
$$
$$
\vec{E}_{\mathcal{G}}(x)=-\frac{e}{2cR(1-\vec{n}\cdot\vec{\beta})^{3}}
\left\{\vec{n}\times(\dot{\vec{\beta}}\times\vec{n})
+\vec{n}\times(\vec{\beta}\times\dot{\vec{\beta}})
+2(\vec{n}\cdot\dot{\vec{\beta}})\vec{n}\right\}.
\eqno (2.11)
$$

On the other hand, substituting (2.9) into (1.20) and (1.23) gives the following relations for the field strengths $\vec{E}_{\mathcal F}(x)$, $\vec{E}_{\mathcal G}(x)$, and $\vec{E}_{A}(x)$, equivalent to (1.50) and (1.55) in the radiation:
$$
\vec{E}_{\mathcal F}(x)=\frac{1}{2}\vec{E}_{A}^{\,t}(x),
\eqno (2.12)
$$
$$
\vec{E}_{\mathcal G}(x)=\frac{1}{2}\vec{E}_{A}^{\,t}(x)+\vec{E}_{A}^{\,\ell}(x):=
\vec{E}_{\mathcal G^{t}}(x)+\vec{E}_{\mathcal G^{\ell}}(x).
\eqno (2.13)
$$

Equations (2.13) and (2.11) demonstrate that, in the general case of an arbitrary radiation direction, the electric-field strength accompanying the jeitonic field in the electrojeitonic radiation of a point electric charge contains both a transverse component equal to $\vec{E}_{\mathcal F}(x)$ and a longitudinal component equal to $\vec{E}_{A}^{\,\ell}(x)$ \cite{Al2}, \cite{Al8}.

The presence of these two components in $\vec{E}_{\mathcal G}(x)$ is consistent with the fact that the jeitonic-field 3-tensor of the field system under consideration also has two components,
$\mathcal G^{\,t}(x):=\vec{n}\overset{s}{\otimes}\vec{E}_{A}^{\,t}(x)$
and
$\mathcal G^{\,\ell}(x):=\vec{n}\overset{s}{\otimes}\vec{E}_{A}^{\,\ell}(x)$,
which behave, with respect to the direction of radiation, identically to the components $\vec{E}_{\mathcal G^{t}}(x)$ and $\vec{E}_{\mathcal G^{\ell}}(x)$, respectively.

Thus, the \emph{electrojeitonic} radiation of a point electric charge is represented by two fundamentally different field complexes,
$\{\vec{E}_{\mathcal G^{t}}(x),\mathcal G^{\,t}(x)\}$
and
$\{\vec{E}_{\mathcal G^{\ell}}(x),\mathcal G^{\,\ell}(x)\}$,
corresponding, respectively, to the field complexes of the transverse and longitudinal \emph{electrojeitonic} waves generated by the charge.

The field complex $\{\vec{E}_{\mathcal F}(x),\mathcal F(x)\}$ represents the electromagnetic component of the radiation from the given source, in which the electric-field strength is transversely polarized and is expressed, by means of (2.12), in terms of the transverse component of $\vec{E}_{A}(x)$.

The collinearity of $\vec{E}_{\varphi}(x)$ and $\vec{n}$, established by (2.8), together with the identity
$$
\vec{a}\overset{a}{\otimes}\vec{b}
=\frac{1}{2}\varepsilon\cdot(\vec{a}\times\vec{b}),
\eqno (2.14)
$$
leads to the identity
$$
\vec{n}\overset{a}{\otimes}\vec{E}_{\varphi}(x)\equiv\tilde{0},
\eqno (2.15)
$$
and, consequently, relation (1.34) takes the form of a field equation relating the wave field variables $\mathcal F(x)$ and $\vec{E}_{\mathcal F}(x)$:
$$
\mathcal F(x)=2\vec{n}\overset{a}{\otimes}\vec{E}_{\mathcal F}(x).
\eqno (2.16)
$$

In turn, (2.16) and (2.12) demonstrate that, like $\vec{E}_{\mathcal F}(x)$, the 3-tensor $\mathcal F(x)$ is likewise expressed in terms of the transverse component of $\vec{E}_{A}(x)$.

Thus, to fully determine the \emph{electromagnetic} field of the source under consideration in its wave zone, it is sufficient to calculate only the electric-field strength determined solely by the vector potential and, moreover, only its transverse component.

Accordingly, to fully determine the \emph{electrojeitonic} field of the source under consideration in this zone, it is sufficient to calculate only the electric-field strength determined solely by the vector potential and, moreover, both its transverse and longitudinal components.

In turn, substituting (2.2) into (1.34) gives
$$
\mathcal F(x)=-\frac{e}{cR(1-\vec{n}\cdot\vec{\beta})^{3}}
\left\{(1-\vec{n}\cdot\vec{\beta})
(\vec{n}\overset{a}{\otimes}\dot{\vec{\beta}})
+(\vec{n}\cdot\dot{\vec{\beta}})
(\vec{n}\overset{a}{\otimes}\vec{\beta})\right\}.
\eqno (2.17)
$$
Taking the scalar product of both sides of (2.16) twice with $-\frac{1}{2}\varepsilon$ and then using (1.25), together with the identity
$$
rot\,\vec{a}=-\varepsilon\cdot\cdot\widetilde{rot}\,\vec{a},
\eqno (2.18)
$$
leads to the well-known relation between the field variables $\vec{\mathcal H}(x)$ and $\vec{E}_{\mathcal F}(x)$:
$$
\vec{\mathcal H}(x)=\vec{n}\times\vec{E}_{\mathcal F}(x).
\eqno (2.19)
$$
Substituting (2.2) into the field equation relating the wave field variables $\mathcal G(x)$ and $\vec{E}_{A}(x)$, given by (1.33), we obtain the following expression for the jeitonic-field 3-tensor:
$$
\mathcal G(x)=-\frac{e}{cR(1-\vec{n}\cdot\vec{\beta})^{3}}
\left\{(1-\vec{n}\cdot\vec{\beta})
(\vec{n}\overset{s}{\otimes}\dot{\vec{\beta}})
+(\vec{n}\cdot\dot{\vec{\beta}})
(\vec{n}\overset{s}{\otimes}\vec{\beta})\right\}.
\eqno (2.20)
$$

Using (1.4) in (1.10) gives the field equation relating the wave field variables $\mathcal G_{\ast}(x)$ and $\vec{E}_{A}(x)$:
$$
\mathcal G_{\ast}(x)=\, \frac{1}{tr\, g}\, (\vec{n}\cdot \vec{E}_{A}(x)\, )\, g\, .   \eqno (2.21)
$$
Substituting (2.4) into this relation yields the following expression for the 3-tensor of the dilatational jeitonic field:
$$
\mathcal \mathcal \mathcal G_{\ast}(x)=\, -\, \frac{e}{c\, R\, (1-\vec{n}\cdot\vec{\beta}\, )^{\, 3}}\, \, \frac{1}{tr\, g}\, (\vec{n}\cdot \dot{\vec{\beta}}\, )\, g\, .                                                           \eqno (2.22)
$$

In turn, using the definition of the linear invariant of the 3-tensor $\mathcal G(x)$ \cite{Al0},
$$
 \mathcal G^{\ast}(x)\, \eqdef \, tr\, \mathcal G(x)=\, tr\, \partial A(x)=\, div\, \vec{A}(x)\, ,       \eqno (2.23)
$$
it follows from (2.21) that the wave field variables $\mathcal G^{\ast}(x)$ and $\vec{E}_{A}(x)$ satisfy the field equation
$$
\mathcal G^{\ast}(x)=\vec{n}\cdot\vec{E}_{A}(x).
\eqno (2.24)
$$
Substituting (2.4) into (2.24) gives
$$
\mathcal G^{\ast}(x)=-\frac{e}{cR(1-\vec{n}\cdot\vec{\beta})^{3}}
(\vec{n}\cdot\dot{\vec{\beta}}).
\eqno (2.25)
$$

The field equation relating the 3-tensor $\mathcal{G}_{\ast}(x)$ and the linear invariant $\mathcal{G}^{\ast}(x)$ of the 3-tensor $\mathcal{G}(x)$, according to (1.10) and (2.23), takes the form
$$
\mathcal{G}_{\ast}(x)
=
\frac{\mathcal{G}^{\ast}(x)}{tr\,g}\,g.
\eqno (2.26)
$$

Using (1.32) and (2.9), we obtain
$$
\partial_{0}A_{0}(x)
=
-\bigl(\vec{n}\cdot\vec{E}_{A}(x)\bigr),
\eqno (2.27)
$$
which, in view of (2.23) and (2.24), leads to the Lorenz relation
$$
\partial_{0}A_{0}(x)
=
-div\,\vec{A}(x).
\eqno (2.28)
$$
However, in the present context, the latter does not serve as the traditional \emph{additional} condition imposed on the 4\nobreakdash-potential $A(x)$, but rather as a relation defining the ``universal'' connection between the field variables $\partial_{0}A_{0}(x)$ and $div\,\vec{A}(x)$ \emph{inherent} in the field system under consideration. It therefore constitutes an integrable differential relation between the components of the 4\nobreakdash-potential $A(x)$ of this field system, reducing the number of mutually linearly independent components of the potential from four to three, \cite{BSh}.

The corresponding representations of the 3-tensor of the deviatoric jeitonic field, $\mathcal{G}_{d}(x)$, defined by (1.9), are most easily obtained from the unique decomposition
$$
\mathcal{G}(x)
=
\mathcal{G}_{d}(x)
+
\mathcal{G}_{\ast}(x),
\eqno (2.29)
$$
which immediately yields the desired representations:
$$
\mathcal{G}_{d}(x)
=
\vec{n}\overset{s}{\otimes}\vec{E}_{A}(x)
-
\frac{1}{tr\,g}
\bigl(\vec{n}\cdot\vec{E}_{A}(x)\bigr)g,
\eqno (2.30)
$$
$$
\mathcal{G}_{d}(x)
=
-\frac{e}{cR(1-\vec{n}\cdot\vec{\beta})^{3}}
\left\{
(1-\vec{n}\cdot\vec{\beta})
(\vec{n}\overset{s}{\otimes}\dot{\vec{\beta}})
+
(\vec{n}\cdot\dot{\vec{\beta}})
(\vec{n}\overset{s}{\otimes}\vec{\beta})
-
\frac{1}{tr\,g}
(\vec{n}\cdot\dot{\vec{\beta}})g
\right\}.
\eqno (2.31)
$$

The decomposition of the electric-field strength $\vec{E}_{A}(x)$ into the components $\vec{E}_{A}^{\,t}(x)$ and $\vec{E}_{A}^{\,\ell}(x)$, defined by (2.4), induces corresponding decompositions of both the field variables $\vec{E}_{\mathcal{F}}(x)$ and $\vec{E}_{\mathcal{G}}(x)$, given by (2.12) and (2.13), and the field variables $\mathcal{F}(x)$, $\mathcal{G}(x)$, $\mathcal{G}_{d}(x)$, and $\mathcal{G}_{\ast}(x)$, considered below.

The decomposition of the 3-tensor of the magnetic field under consideration
$$
\mathcal{F}(x)
=
\vec{n}\overset{a}{\otimes}\vec{E}_{A}^{\,t}(x)
+
\vec{n}\overset{a}{\otimes}\vec{E}_{A}^{\,\ell}(x)
:=
\mathcal{F}^{\,t}(x)
+
\mathcal{F}^{\,\ell}(x),
\eqno (2.32)
$$
upon using the identity
$$
\mathcal{F}^{\,\ell}(x)
\equiv
\tilde{0},
\eqno (2.33)
$$
takes the form
$$
\mathcal{F}(x)
=
\vec{n}\overset{a}{\otimes}\vec{E}_{A}^{\,t}(x)
=
\mathcal{F}^{\,t}(x).
\eqno (2.34)
$$

The corresponding decomposition of the 3-tensor of the jeitonic field
$$
\mathcal{G}(x)
=
\vec{n}\overset{s}{\otimes}\vec{E}_{A}^{\,t}(x)
+
\vec{n}\overset{s}{\otimes}\vec{E}_{A}^{\,\ell}(x)
:=
\mathcal{G}^{\,t}(x)
+
\mathcal{G}^{\,\ell}(x),
\eqno (2.35)
$$
consists, in general, of two terms, both different from $\tilde{0}$.

Using (2.34) and (2.35) when calculating the field strengths $\vec{E}_{\mathcal{F}}(x)$ and $\vec{E}_{\mathcal{G}}(x)$ from (1.44) and (1.35), respectively, also yields (2.12) and (2.13), thus demonstrating once again the equivalence of the methods for calculating these field strengths discussed in \S\,1.

For the 3-tensor of the dilatational jeitonic field, $\mathcal{G}_{\ast}(x)$, we have, according to (2.21), the obvious relation
$$
\mathcal{G}_{\ast}(x)
=
\frac{1}{tr\,g}
\bigl(\vec{n}\cdot\vec{E}_{A}^{\,\ell}(x)\bigr)g
:=
\mathcal{G}_{\ast}^{\,\ell}(x).
\eqno (2.36)
$$

Using (2.29), (2.35), and (2.36) yields the corresponding representation of the 3-tensor of the deviatoric jeitonic field
$$
\mathcal{G}_{d}(x)
=
\vec{n}\overset{s}{\otimes}\vec{E}_{A}^{\,t}(x)
+
\vec{n}\overset{s}{\otimes}\vec{E}_{A}^{\,\ell}(x)
-
\frac{1}{tr\,g}
\bigl(\vec{n}\cdot\vec{E}_{A}^{\,\ell}(x)\bigr)g
:=
\mathcal{G}_{d}^{\,t}(x)
+
\mathcal{G}_{d}^{\,\ell}(x),
\eqno (2.37)
$$
whose terms on the right-hand side, in general, are also different from $\tilde{0}$, as in (2.35).

\subsection*{\S \, 3. Energy Density and Energy Flux Density of the Electrodynamic Radiation of a Point Electric Charge}
\addcontentsline{toc}{subsection}{\S\, 3. Energy Density and Energy Flux Density of the Electrodynamic Radiation of a Point Electric Charge}
\setcounter{section}{1}
\setcounter{figure}{0}

At any given time, the electrodynamic radiation from a point electrically charged particle undergoing arbitrary motion, as registered by a detector located at point $N$ in configuration space and stationary with respect to the chosen inertial reference frame, is considered as a superposition of coherent radiations associated with the transverse and longitudinal components of the particle's acceleration relative to $\vec{n}$ \cite{Jack}.

\subsubsection*{3.1. Radiation Due to the Longitudinal Component of Acceleration}
\addcontentsline{toc}{subsection}{3.1. Radiation Due to the Longitudinal Component of Acceleration}
\setcounter{section}{1}
\setcounter{figure}{0}

The expressions for the instantaneous values of the energy density and the energy-flux density vector of the radiation detected at the point $N$ under consideration and due to the longitudinal component of the acceleration, $\dot{\vec{\beta}}^{\,\ell}$, as well as the corresponding continuity equation for the energy-flux density of the associated electrojeitonic field, are given, according to \cite[(17.2)--(17.4)]{Al0}, by the relations\footnote[1]{Relation (3.2), together with the formula for the radiation pattern of the energy-flux density of the longitudinal electrojeitonic radiation due to the electric dipole moment of the system, was obtained in \cite{Al6}.}
$$
W_{\mathcal G^{\ell}}(x)=\frac{1}{8\pi}
\left(E_{A}^{\,\ell}(x)^{2}+\mathcal G^{\ell}(x)^{2}\right),
\eqno (3.1)
$$
$$
\vec{S}_{\mathcal G^{\ell}}(x)=\frac{c}{4\pi}\,
\mathcal G^{\ell}(x)\cdot\vec{E}_{A}^{\,\ell}(x),
\eqno (3.2)
$$
$$
\partial_{t}W_{\mathcal G^{\ell}}(x)+
div\,\vec{S}_{\mathcal G^{\ell}}(x)=0.
\eqno (3.3)
$$

For notational convenience, the arguments of the field and dynamical variables will henceforth be omitted from the formulas.

According to (2.35), we have
$$
\mathcal G^{\, \ell}\, =\, \vec{n}\, \overset{s}{\otimes}\, \vec{E}_{A}^{\, \ell}\, ,         \eqno (3.4)
$$
which, in view of the identities
$$
\vec{a}\overset{s}{\otimes}\vec{b}
=
\vec{a}\otimes\vec{b},
\quad \mbox{for } \vec{a}\, coll\, \vec{b},
\eqno (3.5)
$$
$$
(\vec{a}\otimes\vec{b})\cdot\vec{c}
=
\vec{a}(\vec{b}\cdot\vec{c}),
\eqno (3.6)
$$
rewrites (3.2) as
$$
\vec{S}_{\mathcal G^{\, \ell}}\, =\, \frac{c}{4\, \pi}\, \, E_{A}^{\, \ell}\, ^{2}\, \vec{n}\, .
\eqno (3.7)
$$

Using the relation
$$
(\vec{a}\otimes\vec{b})^{2}
:=
(\vec{a}\otimes\vec{b})\mathbin{\cdot\cdot}
(\vec{a}\otimes\vec{b})^{T}
=
a^{2}b^{2},
\eqno (3.8)
$$

which, according to (3.4), implies
$$
\mathcal G^{\, \ell}\, ^{2}\, =\, E_{A}^{\, \ell}\, ^{2}\, ,
\eqno (3.9)
$$
we obtain the following compact expressions for the energy density of the electrojeitonic radiation represented by the ``longitudinal'' field complex
$\{\vec{E}_{A}^{\, \ell}(x), \mathcal G^{\, \ell}(x)\}$:
$$
W_{\mathcal G^{\ell}}
=
\frac{1}{4\pi}E_{A}^{\ell\,2}
=
\frac{1}{4\pi}\mathcal G^{\ell\,2}.
\eqno (3.10)
$$

This allows (3.7) to be rewritten in the standard form \cite[(4.75), \S47]{La}:
$$
\vec{S}_{\mathcal G^{\, \ell}}\, =\, c\, W_{\mathcal G^{\ell}}\, \vec{n}\, =\, W_{\mathcal G^{\, \ell}}\, \vec{c}\, .
\eqno (3.11)
$$

Subsequent use of (3.11) in (3.3) yields the differential form of the energy conservation law for this type of radiation:
$$
\partial_{t}W_{\mathcal G^{\ell}}
+
div\left(W_{\mathcal G^{\ell}}\vec{c}\right)
=
0.
\eqno (3.12)
$$

In turn, (3.9) allows (3.7) to be expressed as
$$
\vec{S}_{\mathcal G^{\ell}}
=
\frac{c}{4\pi}\mathcal G^{\ell\,2}\vec{n}.
\eqno (3.13)
$$

After decomposing the right-hand side of (3.2) into the components defining the energy-flux density vectors associated with the $\mathcal G_{d}^{\ell}(x)$ and $\mathcal G_{\ast}(x)$ fields,
$$
\vec{S}_{\mathcal G^{\ell}}
=
\frac{c}{4\pi}\mathcal G_{d}^{\ell}\cdot\vec{E}_{A}^{\ell}
+
\frac{c}{4\pi}\mathcal G_{\ast}\cdot\vec{E}_{A}^{\ell}
:=
\vec{S}_{\mathcal G_{d}^{\ell}}
+
\vec{S}_{\mathcal G_{\ast}},
\eqno (3.14)
$$
and subsequently using (2.21), together with the relations
$$
g\cdot\vec{a}
=
\vec{a},
$$
$$
(\vec{a}\cdot\vec{b})\vec{b}
=
b^{2}\vec{a},
\quad \mbox{for } \vec{a}\, coll\, \vec{b},
$$
$$
tr\,g=3,
$$
the second term on the right-hand side of (3.14) takes the form
$$
\vec{S}_{\mathcal G_{\ast}}
=
\frac{1}{3}\frac{c}{4\pi}
E_{A}^{\ell\,2}\vec{n}.
\eqno (3.15)
$$

This provides a compact expression for the energy-flux density vector associated with the \emph{dilatational} jeitonic field.

The first term on the right-hand side of (3.14), using (3.7), (3.14), and (3.15), can be expressed as
$$
\vec{S}_{\mathcal G_{d}^{\ell}}
=
\frac{2}{3}\frac{c}{4\pi}
E_{A}^{\ell\,2}\vec{n}.
\eqno (3.16)
$$
This provides a compact expression for the energy-flux density vector associated with the \emph{deviatoric} jeitonic field.

Comparison of (3.15) and (3.16) yields the following relation between the energy-flux density vectors under consideration:
$$
\vec{S}_{\mathcal G_{d}^{\ell}}
=
2\vec{S}_{\mathcal G_{\ast}}.
\eqno (3.17)
$$

On the other hand, using (2.36), together with the collinearity of the vectors $\vec{E}_{A}^{\,\ell}$ and $\vec{n}$, expressed by (2.6), we find
$$
\mathcal G_{\ast}^{\,2}
= \frac{1}{(tr\, g)^{2}}
(\vec{n}\cdot\vec{E}_{A}^{\,\ell})^{2}\,g\cdot\cdot g
= \frac{1}{3}\,E_{A}^{\,\ell\,2}.
\eqno (3.18)
$$
In turn, using the orthogonality of the terms on the right-hand side of the decomposition
$$
\mathcal G^{\,\ell}
= \mathcal G_{d}^{\,\ell} + \mathcal G_{\ast},
\eqno (3.19)
$$
we obtain the relation
$$
\mathcal G_{d}^{\,\ell\,2}
= \mathcal G^{\,\ell\,2} - \mathcal G_{\ast}^{\,2},
\eqno (3.20)
$$
which, upon using (3.9) and (3.18), gives
$$
\mathcal G_{d}^{\,\ell\,2}
= \frac{2}{3}\,E_{A}^{\,\ell\,2}.
\eqno (3.21)
$$

Separating out in (3.1) the energy densities associated with the $\mathcal G_{d}^{\,\ell}$- and $\mathcal G_{\ast}$-fields gives
$$
W_{\mathcal G^{\ell}}
= \frac{1}{8\pi}\,E_{A}^{\,\ell\,2}
+ \frac{1}{8\pi}\,\mathcal G_{d}^{\,\ell\,2}
+ \frac{1}{8\pi}\,\mathcal G_{\ast}^{\,2}
:= W_{E_{A}^{\,\ell}} + W_{\mathcal G_{d}^{\,\ell}} + W_{\mathcal G_{\ast}},
\eqno (3.22)
$$
which demonstrates that the volume energy density carried by the longitudinal electrojeitonic wave is the sum of the volume energy densities of the electric and jeitonic fields of the longitudinal field complex
$\{\vec{E}_{A}^{\,\ell}(x),\mathcal G_{d}^{\,\ell}(x),\mathcal G_{\ast}(x)\}$.

Using (3.18) and (3.21), we obtain the following relations for the energy densities represented by the second and third terms on the right-hand side of (3.22):
$$
W_{\mathcal G_{d}^{\,\ell}}
= \frac{2}{3}\,W_{E_{A}^{\,\ell}},
\eqno (3.23)
$$
$$
W_{\mathcal G_{\ast}}
= \frac{1}{3}\,W_{E_{A}^{\,\ell}}.
\eqno (3.24)
$$
Comparison of (3.23) and (3.24), in accordance with (3.17), yields the relation between the energy densities of the deviatoric and dilatational jeitonic fields in the longitudinal electrojeitonic radiation of a point electric charge:
$$
W_{G_{d}^{\ell}}\, =2\, W_{\mathcal G_{\ast}}\, .                         \eqno (3.25)
$$

In turn, (3.18) and (3.21) allow (3.15) and (3.16) to be expressed in the form
$$
\vec{S}_{\mathcal G_{\ast}}\, =\, \frac{c}{4\, \pi}\, \, \mathcal G_{\ast}^{\, 2}\, \vec{n}\, , 
                                                                                    \eqno (3.26)
$$
$$
\vec{S}_{\mathcal G_{d}^{\ell}}\, =\, \frac{c}{4\, \pi}\, \, \mathcal G_{d}^{\ell}\, ^{2}\, \vec{n}\, .
                                                                                    \eqno (3.27)
$$

These relations can also be obtained directly by substituting (3.19) into (3.13) and then taking into account the orthogonality of the decomposition (3.19).

Using the relation
$$
\mathcal G^{\ast\,2}
= E_{A}^{\,\ell\,2},
\eqno (3.28)
$$
which follows from (2.24), the energy-flux density vectors (3.15) and (3.16) can likewise be expressed as
$$
\vec{S}_{\mathcal G_{\ast}}\, =\, \frac{1}{3}\, \, \frac{c}{4\, \pi}\, \, \mathcal G^{\ast}\, ^{2}\, \vec{n}\, , 
                                                                                    \eqno (3.29)
$$
$$
\vec{S}_{\mathcal G_{d}^{\ell}}\, =\, \frac{2}{3}\, \, \frac{c}{4\, \pi}\, \, \mathcal G^{\ast}\, ^{2}\, \vec{n}\, .
                                                                                    \eqno (3.30)
$$

Relation (3.28) also allows (3.7) and (3.10) to be expressed in terms of the linear invariant of the 3-tensor $\mathcal G(x)$:
$$
\vec{S}_{\mathcal G^{\, \ell}}\, =\, \frac{c}{4\, \pi}\, \, \mathcal G^{\ast}\, ^{2}\, \vec{n}\, , 
                                                                                    \eqno (3.31)
$$
$$
W_{\mathcal G^{\ell}}\, =\, \frac{1}{4\, \pi}\, \mathcal G^{\ast}\, ^{2}\, .               \eqno (3.32)
$$

Equations (3.31) and (3.32) show that the total energy-flux density and the total energy density of the longitudinal electrojeitonic radiation of a point electrically charged particle can also be calculated using these relations.

On the other hand, using (2.6) in (3.7) gives the following expression for the energy-flux density vector of the longitudinal electrojeitonic radiation of a point electrically charged particle, represented by the above-mentioned longitudinal field complex:
$$
\vec{S}_{\mathcal G^{\, \ell}}\, =\, \frac{e^{2}}{4\pi c\, R^{\, 2}\, (1-\vec{n}\cdot\vec{\beta}\, )^{\, 6}}\, \, |\, \vec{n}\cdot \dot{\vec{\beta}}\, |^{\, 2}\, \vec{n}\, .
                                                                                            \eqno (3.33)
$$

In turn, the momentum density of the electrojeitonic field under consideration is given by
$$
\vec{p}_{\mathcal G^{\ell}}\, =\, \frac{1}{c^{\, 2}}\, \vec{S}_{\mathcal  G^{\, \ell}}\, =\, \frac{1}{c}\, W_{\mathcal G^{\ell}}\, \vec{n}\, .                    \eqno (3.34)
$$

As in \cite{La}, we note that the relation between the volume energy density and volume momentum density of this field is the same as that for particles moving at the speed of light.

The presence of momentum and momentum-flux density in the wave electrojeitonic field under consideration implies that this radiation exerts pressure on a surface that reflects or absorbs this type of radiation \cite{La}.

\subsubsection*{3.2. Radiation due to the Transverse Component of Acceleration}
\addcontentsline{toc}{subsection}{3.2. Radiation due to the Transverse Component of Acceleration}
\setcounter{section}{1}
\setcounter{figure}{0}

The expressions for the instantaneous energy density and energy-flux density vector of the radiation observed at the point $N$ under consideration and due to the transverse component of acceleration, $\dot{\vec{\beta}}^{\,t}$, as well as the corresponding continuity equation for the energy flux density of the electromagnetic-jeitonic field, are given, according to \cite[(13.12)--(13.14)]{Al0}, by
$$
W_{\mathcal F \, \mathcal G_{d}^{\, t}}\, =\, \frac{1}{8\, \pi}\, (E_{A}^{\, \, t}\, ^{\, 2}\, +\, \mathcal F^{\, 2}\, +\mathcal G_{d}^{\, t}\, ^{\, 2})\, ,                                      \eqno (3.35)
$$
$$
\vec{S}_{\mathcal F \, \mathcal G_{d}^{\, t}}\, =\, \frac{c}{4\, \pi}\, \, \mathcal F \cdot\vec{E}_{A}^{\, t}\, +\, \frac{c}{4\, \pi}\, \, \mathcal G_{d}^{\, t} \cdot\vec{E}_{A}^{\, t}:=\, \vec{S}_{\mathcal F}\, +\, \vec{S}_{\mathcal G_{d}^{\, t}}\, , \eqno (3.36)
$$
$$
\partial_{\, t}\, W_{\mathcal F \, \mathcal G_{d}^{\, t}}\, +\, div\, \vec{S}_{\mathcal F \, \mathcal G_{d}^{\, t}}\, =\, 0\, .                                                                      \eqno (3.37)
$$

Equation (3.35) shows that the volume energy density carried by the transverse electromagnetic-jeitonic wave is the sum of the energy densities of the electric, magnetic, and jeitonic fields defined by the transverse wave complex
$\{\vec{E}_{A}^{\,t}(x),\mathcal F(x),\mathcal G_{d}^{\,t}(x)\}$.

Using (2.34) and (2.37), followed by the use of the identities
$$
\vec{a}\cdot\, (\vec{b}\overset{a}{\otimes}\vec{c}\, )=
-\, \frac{1}{2}\, (\vec{b}\, (\vec{a}\cdot\vec{c}\, )-\vec{c}\, (\vec{a}\cdot\vec{b}\, ))\, ,   \eqno (3.38)
$$
$$
\vec{a}\cdot\, (\vec{b}\overset{s}{\otimes}\vec{c}\, )=
\, \frac{1}{2}\, (\vec{b}\, (\vec{a}\cdot\vec{c}\, )+\vec{c}\, (\vec{a}\cdot\vec{b}\, ))\, ,   \eqno (3.39)
$$
yields the following expressions for the energy-flux density vectors corresponding to the terms on the right-hand side of (3.36):
$$
\vec{S}_{\mathcal F}\, =\, \frac{c}{8\, \pi}\, E_{A}^{\, \, t}\, ^{\, 2}\, \vec{n}\, =\, \vec{S}_{\mathcal G_{d}^{\, t}}\, .                                                                            \eqno (3.40)
$$

Using (2.5) further, (3.40) can be written as
$$
\vec{S}_{\mathcal F}\, =\frac{e^{\, 2}}{8\, \pi c\, R^{\, 2}\, (1-\vec{n}\cdot\vec{\beta}\, )^{\, 6}}\, 
|\, \vec{n}\times (\dot{\vec{\beta}}\times \vec{n}\, )\, +\, \vec{n}\times (\vec{\beta}\times \dot{\vec{\beta}}\, )|^{\, 2}\, \vec{n}\, =\, \vec{S}_{\mathcal G_{d}^{\, t}}\, .                  \eqno (3.41)
$$

In turn, successive substitution of (3.40) and (3.41) into (3.36) yields the following expressions for the energy-flux density vector carried by the transverse field complex
$\{\vec{E}_{A}^{\,t}(x),\mathcal F(x),\mathcal G_{d}^{\,t}(x)\}$:
$$
\vec{S}_{\mathcal F \, \mathcal G_{d}^{\, t}}\, =
\, \frac{c}{4\, \pi}\, E_{A}^{\, \, t}\, ^{\, 2}\, \vec{n}\, ,                                     \eqno (3.42)
$$
$$
\vec{S}_{\mathcal F \, \mathcal G_{d}^{\, t}}\, =\, \frac{e^{\, 2}}{4\, \pi c\, R^{\, 2}\, (1-\vec{n}\cdot\vec{\beta}\, )^{\, 6}}\, 
|\, \vec{n}\times (\dot{\vec{\beta}}\times \vec{n}\, )\, +\, \vec{n}\times (\vec{\beta}\times \dot{\vec{\beta}}\, )|^{\, 2}\, \vec{n}\, .                                                   \eqno (3.43)
$$

Using (2.7) in (3.42), we obtain
$$
\vec{S}_{\mathcal F \, \mathcal G_{d}^{\, t}}\, =\, \frac{e^{\, 2}}{4\, \pi c\, R^{\, 2}\, (1-\vec{n}\cdot\vec{\beta}\, )^{\, 6}}\, 
|\, \vec{n}\times \, [(\vec{n}-\vec{\beta})\times \dot{\vec{\beta}}\, ]\, |^{\, 2}\, \vec{n}\, , 
                                                                                             \eqno (3.44)
$$

whose right-hand side coincides with the right-hand side of the relation defining the energy-flux density vector of the electromagnetic radiation from a point electric charge obtained within the framework of the standard model of classical electrodynamics \cite{Jack}.

On the other hand, again using (2.34) and (2.37), followed by the use of the relations
$$
(\vec{a}\otimes\vec{b}\, )\, \cdot\cdot\, (\vec{a}\otimes\vec{b}\, )^{T}\, =\, a^{\, 2}\, b^{\, 2}\, ,  \eqno (3.45)
$$
$$
(\vec{a}\otimes\vec{b}\, )\, \cdot\cdot\, (\vec{a}\otimes\vec{b}\, )\, =\, (\vec{a}\cdot\vec{b}\, )^{\, 2}\, , 
                                                                                       \eqno (3.46)
$$
we obtain
$$
\mathcal F^{\, 2}\, =\, \frac{1}{2}\, E_{A}^{\, \, t}\, ^{\, 2}\, =\, \mathcal G_{d}^{\, t}\, ^{\, 2}\, , \eqno (3.47)
$$

as a result of which (3.35) takes the following form, convenient for calculations:
$$
W_{\mathcal F \, \mathcal G_{d}^{\, t}}\, =\, \frac{1}{4\, \pi}\, E_{A}^{\, \, t}\, ^{\, 2}\, .      \eqno (3.48)
$$

Further use of (3.48) in (3.42) leads to the standard relation analogous to (3.11):
$$
\vec{S}_{\mathcal F \, \mathcal G_{d}^{\, t}}\, =\, c\, W_{\mathcal F \, \mathcal G_{d}^{\, t}}\, \vec{n}\, =\, W_{\mathcal F \, \mathcal G_{d}^{\, t}}\, \vec{c}\, .
                                                                                             \eqno (3.49)
$$

Further use of (3.49) in (3.37) leads to the differential form of the energy conservation law for this type of radiation:
$$
\partial_{\, t}\, W_{\mathcal F \, \mathcal G_{d}^{\, t}}\, +\, div\, (W_{\mathcal F \, \mathcal G_{d}^{\, t}}\, \vec{c}\, )\, =\, 0\, .                                                          \eqno (3.50)
$$

\subsubsection*{3.3. Dynamic Variables and Corresponding Conservation Laws for the Total Radiation of a Point Electric Charge}
\addcontentsline{toc}{subsection}{3.3. Dynamic Variables and Corresponding Conservation Laws for the Total Radiation of a Point Electric Charge}
\setcounter{section}{1}
\setcounter{figure}{0}

Summing (3.7) and (3.42) gives the expression for the total energy-flux density of the radiation under consideration:
$$
\vec{S}:=\, \vec{S}_{\mathcal G^{\, \ell}}\, +\, \vec{S}_{\mathcal F \, \mathcal G_{d}^{\, t}}\, =\, \frac{c}{4\, \pi}\, \, E_{A}^{\, \, \, 2}\, \, \vec{n}\, , 
                                                                                    \eqno (3.51)
$$

Using (2.4), this takes the form
$$
\vec{S}\, =\, \frac{e^{\, 2}}{4\, \pi c\, R^{\, 2}\, (1-\vec{n}\cdot\vec{\beta}\, )^{\, 6}}\, 
|\, \vec{n}\times (\dot{\vec{\beta}}\times \vec{n}\, )\, +\, \vec{n}\times (\vec{\beta}\times \dot{\vec{\beta}}\, )\, +\, (\vec{n}\cdot \dot{\vec{\beta}}\, )\, \vec{n}\, |^{\, 2}\, \vec{n}\, , 
\eqno (3.52)
$$

It can also be written in the more compact form as
$$
\vec{S}\, =\, \frac{e^{\, 2}}{4\, \pi c\, R^{\, 2}\, (1-\vec{n}\cdot\vec{\beta}\, )^{\, 6}}\, 
|\, \dot{\vec{\beta}}\, +\, \vec{n}\times (\vec{\beta}\times \dot{\vec{\beta}}\, )\, |^{\, 2}\, \vec{n}\, .
\eqno (3.53)
$$

The right-hand side of (3.51) appears to coincide with the expression for the energy-flux density of plane electromagnetic waves obtained in \cite{La} under the corresponding additional conditions imposed on the components of the 4\nobreakdash-potential of such waves.

Summing (3.10) and (3.48) gives the expression for the total energy density of the radiation under consideration:
$$
W:=\, W_{\mathcal G^{\ell}}\, +\, W_{\mathcal F \, \mathcal G_{d}^{\, t}}\, =\, \frac{1}{4\, \pi}\, E_{A}^{\, \, 2}\, .
\eqno (3.54)
$$

This allows (3.51) to be written in the form
$$
\vec{S}\, =\, c\, W\, \vec{n}\, =\, W\, \vec{c}\, .
\eqno (3.55)
$$
Further use of (3.12) and (3.50) leads to the differential form of the energy conservation law for the field system under consideration:
$$
\partial_{\, t}\, W\, +\, div\, (W\, \vec{c}\, )\, =\, 0\, .  \eqno (3.56)
$$

As in \cite{La}, we integrate (3.56) over a closed, fixed region $\bar{\mathfrak{D}}$ of configuration space, at every point of which this equation holds. Applying the Gauss--Ostrogradsky theorem to the second term of the resulting equation and then using relation (9) from \cite{Nov}, we obtain the standard integral form of the energy-change law \cite{Nev} for the radiation under consideration within the fixed region $\bar{\mathfrak{D}}$:
$$
\frac{d}{d\, t}\, W_{\bar{\mathfrak{D}}}\, :=\, \frac{d}{d\, t}\, \int_{\bar{\mathfrak{D}}}W\, d\, V\, =
\, -\, \int_{\partial\, \bar{\mathfrak{D}}}\vec{S}\cdot\vec{n}\, \, dS\, , 
$$
with its corresponding standard interpretation.

On the other hand, using the definition of the integral invariant of a dynamical system \cite{Nem}, according to which the density of the integral invariant is the volume energy density $W(x)$, which identically satisfies (3.56), we obtain the Poincaré form of the integral energy conservation law:
$$
W_{\bar{\mathfrak{D}}_{t}}\, :=\, \int_{\bar{\mathfrak{D}}_{t}}W\, d\, V\, =
\, \int_{\bar{\mathfrak{D}}_{t_{0}}}W\, d\, V\, .
$$

In this relation, $\bar{\mathfrak{D}}_{t}$ is the closed region of configuration space occupied by the continuum of points at time $t$, which occupied the region $\bar{\mathfrak{D}}_{t_{0}}$ at the initial time $t_{0}$ \cite{Nem}.

The motion of the continuum points is determined by the solutions of the corresponding normal system of differential equations of the dynamical system describing the propagation of the classical field continuum in configuration space \cite{Sed}.

In the case of a continuum modeling a classical mechanical medium, such a region is referred to as a material region or an individual ``volume'' \cite{Sed}.

Thus, this form of the conservation law can be interpreted as the conservation of the energy ``contained'' in the moving region $\bar{\mathfrak{D}}_{t}$, which accompanies the field under consideration as it propagates through configuration space.

Analogous integral forms of the energy-change and energy-conservation laws for the longitudinal electrojeitonic and transverse electromagnetic-jeitonic radiation follow from the corresponding differential forms (3.12) and (3.50).

Using (2.3), (3.6), and (3.45) yields
$$
\partial A\cdot\vec{E}_{A}\, =\, E_{A}^{\, \, 2}\, \, \vec{n}\, ,                  \eqno (3.57)
$$
$$
\partial A ^{\, 2}\, =\, E_{A}^{\, \, 2}\, .                                  \eqno (3.58)
$$

These relations allow (3.51) and (3.54) to be written as
$$
\vec{S}\, =\, \frac{c}{4\, \pi}\, \, \partial A\cdot\vec{E}_{A}\, ,                    \eqno (3.59)
$$
$$
W\, =\, \frac{1}{8\, \pi}\, (\, E_{A}^{\, \, 2}\, +\, \partial A ^{\, 2}\, )\, .               \eqno (3.60)
$$

In turn, they show that the dynamic variables $\vec{S}(x)$ and $W(x)$ can be regarded as dynamic variables of the wave field complex $\{\vec{E}_{A}(x),\partial A(x)\}$.

To further justify this interpretation, we derive (3.59) and (3.60) by the conventional method \cite{La}, using the field equations involving the field variables of this complex.

These equations are the inhomogeneous d'Alembert equation for the vector potential, which, upon using relation (1.37), takes the form
$$
\partial^{\, 0}\vec{E}_{A}\, +\, \vec{\bigtriangledown}\cdot\partial A\, =\, -\, \frac{4\, \pi}{c}\, \vec{\jmath}\, ,                                       \eqno (3.61)
$$
and the differential form of the law of electrodynamic induction, given by \cite{Al0}
$$
\partial^{\, 0}\partial A\, =\, -\, \partial E_{A}\, ,                         \eqno (3.62)
$$
where $\partial E_{A}(x):=\vec{\bigtriangledown}\otimes\vec{E}_{A}(x)$ is the 3-gradient of the 3-vector field $\vec{E}_{A}(x)$.

Taking the scalar product of both sides of (3.61) with $\vec{E}_{A}(x)$, and of both sides of (3.62) with the 3-tensor $\partial A(x)$, and then adding the left- and right-hand sides of the resulting relations, we obtain
$$
\partial^{\, 0}\, \frac{ E_{A}^{\, 2}+\partial A^{\, 2}}{2}\, +\, (\vec{\bigtriangledown}\cdot\partial A\, )\cdot\vec{E}_{A}\, +\, (\, \partial E_{A}, \, \partial A\, )=\, -\, \frac{4\, \pi}{c}\, \vec{\jmath}\cdot\vec{E}_{A}\, .                  \eqno (3.63)
$$

Further use of the well-known tensor-analysis identity
$$
\vec{\bigtriangledown}\cdot(T\cdot\vec{a})
=(\vec{\bigtriangledown}\cdot T)\cdot\vec{a}
+(T,\vec{\bigtriangledown}\otimes\vec{a}),
\eqno (3.64)
$$
leads to
$$
\partial_{\, t}\, \frac{ E_{A}^{\, 2}+\partial A^{\, 2}}{8\, \pi}\, +\, \frac{c}{4\, \pi}\, \vec{\bigtriangledown}\cdot(\, \partial A \cdot\vec{E}_{A})=\, -\, \vec{\jmath}\cdot\vec{E}_{A}\, ,                  \eqno (3.65)
$$
which, in particular, defines the corresponding dynamic variables of the field complex $\{\vec{E}_{A}(x),\partial A(x)\}$:
$$
\vec{S}_{\partial A}
=\frac{c}{4\pi}\partial A\cdot\vec{E}_{A},
\eqno (3.66)
$$
$$
W_{\partial A}
=\frac{1}{8\pi}(E_{A}^{2}+\partial A^{2}),
\eqno (3.67)
$$
whose right-hand sides coincide with those of (3.59) and (3.60), respectively.

In view of the mutual orthogonality of the terms on the right-hand side of the decomposition (1.16), (3.60) takes the form
$$
W_{\partial A}\, =\, \frac{1}{8\, \pi}\, (\, E_{A}^{\, \, 2}\, +\, \mathcal {F} ^{\, 2}+\, \mathcal G_d\, ^{\, 2}+\, \mathcal G_*\, ^{\, 2}\, )\, .                                                                             \eqno (3.68)
$$

This explicitly demonstrates that, in the general case, the instantaneous energy density of the radiation from a point electric charge consists not only of the energy density of the electric field defined by $\vec{E}_{A}(x),$ but also of the sum of the energy densities of the magnetic, deviatoric jeitonic, and dilatational jeitonic fields defined by the wave field complex $\{\mathcal{F}(x),\mathcal{G}_{d}(x),\mathcal{G}_{\ast}(x)\}$. Accordingly, within the context considered here, this radiation is referred to as \emph{electromagnetic-jeitonic} radiation.

Using (1.16) in (3.59) gives the instantaneous energy-flux density vector of the radiation field under consideration in the form
$$
\vec{S}_{\partial A}\, =\, \frac{c}{4\, \pi}\, (\mathcal {F}\cdot\vec{E}_{A}+
\mathcal G_d\cdot\vec{E}_{A}+\mathcal G_*\cdot \vec{E}_{A})\, .                   \eqno (3.69)
$$

This expression explicitly reveals the field content of the energy-flux density of the radiation from a point electric charge.

In view of the linear dependence of the terms on the right-hand sides of (3.68) and (3.69), these relations simplify to (3.54) and (3.51), respectively, providing a simpler, \emph{quantitatively equivalent way} of calculating these dynamic variables.

To simplify the subsequent transformations, it is convenient to rewrite (3.69) as
$$
\vec{S}_{\partial A}\, =\, \frac{c}{4\, \pi}\, (\mathcal {F}\cdot\vec{E}_{A}+
\mathcal G\cdot\vec{E}_{A})\, .                                                \eqno (3.70)
$$

By virtue of (2.33), the first term in the parentheses on the right-hand side of this relation can be written as
$$
\mathcal {F}\cdot\vec{E}_{A}=\, \mathcal {F}^{\, t}\cdot\vec{E}_{A}^{\, \, t}+\, \mathcal {F}^{\, t}\cdot\vec{E}_{A}^{\, \ell}\, .                                           \eqno (3.71)
$$

The second term in the parentheses, upon using the identity
$$
\mathcal G^{\, \ell}\cdot\vec{E}_{A}^{\, \, t}\equiv\vec{0}\, ,                    \eqno (3.72)
$$
takes the form
$$
\mathcal G\cdot\vec{E}_{A}=\, \mathcal G_{d}^{\, \, t}\cdot\vec{E}_{A}^{\, \, t}+\mathcal G_{d}^{\, \, t}\cdot\vec{E}_{A}^{\, \, \ell}+\mathcal G_{d}^{\, \, \ell}\cdot\vec{E}_{A}^{\, \, \ell}+\mathcal G_{\ast}\cdot\vec{E}_{A}^{\, \, \ell}\, .                                           \eqno (3.73)
$$

Equations (3.71) and (3.73), together with the readily verified relation
$$
\mathcal G_{d}^{\, \, t}\cdot\vec{E}_{A}^{\, \, \ell}=\, -\, \mathcal {F}^{\, t}\cdot\vec{E}_{A}^{\, \ell},
\eqno (3.74)
$$
allow (3.70) to be written as
$$
\vec{S}_{\partial A}\, =\, \frac{c}{4\, \pi}\, (\mathcal {F}\cdot\vec{E}_{A}^{\, \, t}+\, \mathcal G_{d}^{\, \, t}\cdot\vec{E}_{A}^{\, \, t})+\, \frac{c}{4\, \pi}\, (\mathcal G_{d}^{\, \, \ell}\cdot\vec{E}_{A}^{\, \, \ell}+\mathcal G_{\ast}\cdot\vec{E}_{A}^{\, \, \ell})=\, \vec{S}_{\mathcal F \, \mathcal G_{d}^{\, t}}+\vec{S}_{\mathcal G^{\, \ell}}\, .
\eqno (3.75)
$$
The terms on the right-hand side coincide with those in (3.14) and (3.36), respectively, which were obtained independently from the systems of equations (1.30) and (1.42).

Equation (3.75) also shows that the fluxes defined by (3.14) and (3.36) constitute components of the energy-flux density defined by the field complex $\{\vec{E}_{A}(x),\partial A(x)\}$.

At the same time, the radiations represented by these components, $\vec{S}_{\mathcal F \, \mathcal G_{d}^{\, t}}(x)$ and $\vec{S}_{\mathcal G^{\ell}}(x)$, have three fundamentally important distinguishing features: the direction of the radiation-pattern maximum, the polarization of the electric field, and the field content. Consequently, each of these radiations has an \emph{independent physical} significance, expressed, in particular, in the possibility of their\emph{ independent} technical use \cite{Al6}, \cite{Al8}.

On the other hand, (3.68) and (3.75) also demonstrate that \emph{all three} components of the 3-tensor $\partial A(x)$, represented by the terms on the right-hand side of (1.16), contribute to determining the dynamic variables of the radiation from spatially bounded systems of electrically charged particles. This ensures the mathematical completeness of the theory with respect to the accounting for \emph{all field types} defined by the terms on the right-hand side of (1.16).

To conclude this section, it is also appropriate to present a method for calculating the dynamic variables under consideration, starting from equation (3.61), written in the form \cite{Al0}
 $$
 \partial^{\, 0}\vec{E}_{A}+\, \vec{\bigtriangledown}\cdot\mathcal F+\, \vec{\bigtriangledown}\cdot\mathcal G_{d}+\, \vec{\bigtriangledown}\cdot\mathcal G_{\ast}\, =\, -\frac{4\, \pi}{c}\, \vec{\jmath}\, .
\eqno (3.76)
$$

This equation explicitly shows from the outset that the conduction current and the corresponding displacement current generate not only the magnetic field represented by the 3-tensor $\mathcal F(x)$, but also the jeitonic fields represented by the 3-tensors $\mathcal G_{d}(x)$ and $\mathcal G_{\ast}(x)$.

Taking the scalar product of both sides of (3.76) with $\vec{E}_{A}(x)$, we obtain
$$
\partial^{\, 0}\, \frac{E_{A}^{\, \, 2}}{2}+\, 
(\, \vec{\bigtriangledown}\cdot\mathcal F\, )\cdot\vec{E}_{A}+\, (\, \vec{\bigtriangledown}\cdot\mathcal G_{d}\, )\cdot\vec{E}_{A}+\, (\, \vec{\bigtriangledown}\cdot\mathcal G_{\ast}\, )\cdot\vec{E}_{A}=\, -\frac{4\, \pi}{c}\, \vec{\jmath}\, \cdot\vec{E}_{A}\, .
\eqno (3.77)
$$

According to (3.64), the second term on the left-hand side of (3.77) takes the form
$$
(\vec{\bigtriangledown}\cdot\mathcal F)\cdot\vec{E}_{A}
= \vec{\bigtriangledown}\cdot(\mathcal F\cdot\vec{E}_{A})
- (\mathcal F,\partial E_{A}).
\eqno (3.78)
$$

Using (3.62), we obtain
$$
(\mathcal F,\partial E_{A})
= -(\mathcal F,\partial^{0}\mathcal F)
= -\partial^{0}\frac{\mathcal F^{2}}{2},
\eqno (3.79)
$$
after which (3.78) takes the form
$$
(\vec{\bigtriangledown}\cdot\mathcal F)\cdot\vec{E}_{A}
= \vec{\bigtriangledown}\cdot(\mathcal F\cdot\vec{E}_{A})
+ \partial^{0}\frac{\mathcal F^{2}}{2}.
\eqno (3.80)
$$

By an analogous procedure, we obtain expressions for the third and fourth terms on the left-hand side of (3.77):
$$
(\vec{\bigtriangledown}\cdot\mathcal G_{d})\cdot\vec{E}_{A}
= \vec{\bigtriangledown}\cdot(\mathcal G_{d}\cdot\vec{E}_{A})
+ \partial^{0}\frac{\mathcal G_{d}^{2}}{2},
\eqno (3.81)
$$
$$
(\vec{\bigtriangledown}\cdot\mathcal G_{\ast})\cdot\vec{E}_{A}
= \vec{\bigtriangledown}\cdot(\mathcal G_{\ast}\cdot\vec{E}_{A})
+ \partial^{0}\frac{\mathcal G_{\ast}^{2}}{2}.
\eqno (3.82)
$$

As a result, (3.77) takes the form of the relation (theorem) representing the differential form of the energy-change law for the fields under consideration:
$$
\partial_{\, t}\, \frac{ E_{A}^{\, 2}+\mathcal F^{\, 2}+\mathcal G_{d}^{\, 2}+\mathcal G_{\ast}^{\, 2}}{8\, \pi}\, +\, \frac{c}{4\, \pi}\, \vec{\bigtriangledown}\cdot(\mathcal F\cdot\vec{E}_{A}+\mathcal G_{d}\cdot\vec{E}_{A}+\mathcal G_{\ast}\cdot\vec{E}_{A})=\, -\, \vec{\jmath}\, \cdot\vec{E}_{A}\, .
\eqno (3.83)
$$

This equation simultaneously defines the dynamic variables of the field complex $\{\vec{E}_{A}(x), \, \mathcal F(x), \, \mathcal G_{d}(x), \, \mathcal G_{\ast}(x)\}$ given by (3.68) and (3.69).

Equation (3.83) immediately reduces to (3.65), which defines the dynamic variables (3.66) and (3.67).

\section{Angular Distributions of the Instantaneous Radiation Powers of a Point Electric Charge}

\subsection*{\S \, 4. Angular Distributions of the Instantaneous Radiation Powers of a Point Charge in the Nonrelativistic Motion Regime. Larmor Formulas.}
\addcontentsline{toc}{subsection}{\S\, 4. Angular Distributions of the Instantaneous Radiation Powers of a Point Charge in the Nonrelativistic Motion Regime. Larmor Formulas.}

\setcounter{section}{4}
\setcounter{figure}{0}

The elementary instantaneous power (elementary instantaneous intensity \cite{La}) of the radiation passing through an elementary surface element of a sphere of radius $R(t')$, centered at the position of the charge at the time $t'=t-R(t')/c$ and measured at the observation point at time $t$, is given by
$$
dP(t)\, =\, (\, \vec{S}\cdot\vec{n})\, R^{\, 2}\, d\, \Omega\, .         \eqno (4.1)
$$

For simplicity, the argument $t$ in the symbols $dP(t)$ will be omitted below.

Successive substitution of (3.33), (3.43), and (3.53) into (4.1), followed by use of (3.41), yields, respectively, the following relations, which, in particular, determine the angular distributions of the instantaneous radiation powers of a point electric charge undergoing nonrelativistic motion:
$$
dP_{\, \mathcal{G}^{\ell}}\, =\, \frac{e^{\, 2}\dot{\vec{\beta}}^{\, \, 2}}{4\, \pi\, c}\, \cos^{\, 2}\theta\, \, d\, \Omega\, , 
                                                                                 \eqno (4.2)
$$
$$
dP_{\, \mathcal{F}\mathcal{G}_{d}^{\, t}}\, =
\, \frac{e^{\, 2}\dot{\vec{\beta}}^{\, \, 2}}{4\, \pi\, c}\, \sin^{\, 2}\theta\, \, d\, \Omega\, , 
                                                                                 \eqno (4.3)
$$
$$
dP\, =\, \frac{e^{\, 2}\dot{\vec{\beta}}^{\, \, 2}}{4\, \pi\, c}\, \, d\, \Omega\, , 
                                                                                 \eqno (4.4)
$$
$$
dP_{\mathcal{\, F}}\, =\, \frac{1}{2}\, dP_{\, \mathcal{F}\mathcal{G}_{d}^{\, t}}\, =
\, dP_{\, \mathcal{G}_{d}^{\, t}}\, ,                                                \eqno (4.5)
$$
where $\theta$ is the angle between the vectors $\dot{\vec{\beta}}$ and $\vec{n}$.

The right-hand side of (4.3), which determines the elementary instantaneous power of the transverse electromagnetic-jeitonic radiation from a point electric charge undergoing nonrelativistic motion, exactly coincides with the right-hand side of the well-known expression for the elementary instantaneous power of the electromagnetic radiation from such a charge, as obtained within the framework of conventional classical electromagnetic theory \cite{Jack}.

Figure 4.1 shows the meridional cross-sections of the radiation patterns for the elementary instantaneous powers of longitudinal electrojeitonic and transverse electromagnetic-jeitonic radiations from a point electric charge in nonrelativistic motion, as given by (4.2) and (4.3).

\begin{figure}[!h]
\begin{center}
\vspace{-6pt}
\includegraphics[width=65mm]{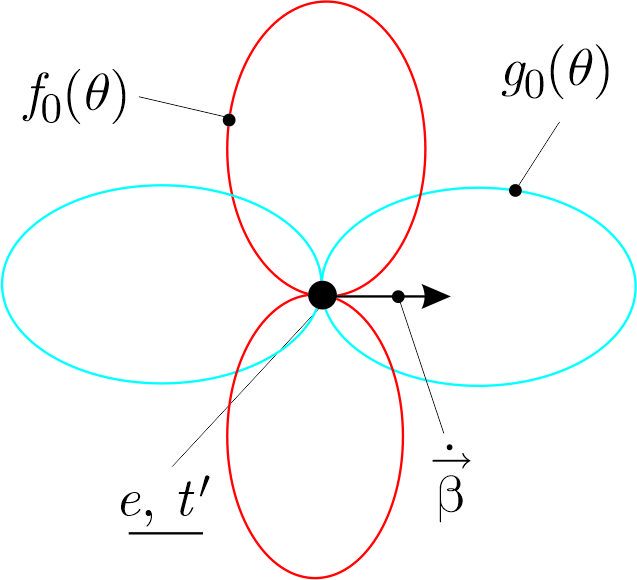}
\vspace{+10pt}
\caption{Meridional cross-sections of the radiation patterns of the elementary instantaneous powers of longitudinal electrojeitonic and transverse electromagnetic-jeitonic radiations from a point electric charge in nonrelativistic motion.}
\end{center}
\vspace{-1mm}
\end{figure}

The symbols $g_{0}(\theta)$ and $f_{0}(\theta)$ in Figure 4.1 denote the radiation-pattern functions for the powers of the respective radiations, given by the second factors on the right-hand sides of (4.2) and (4.3).

The existence of longitudinal electrojeitonic radiation, represented in this case by (4.2), implies, in particular, the need to account for this radiation when calculating the characteristics of existing antenna systems in which conventional electromagnetic radiation serves as the useful signal. At the same time, it opens up the possibility of developing electrical systems in which longitudinal electrojeitonic radiation is used as the useful signal.

The isotropy of the \emph{total} radiation described by (4.4) also implies the isotropy of the \emph{total} radiation from classical \emph{elementary} electric radiators of antennas.

For extended antennas, the emergence of a nontrivial radiation-pattern factor (the array factor \cite{Eis}), resulting from the antenna geometry and the amplitude and phase relationships between the excitation currents of the antenna's elementary radiators, also makes the total radiation anisotropic.

Successive integration of (4.2)--(4.5) over the full solid angle yields the following expressions for the total instantaneous powers of the radiations under consideration from a nonrelativistic point electric charge.
$$
P_{\, \mathcal{G}^{\ell}}\, =\, \frac{1}{3}\, \frac{e^{\, 2}\dot{\vec{\beta}}^{\, \, 2}}{c}\, ,   \eqno (4.6)
$$
$$
P_{\, \mathcal{F}\mathcal{G}_{d}^{\, t}}\, =
\, \frac{2}{3}\, \frac{e^{\, 2}\dot{\vec{\beta}}^{\, \, 2}}{c}\, ,                       \eqno (4.7)
$$
$$
P\, =\, \frac{e^{\, 2}\dot{\vec{\beta}}^{\, \, 2}}{c}\, ,                     \eqno (4.8)
$$
$$
P_{\mathcal{\, F}}\, =\, \frac{1}{2}\, P_{\, \mathcal{F}\mathcal{G}_{d}^{\, t}}\, =
\, P_{\, \mathcal{G}_{d}^{\, t}}\, .                                                 \eqno (4.9)
$$

As can be seen, the right-hand side of (4.7) is the right-hand side of the well-known Larmor formula.

However, the left-hand side of this relation shows that the formula determines the instantaneous power as the \emph{sum} of the instantaneous powers of transverse electromagnetic and transverse electrojeitonic radiation generated by the electric charge under consideration.

In turn, comparison of (4.7) and (4.8) reveals the physical meaning of the ``ubiquitous'' numerical coefficient $2/3$ in the Larmor formula: this coefficient now indicates that only $2/3$ of the total instantaneous radiation power given by relation (4.8) is carried by the transverse field complex $\{\vec{E}_{A}^{t}(x), \mathcal{F}(x), \mathcal{G}_{d}^{t}(x)\}$ (the transverse electromagnetic-jeitonic wave).

Similarly, comparison of relations (4.6) and (4.8) reveals the physical significance of the coefficient $1/3$ in (4.6): this coefficient indicates that $1/3$ of the total instantaneous radiation power given by relation (4.8) is carried by the longitudinal field complex $\{\vec{E}_{A}^{\ell}(x), \mathcal{G}_{d}^{\ell}(x), \mathcal{G}_{\ast}(x)\}$ (the longitudinal electrojeitonic wave).

\subsection*{ 4.1 Angular Distributions of the Instantaneous Powers of Dipole Radiation from a System of Electric Charges}
\addcontentsline{toc}{subsection}{4.1. Angular Distributions of the Instantaneous Powers of Dipole Radiation from a System of Electric Charges.}

The analysis of radiation from a point electrically charged particle presented above can be readily extended to the dipole radiation from a \emph{system} of nonrelativistic electric charges.

At distances much greater than the dimensions of the system of moving nonrelativistic electrically charged particles, the vector and scalar potentials generated by the electric dipole moment of such a system are given by the well-known relations \cite{La}
$$
\vec{A}=\frac{\dot{\vec{d}}}{cr},
\eqno (4.10)
$$
$$
\varphi=-\operatorname{div}\frac{\vec{d}}{r},
\eqno (4.11)
$$
where $\vec{d}$ is the electric dipole moment of the system of these particles at the retarded time $t'=t-r/c$.

Carrying out the corresponding differentiations of these potentials, we obtain the following expressions for the terms of the field variables depending on $\ddot{\vec{d}}(t')$:
$$
\partial A=-\, \frac{1}{c^{\, 2}\, r}\, (\, \vec{n}\otimes \ddot{\vec{d}}\, \, )\, ,   \eqno (4.12)
$$
$$
\vec{E}_{A}=-\, \frac{1}{c^{\, 2}\, r}\, \ddot{\vec{d}}=-\frac{1}{c^{\, 2}\, r}\, 
\{\, \vec{n}\times(\, \ddot{\vec{d}}\times\vec{n})+(\, \vec{n}\cdot\ddot{\vec{d}}\, \, )\, \vec{n}\, \}\, , 
                                                                                             \eqno (4.13)
$$
$$
\vec{E}_{\, \varphi}\, =\, \frac{1}{c^{\, 2}\, r}\, (\, \vec{n}\cdot\ddot{\vec{d}}\, \, )\, \vec{n}\, , 
                                                                                            \eqno (4.14)
$$
where $\vec{n}=\vec{r}/r$.

In turn, substituting (4.13) and (4.14) into (1.20) and (1.23) yields the following expressions for the electric-field strengths of the dipole radiation generated by the classical system of point electrically charged particles:
$$
\vec{E}_{\, \mathcal{F}}=-
\frac{1}{2\, c^{\, 2}\, r}\, \{\, (\vec{n}\times\ddot{\vec{d}}\, )\times\vec{n}\, \}\, ,      \eqno (4.15)
$$
$$
\vec{E}_{\, \mathcal{G}}=-
\frac{1}{2\, c^{\, 2}\, r}\, \{\, (\vec{n}\times\ddot{\vec{d}}\, )\times\vec{n}\, +
\, 2\, (\, \vec{n}\cdot\ddot{\vec{d}}\, \, )\, \vec{n}\, \}\, .                                     \eqno (4.16)
$$

Using (4.13), we obtain the following expressions for the instantaneous values of the corresponding energy-flux density vectors:
$$
\vec{S}_{\, \mathcal{G}^{\, \ell}}\, =
\, \frac{1}{4\, \pi\, c^{\, 3}\, r^{\, 2}}\, \, |\, (\, \vec{n}\cdot\ddot{\vec{d}}\, \, )\, \vec{n}\, |^{\, 2}\, \vec{n}\, , 
                                                                                         \eqno (4.17)
$$
$$
\vec{S}_{\, \mathcal{F}\mathcal{G}_{d}^{\, \, t}}\, =
\, \frac{1}{4\, \pi\, c^{\, 3}\, r^{\, 2}}\, \, |\, (\vec{n}\times\ddot{\vec{d}}\, )\times\vec{n}\, |^{\, 2}\, \vec{n}\, , 
                                                                                         \eqno (4.18)
$$
$$
\vec{S}\, =
\, \frac{1}{4\, \pi\, c^{\, 3}\, r^{\, 2}}\, \, \ddot{\vec{d}}\, ^{\, \, 2}\, \vec{n}\, , 
                                                                                         \eqno (4.19)
$$
$$
\vec{S}_{\mathcal{F} }\, =\, \frac{1}{2}\, \vec{S}_{\, \mathcal{F}\mathcal{G}_{d}^{\, \, t}}\, =
\, \vec{S}_{\mathcal{G}_{d}^{\, t}}\, .                                                    \eqno (4.20)
$$

Relations (4.17)--(4.20), in turn, together with (4.1), yield the following expressions for the angular distributions of the elementary instantaneous powers of dipole radiation from the system of electrically charged particles under consideration:
$$
dP_{\, \mathcal{G}^{\ell}}\, =
\, \frac{\ddot{\vec{d}}\, ^{\, \, 2}}{4\, \pi\, c^{\, 3}}\, \cos^{\, 2}\theta\, \, d\, \Omega\, ,       \eqno (4.21)
$$
$$
dP_{\, \mathcal{F}\mathcal{G}_{d}^{\, t}}\, =
\, \frac{\ddot{\vec{d}}\, ^{\, \, 2}}{4\, \pi\, c^{\, 3}}\, \sin^{\, 2}\theta\, \, d\, \Omega\, ,       \eqno (4.22)
$$
$$
dP\, =\, \frac{\ddot{\vec{d}}\, ^{\, \, 2}}{4\, \pi\, c^{\, 3}}\, \, d\, \Omega\, ,                 \eqno (4.23)
$$
$$
dP_{\, \mathcal{F}}\, =\, \frac{1}{2}\, dP_{\, \mathcal{F}\mathcal{G}_{d}^{\, t}}\, =
\, dP_{\, \mathcal{G}_{d}^{\, t}}\, ,                                                          \eqno (4.24)
$$
where $\theta$ is the angle between the vectors $\ddot{\vec{d}}$ and $\vec{n}$.

We again note that the right-hand side of relation (4.22) exactly coincides with the right-hand side of formula (67.7) in \cite{La}.

At the same time, this relation shows that this formula determines the elementary instantaneous intensity as the \emph{sum} of the elementary instantaneous intensities of transverse electromagnetic and transverse electrojeitonic radiation generated by the electric dipole moment of the system of electrically charged particles under consideration.

According to (4.21) and (4.22), the meridional cross-section of the radiation patterns of the instantaneous powers of longitudinal electrojeitonic and transverse electromagnetic-jeitonic radiations generated by the electric dipole moment of the system of particles under consideration has the form shown in Figure 4.2.

\begin{figure}[!h]
\begin{center}
\vspace{-6pt}
\includegraphics[width=65mm]{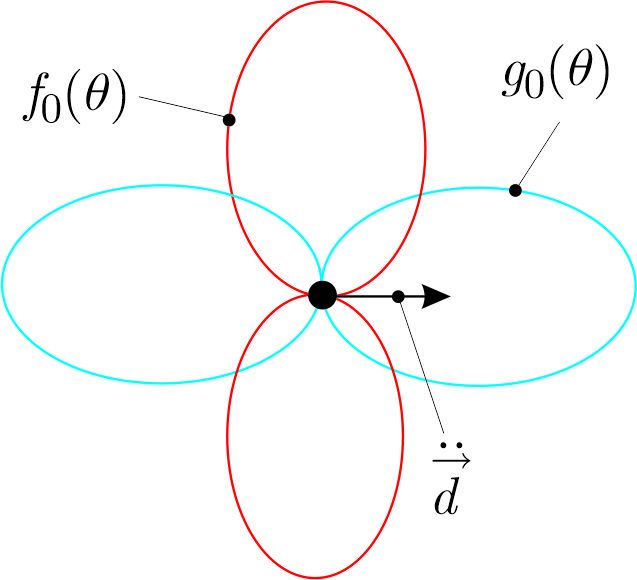}
\vspace{+10pt}
\caption{Meridional cross-section of the radiation patterns of the elementary instantaneous powers of longitudinal electrojeitonic and transverse electromagnetic-jeitonic radiations generated by the electric dipole moment of the system.}
\end{center}
\vspace{-1mm}
\end{figure}

Successive integration of relations (4.21)--(4.24) over the full solid angle yields the corresponding expressions for the total instantaneous powers of radiation generated by the electric dipole moment of the system of charges under consideration:
$$
P_{\, \mathcal{G}^{\ell}}\, =\, \frac{1}{3}\, \, \frac{\ddot{\vec{d}}\, ^{\, \, 2}}{c^{\, \, 3}}\, , \eqno (4.25)
$$
$$
P_{\, \mathcal{F}\mathcal{G}_{d}^{\, t}}\, =\, \frac{2}{3}\, \, \frac{\ddot{\vec{d}}\, ^{\, \, 2}}{c^{\, 3}}\, , 
                                                                                              \eqno (4.26)
$$
$$
P\, =\, \frac{\ddot{\vec{d}}\, ^{\, \, 2}}{c^{\, 3}}\, ,                       \eqno (4.27)
$$
$$
P_{\, \mathcal{F}}\, =\, \frac{1}{2}\, P_{\, \mathcal{F}\mathcal{G}_{d}^{\, t}}\, =
\, P_{\, \mathcal{G}_{d}^{\, t}}\, .                                                           \eqno (4.28)
$$

The comments on relations (4.25)--(4.28) are identical to those on relations (4.6)--(4.9).

\subsection*{\S \, 5. Angular Distributions of the Instantaneous Powers of Radiation from a Point Electric Charge in Relativistic Motion. The Dominance of Longitudinal Electrojeitonic Radiation in Ultrarelativistic Particles.}
\addcontentsline{toc}{subsection}{\S\, 5. Angular Distributions of the Instantaneous Powers of Radiation from a Point Electric Charge in Relativistic Motion. The Dominance of Longitudinal Electrojeitonic Radiation in Ultrarelativistic Particles.}
\setcounter{section}{5}
\setcounter{figure}{0}

The elementary instantaneous radiation power as a function of the proper time of the charge under consideration is determined by relation \cite{Jack3}
$$
dP(t\, \acute{}\, )\, =
\, (\, 1-\, \vec{n}\cdot\vec{\beta}\, )\, (\, \vec{S}\cdot\vec{n}\, )\, R^{\, 2}\, d\, \Omega\, .         \eqno (5.1)
$$

Again, for simplicity of notation, the argument $t'$ in the symbols $dP(t')$ will be omitted below.

Successive use of (3.33), (3.43), and (3.53) in (5.1), followed by taking (3.41) into account, yields, respectively, the following relations:
$$
dP_{\mathcal{G}^{\, \ell}}\, =\, \frac{e^{\, 2}}{4\, \pi\, c\, (1-\vec{n}\cdot\vec{\beta}\, )^{5}}\, 
|\, \vec{n}\cdot\dot{\vec{\beta}}\, |^{\, 2}\, d\, \Omega\, ,                                        \eqno (5.2)
$$
$$
dP_{\mathcal{F}\mathcal{G}_{d}^{\, t}}\, =\, \frac{e^{\, 2}}{4\, \pi\, c\, (1-\vec{n}\cdot\vec{\beta}\, )^{5}}\, 
|\, \vec{n}\times(\, \dot{\vec{\beta}}\times\vec{n}\, )+
\, \vec{n}\times(\, \vec{\beta}\times\dot{\vec{\beta}}\, )\, |^{\, 2}\, d\, \Omega\, ,                 \eqno (5.3)
$$
$$
dP\, =\, \frac{e^{\, 2}}{4\, \pi\, c\, (1-\vec{n}\cdot\vec{\beta}\, )^{5}}\, 
|\, \dot{\vec{\beta}}+
\, \vec{n}\times(\, \vec{\beta}\times\dot{\vec{\beta}}\, )\, |^{\, 2}\, d\, \Omega\, ,               \eqno (5.4)
$$
$$
dP_{\mathcal{\, F}}\, =\, \frac{1}{2}\, dP_{\, \mathcal{F}\mathcal{G}_{d}^{\, t}}\, =
\, dP_{\, \mathcal{G}_{d}^{\, t}}\, .                                                             \eqno (5.5)
$$

It should be also noted that the right-hand side of relation (5.3), which determines the elementary instantaneous power of transverse electromagnetic-jeitonic radiation from a point electric charge, coincides with the right-hand side of the relation representing the elementary instantaneous power of electromagnetic radiation from a point charge, obtained within the framework of the traditional theory of the electromagnetic field \cite{Jack}.

Let us consider two practically important special cases: when, at the moment of radiation,  $\dot{\vec{\beta}}\, coll\, \vec{\beta}$, and when, at that moment, $\dot{\vec{\beta}}\, norm \, \vec{\beta}$.

\emph{In the first case}, relations (5.2)--(5.5) yield the following expressions, which determine, in particular, the angular distributions of the elementary instantaneous powers of the radiation under consideration, qualitatively represented by the radiation patterns in Figure 5.1:
$$
 dP_{\mathcal{G}^{\, \ell}}\, =\, \frac{e^{\, 2}\, \dot{\vec{\beta}}^{\, \, 2}}{4\, \pi\, c}\, \, 
 \frac{\cos^{\, \, 2}\theta}{(\, 1-\beta\, \cos\theta\, )^{\, 5}}\, \, d\, \Omega\, ,                \eqno (5.6)
$$
$$
 dP_{\mathcal{F}\mathcal{G}_{d}^{\, t}}\, =\, \frac{e^{\, 2}\, \dot{\vec{\beta}}^{\, \, 2}}{4\, \pi\, c}\, \, 
 \frac{\sin^{\, \, 2}\theta}{(\, 1-\beta\, \cos\theta\, )^{\, 5}}\, \, d\, \Omega\, ,               \eqno (5.7)
$$
$$
 dP\, =\, \frac{e^{\, 2}\, \dot{\vec{\beta}}^{\, \, 2}}{4\, \pi\, c}\, \, 
 \frac{1}{(\, 1-\beta\, \cos\theta\, )^{\, 5}}\, \, d\, \Omega\, ,                   \eqno (5.8)
$$
$$
dP_{\, \mathcal{F}}\, =\, \frac{1}{2}\, dP_{\, \mathcal{F}\mathcal{G}_{d}^{\, t}}\, =
\, dP_{\, \mathcal{G}_{d}^{\, t}}\, .                                                           \eqno (5.9)
$$

Here and below, $\theta$ denotes the angle between the vectors $\vec{\beta}$ and $\vec{n}$.

\begin{figure}[!h]
\begin{center}
\vspace{-6pt}
\includegraphics[width=150mm]{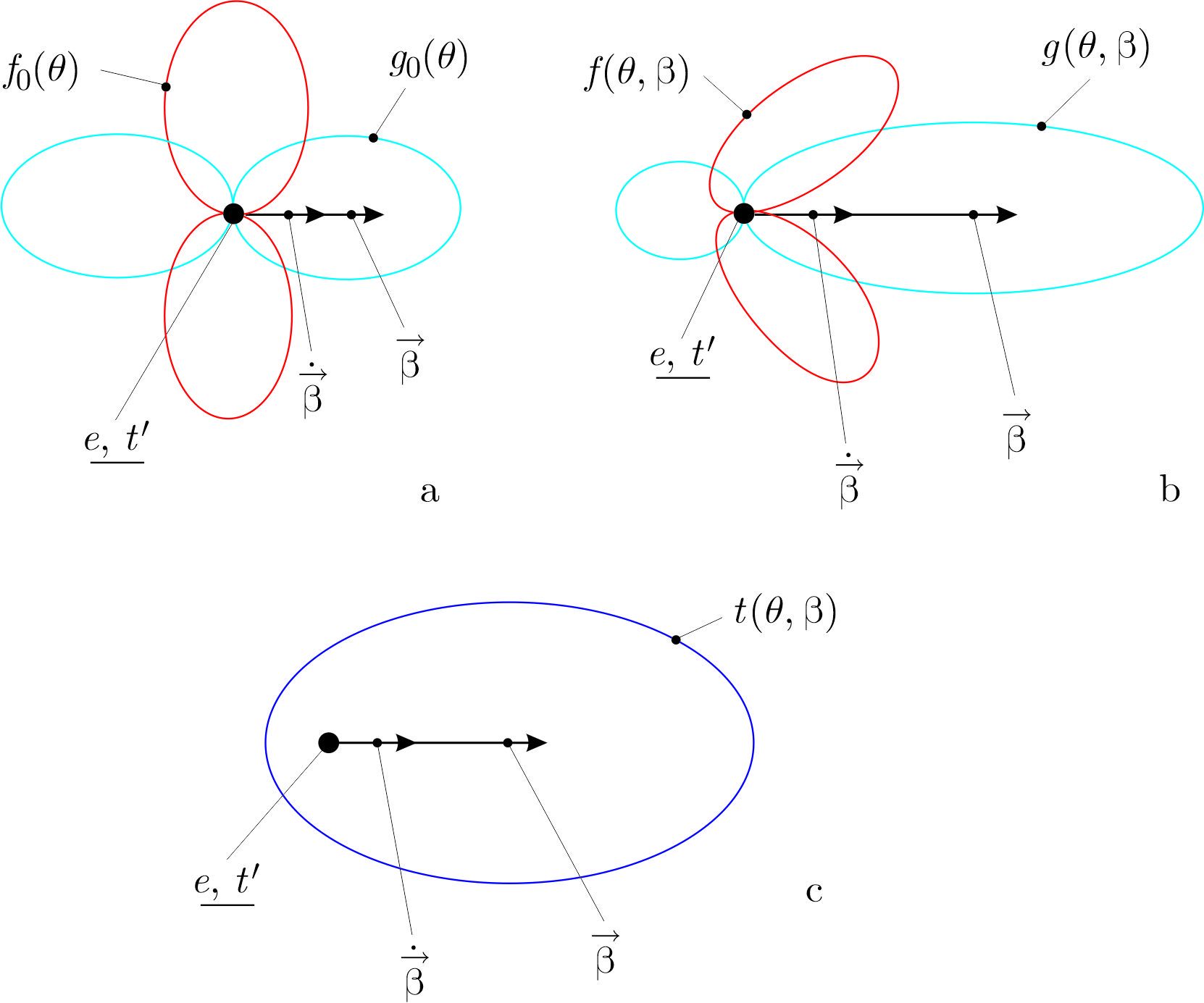}
\vspace{+10pt}
\caption{Meridional cross-sections of the radiation patterns of the elementary instantaneous powers of radiations from a point electric charge with $\dot{\vec{\beta}}\, \uparrow\uparrow\, \vec{\beta}$\, , for nonrelativistic motion (a) and relativistic motion (with $\beta\sim0.3$) (b) and (c).}
\end{center}
\vspace{-1mm}
\end{figure}

The symbols $g(\theta,\beta)$, $f(\theta,\beta)$, and $t(\theta,\beta)$ in Figure 5.1 denote the formulas for the radiation patterns \cite{Eis} of the powers of the respective radiations, defined by the second factors on the right-hand sides of relations (5.6), (5.7), and (5.8), respectively.

The scale of pattern (a) is reduced by approximately a factor of 6 relative to the scales of patterns (b) and (c), for the same value of acceleration.

Pattern (b) in Figure 5.1 \emph{visually} demonstrates the existence of \emph{longitudinal} electrojeitonic radiation (electrojeitonic radiation with \emph{longitudinal} polarization of the electric-field strength), represented in this pattern by the rear and \emph{main} lobes, with the maximum energy-density value in the direction of the vector $\vec{\beta}$ exceeding the maximum energy-density value of the generally known transverse electromagnetic-jeitonic radiation, represented in this pattern by the \emph{side} lobes.

Successive integration of relations (5.6)--(5.9) over the full solid angle yields the following expressions for the total instantaneous powers of the radiation from the point electric charge in this case:
$$
P_{\mathcal{G}^{\, \ell}}\, =\, \frac{1}{3}\, \, \frac{e^{\, 2}\, \dot{\vec{\beta}}^{\, \, 2}}{c}\, \, (\, 1+
5\, \beta^{\, 2}\, )\, \, \gamma^{\, 8}\, ,                                                        \eqno (5.10)
$$
$$
P_{\mathcal{F}G_{d}^{\, t}}\, =\, \frac{2}{3}\, \, \frac{e^{\, 2}\, \dot{\vec{\beta}}^{\, \, 2}}{c}\, \, \gamma^{\, 6}\, , 
                                                                                               \eqno (5.11)
$$
$$
P\, =\, \frac{e^{\, 2}\, \dot{\vec{\beta}}^{\, \, 2}}{c}\, \, (\, 1+\beta^{\, 2}\, )\, \, \gamma^{\, 8}\, , 
                                                                                              \eqno (5.12)
$$
$$
P_{\, \mathcal{F}}\, =\, \frac{1}{2}\, P_{\, \mathcal{F}\mathcal{G}_{d}^{\, t}}\, =
\, P_{\, \mathcal{G}_{d}^{\, t}}\, .                                                           \eqno (5.13)
$$

Comparison of (5.10) and (5.11) yields the relation
$$
P_{\mathcal{G}^{\, \ell}}\, =\, \frac{1+5\, \beta^{\, 2}}{2}\, \gamma^{\, 2}\, P_{\mathcal{F}G_{d}^{\, t}}\, , 
                                                                                           \eqno (5.14)
$$
showing that, in this case, for $\beta>\sqrt{7}/7$, the instantaneous power of \emph{longitudinal electrojeitonic} radiation from a relativistic electrically charged particle \emph{exceeds} the instantaneous power of transverse electromagnetic-jeitonic radiation from this particle by a factor of $(1+5\beta^{2})\gamma^{2}/2$, becoming \emph{dominant} in the ultrarelativistic case.

The difference between the values of the powers under consideration at different values of $\beta$ is due to the different values of the corresponding factors multiplying $e^{2}\dot{\vec{\beta}}^{\,2}/c$ in the right-hand sides of relations (5.10)--(5.12), i.e., to the different values of the following functions
$$
g(\beta)\, :=\, \frac{1}{3}\, \, \frac{1+5\, \beta^{\, 2}}{(\, 1-\beta^{\, \, 2}\, )^{\, 4}}\, ,    \eqno (5.15)
$$
$$
f(\beta)\, :=\, \frac{2}{3}\, \, \frac{1}{(\, 1-\beta^{\, \, 2}\, )^{\, 3}}\, ,    \eqno (5.16)
$$
$$
t(\beta)\, :=\, \frac{1+\beta^{\, 2}}{(\, 1-\beta^{\, \, 2}\, )^{\, 4}}\, .     \eqno (5.17)
$$

Figure 5.2 presents graphical representations of the functions $g(\beta)$ and $f(\beta)$ over the interval of values $\beta\in[0,0.8]$, as well as a graphical representation of the function $k(\beta):=g(\beta)/f(\beta)=P_{\mathcal{G}^{\ell}}/P_{\mathcal{F}G_{d}^{t}}$ over the interval of values $\beta\in[0.90,0.99]$.
\begin{figure}[!h]
\begin{center}
\vspace{-6pt}
\includegraphics[width=90mm]{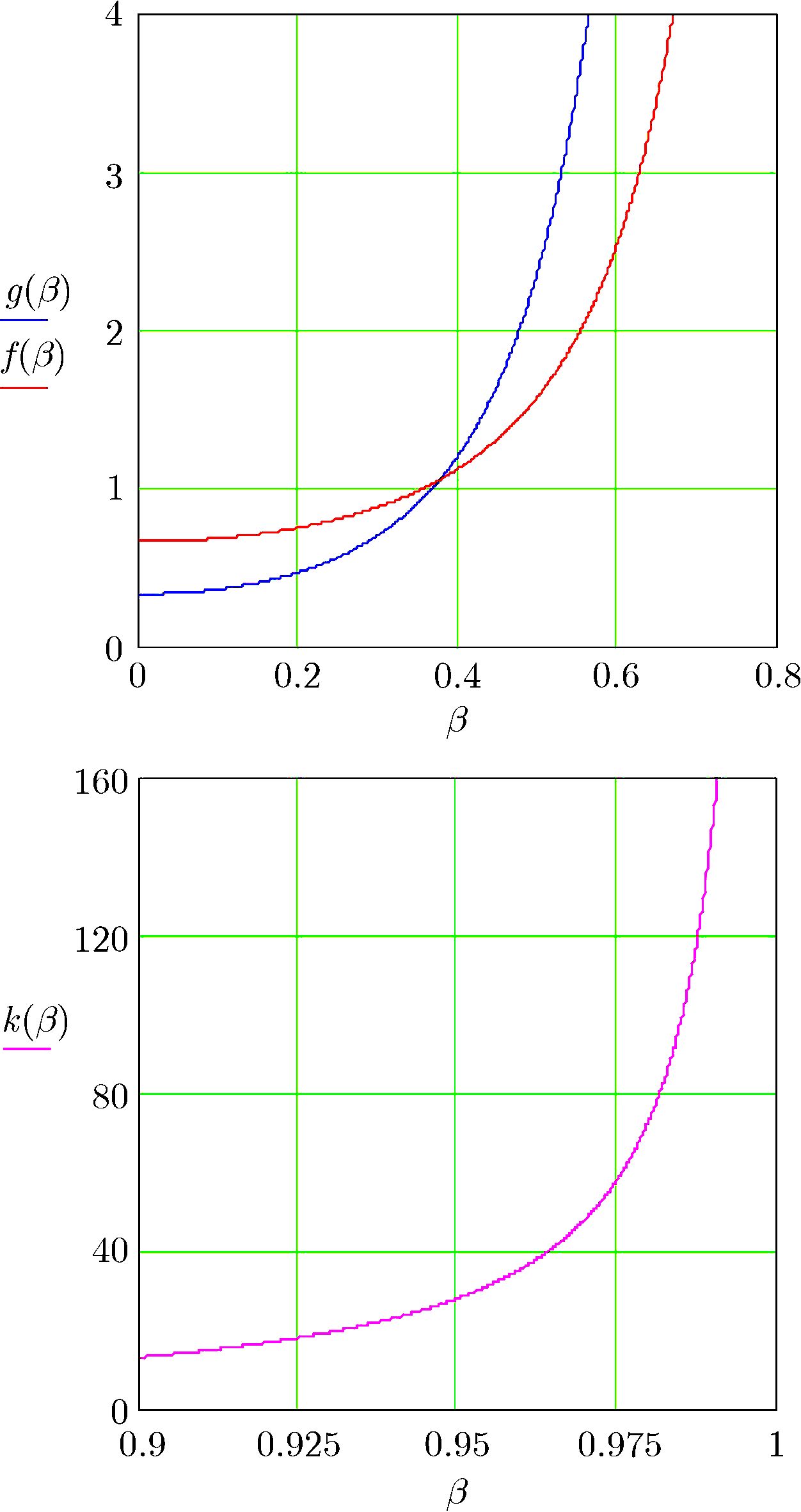}
\vspace{+10pt}
\caption{}
\end{center}
\vspace{-1mm}
\end{figure}

\emph{In the second case}, relations (5.2)--(5.5) yield the following expressions, which determine, in particular, the angular distributions of the elementary instantaneous powers of the radiation under consideration in the \emph{osculating} plane $\mathcal{Q}(t')$ $(\varphi=0)$, qualitatively represented by the radiation patterns in Figure 5.3:
$$
 dP_{\mathcal{G}^{\, \ell}}\, =\, \frac{e^{\, 2}\, \dot{\vec{\beta}}^{\, \, 2}}{4\, \pi\, c}\, \, 
 \frac{\sin^{\, \, 2}\theta}{(\, 1-\beta\, \cos\theta\, )^{\, 5}}\, \, d\, \Omega\, ,                \eqno (5.18)
$$
$$
 dP_{\mathcal{F}\mathcal{G}_{d}^{\, t}}\, =\, \frac{e^{\, 2}\, \dot{\vec{\beta}}^{\, \, 2}}{4\, \pi\, c}\, \, 
 \frac{(\, \cos\theta-\beta)^{\, 2}}{(\, 1-\beta\, \cos\theta\, )^{\, 5}}\, \, d\, \Omega\, , 
                                                                                             \eqno (5.19)
$$
$$
 dP\, =\, \frac{e^{\, 2}\, \dot{\vec{\beta}}^{\, \, 2}}{4\, \pi\, c}\, \, 
 \frac{(\, \cos\theta-\beta)^{\, 2}+\sin^{\, 2}\theta}{(\, 1-\beta\, \cos\theta\, )^{\, 5}}\, \, d\, \Omega\, , 
                                                                                            \eqno (5.20)
$$
$$
dP_{\, \mathcal{F}}\, =\, \frac{1}{2}\, dP_{\, \mathcal{F}\mathcal{G}_{d}^{\, t}}\, =
\, dP_{\, \mathcal{G}_{d}^{\, t}}\, .                                                           \eqno (5.21)
$$
\\
\begin{figure}[!h]
\begin{center}
\vspace{-6pt}
\includegraphics[width=160mm]{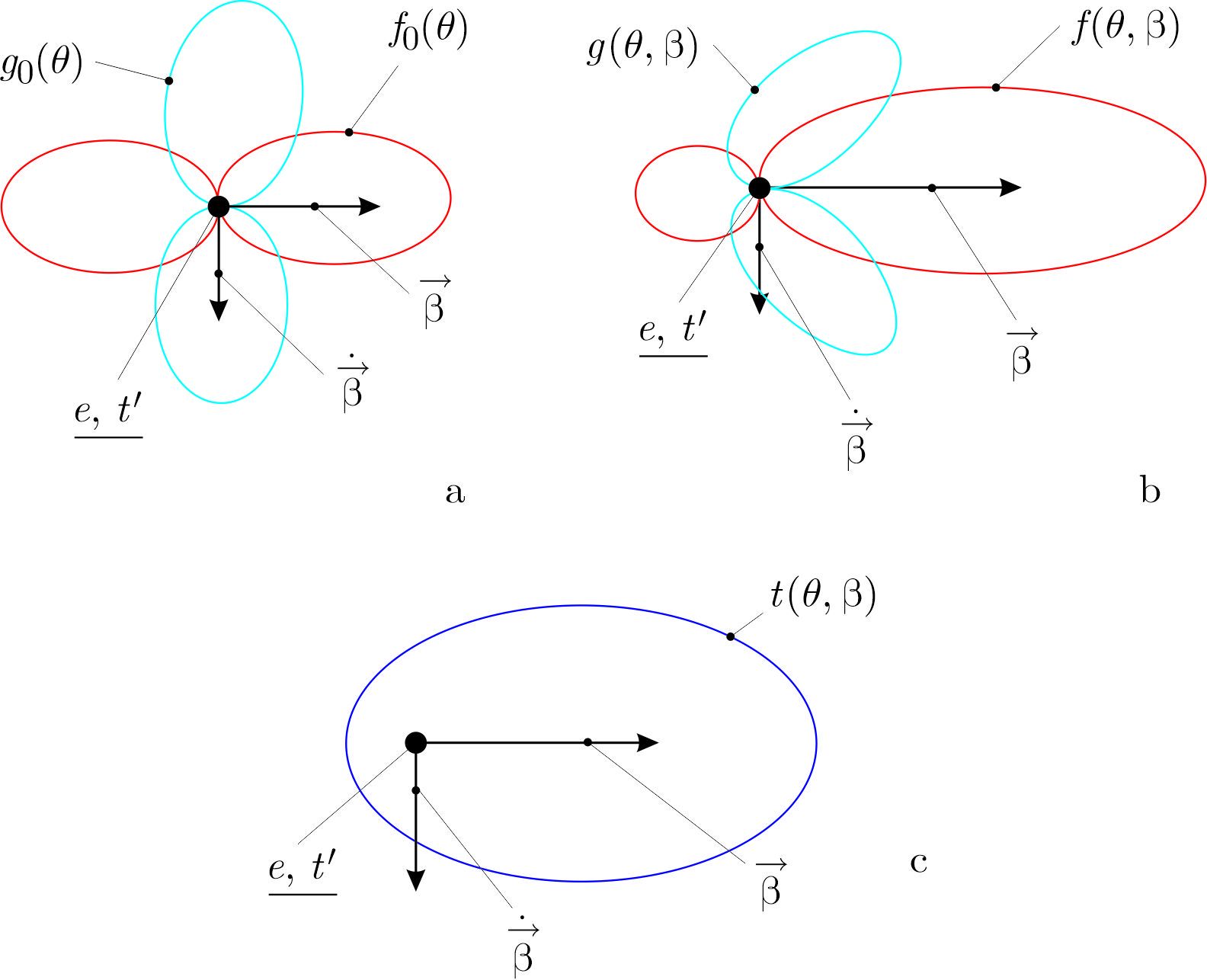}
\vspace{+10pt}
\caption{Cross-sections by the osculating plane of the radiation patterns of the instantaneous powers of radiations from a point electric charge with $\dot{\vec{\beta}}\,norm\,\vec{\beta}$, for nonrelativistic motion (a) and relativistic ($\beta\sim0.3$) motion (b) and (c) of this charge.}
\end{center}
\vspace{-1mm}
\end{figure}

The symbols $g(\theta,\beta)$, $f(\theta,\beta)$, and $t(\theta,\beta)$ in Figure 5.3 denote the formulas for the radiation patterns of the instantaneous powers of radiation, defined by the second factors on the right-hand sides of relations (5.18), (5.19), and (5.20), respectively.

The scale of pattern (a) is reduced by approximately a factor of 3 relative to the scales of patterns (b) and (c), for the same value of acceleration.

Pattern (b) in Figure 5.3 visually demonstrates the presence of the generally known transverse electromagnetic-jeitonic radiation, represented in this pattern by the rear and \emph{main} lobes, with the maximum energy density in the direction of the vector $\vec{\beta}$ exceeding, in this case, the maximum energy density of longitudinal electrojeitonic radiation, represented in this pattern by the \emph{side} lobes.

Upon choosing a comoving orthogonal Cartesian coordinate system as the Cartesian coordinate system, whose unit vectors are defined by the Frenet trihedron [26], and such that $\vec{e}_z=\vec{\tau}=\vec{\beta}/\beta$ (Fig. 14.6 in [6]), we obtain, in this case, the relation
$$
\cos\Theta=\sin\theta\cos\varphi,
\eqno (5.22)
$$
where $\Theta$ is the angle between the vectors $\vec{n}$ and $\dot{\vec{\beta}}$.

Using (5.22) in (5.2)--(5.5) yields the following relations determining the angular distributions of the elementary instantaneous powers of the radiations under consideration for any value of the azimuthal angle $\varphi$:

$$
 dP_{\mathcal{G}^{\, \ell}}\, =\, \frac{e^{\, 2}\, \dot{\vec{\beta}}^{\, \, 2}}{4\, \pi\, c}\, \, 
 \frac{\sin^{\, 2}\theta\, \cos^{\, 2}\varphi}{(\, 1-\beta\, \cos\theta\, )^{\, 5}}\, \, d\, \Omega\, , 
                                                                                         \eqno (5.23)
$$
$$
 dP_{\mathcal{F}\mathcal{G}_{d}^{\, t}}\, =\, \frac{e^{\, 2}\, \dot{\vec{\beta}}^{\, \, 2}}{4\, \pi\, c}\, \, 
 \left(\frac{1}{(\, 1-\beta\, \cos\theta\, )^{\, 3}}-\frac{(1-\beta^{\, 2})\sin^{\, 2}\theta\, \cos^{\, 2}\varphi}{(\, 1-
 \beta\, \cos\theta\, )^{\, 5}}\right)\, d\, \Omega\, , 
                                                                                             \eqno (5.24)
$$
$$
 dP\, =\, \frac{e^{\, 2}\, \dot{\vec{\beta}}^{\, \, 2}}{4\, \pi\, c}\, \, 
 \left(\frac{1}{(\, 1-\beta\, \cos\theta\, )^{\, 3}}+\frac{\beta^{\, 2}\, \sin^{\, 2}\theta\, \cos^{\, 2}\varphi}{(\, 1-
 \beta\, \cos\theta\, )^{\, 5}}\right)\, d\, \Omega\, , 
                                                                                             \eqno (5.25)
$$
$$
dP_{\, \mathcal{F}}\, =\, \frac{1}{2}\, dP_{\, \mathcal{F}\mathcal{G}_{d}^{\, t}}\, =
\, dP_{\, \mathcal{G}_{d}^{\, t}}\, .                                                           \eqno (5.26)
$$

As before, the right-hand side of relation (5.24) coincides with that of the corresponding relation given by formula (14.44) in [6].

Successive integration of relations (5.23)--(5.26) over the full solid angle yields the following expressions for the instantaneous powers of radiations from the point electric charge in this case:

$$
P_{\mathcal{G}^{\, \ell}}\, =\, \frac{1}{3}\, \, \frac{e^{\, 2}\, \dot{\vec{\beta}}^{\, \, 2}}{c}\, \, \gamma^{\, 6}\, , 
                                                                                            \eqno (5.27)
$$
$$
P_{\mathcal{F}\mathcal{G}_{d}^{\, t}}\, =
\, \frac{2}{3}\, \, \frac{e^{\, 2}\, \dot{\vec{\beta}}^{\, \, 2}}{c}\, \, \gamma^{\, 4}\, ,        \eqno (5.28)
$$
$$
P\, =\, \frac{1}{3}\, \, \frac{e^{\, 2}\, \dot{\vec{\beta}}^{\, \, 2}}{c}\, (\, 3-2\, \beta^{\, 2}\, )\, \gamma^{\, 6}\, , 
                                                                                            \eqno (5.29)
$$
$$
P_{\, \mathcal{F}}\, =\, \frac{1}{2}\, P_{\, \mathcal{F}\mathcal{G}_{d}^{\, t}}\, =
\, P_{\, \mathcal{G}_{d}^{\, t}}\, .                                                          \eqno (5.30)
$$

As expected, the right-hand side of relation (5.28) coincides with that of relation (14.46) in [6], which gives the total instantaneous power of electromagnetic radiation from a point electric charge in the case under consideration.

Comparison of (5.27) and (5.28) yields

$$
P_{\mathcal{G}^{\ell}} =
\frac{1}{2}\gamma^{2}P_{\mathcal{F}G_{d}^{t}},
\eqno (5.31)
$$
showing that, in this case, for $\beta>\sqrt{2}/2$, the instantaneous power of \emph{longitudinal electrojeitonic} radiation from a relativistic electrically charged particle \emph{exceeds} the instantaneous power of transverse electromagnetic-jeitonic radiation from the particle by a factor of $\gamma^{2}/2$, becoming \emph{dominant} in the ultrarelativistic case, as in the case $\dot{\vec{\beta}}\,coll\,\vec{\beta}$.

The difference between the values of the powers under consideration at different values of $\beta$ is due to the different values of the corresponding factors multiplying $e^{2}\dot{\vec{\beta}}^{\,2}/c$ in the right-hand sides of relations (5.27)--(5.29), i.e., to the different values of the following functions:

$$
g(\beta):=\, \frac{1}{3}\, \, \frac{1}{(\, 1-\beta^{\, 2}\, )^{3}}\, ,                       \eqno (5.32)
$$
$$
f(\beta):=\, \frac{2}{3}\, \, \frac{1}{(\, 1-\beta^{\, 2}\, )^{2}}\, ,                         \eqno (5.33)
$$
$$
t(\beta):=\, \frac{1}{3}\, \, \frac{3-2\, \beta^{\, 2}}{(\, 1-\beta^{\, 2}\, )^{3}}\, .           \eqno (5.34)
$$

Figure 5.4 presents graphical representations of the functions $g(\beta)$ and $f(\beta)$ over the interval of values $\beta\in[0,0.8]$, as well as a graphical representation of the function $k(\beta):=g(\beta)/f(\beta)=P_{\mathcal{G}^{\ell}}/P_{\mathcal{F}G_{d}^{t}}$ over the interval of values $\beta\in[0.90,0.99]$.

\begin{figure}[!h]
\begin{center}
\vspace{-6pt}
\includegraphics[width=90mm]{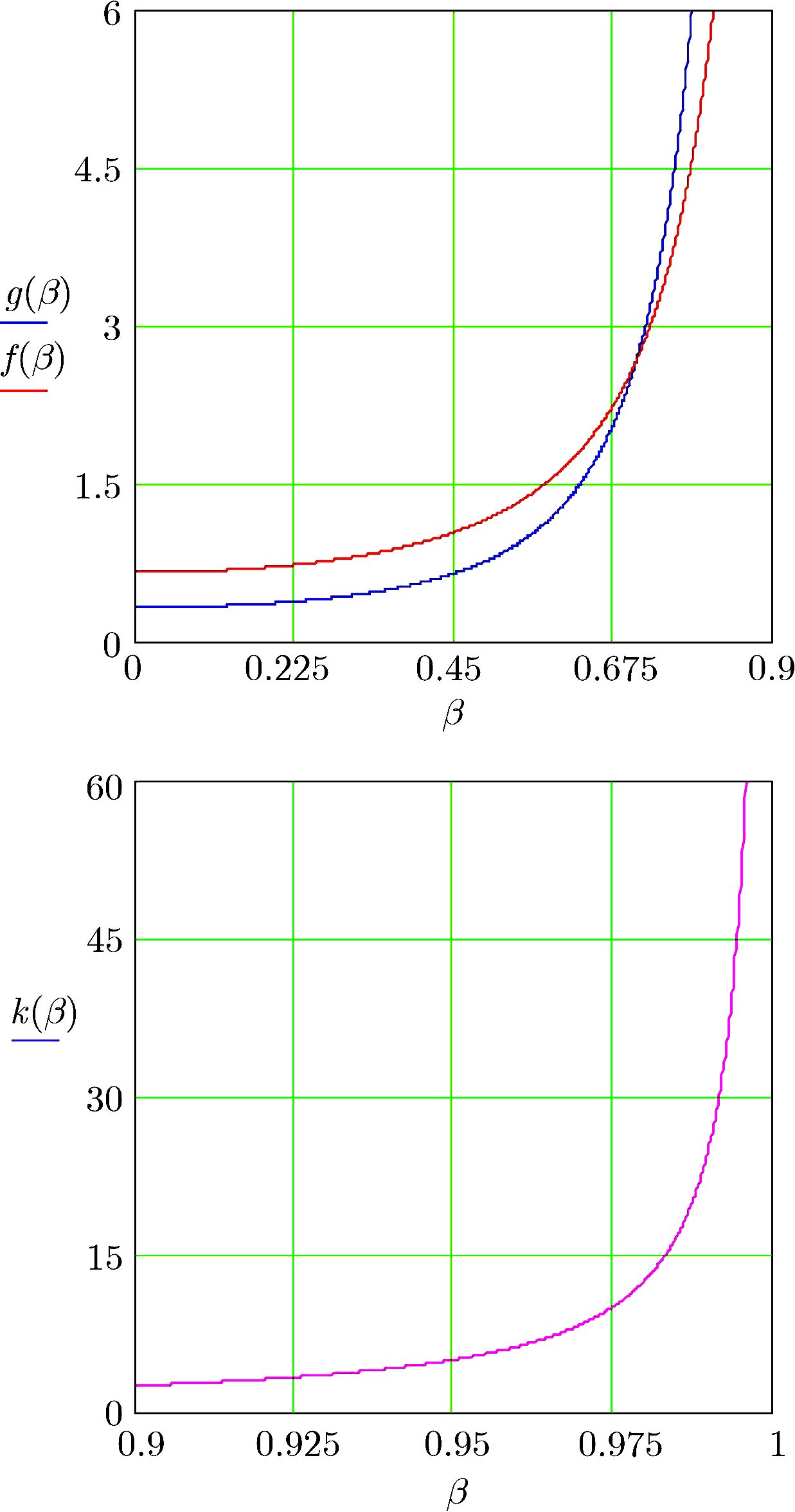}
\vspace{+10pt}
\caption{}
\end{center}
\vspace{-1mm}
\end{figure}

In turn, relations (5.10)--(5.12) and (5.27)--(5.29), together with relations (9.2) and (9.3) from \cite{La}, which establish the connection between the forces acting on a particle and the corresponding accelerations of the particle when its velocity changes only in direction or only in magnitude, respectively, yield the following expressions for the instantaneous powers under consideration in these special cases:
$$
P_{\mathcal{G}^{\, \ell}\, norm}\, =
\, \frac{1}{3}\, \, \frac{e^{\, 2}}{m^{\, 2}\, c^{\, 3}}\, \, \gamma^{\, 4}\, 
\left(\, \frac{d\vec{p}}{dt}\, \right)^{\, 2}\, , 
                                                                                            \eqno (5.35)
$$
$$
P_{\mathcal{G}^{\, \ell}\, coll}\, =
\, \frac{1+5\, \beta^{\, 2}}{3}\, \, \frac{e^{\, 2}}{m^{\, 2}\, c^{\, 3}}\, \, \gamma^{\, 2}\, 
\left(\, \frac{d\vec{p}}{dt}\, \right)^{\, 2}\, , 
                                                                                            \eqno (5.36)
$$
$$
P_{\mathcal{F}\mathcal{G}_{d}^{\, t}\, norm}\, =
\, \frac{2}{3}\, \, \frac{e^{\, 2}}{m^{\, 2}\, c^{\, 3}}\, \, \gamma^{\, 2}\, \left(\, \frac{d\vec{p}}{dt}\, \right)^{\, 2}\, , 
                                                                                            \eqno (5.37)
$$
$$
P_{\mathcal{F}\mathcal{G}_{d}^{\, t}\, coll}\, =
\, \frac{2}{3}\, \, \frac{e^{\, 2}}{m^{\, 2}\, c^{\, 3}}\, \left(\, \frac{d\vec{p}}{dt}\, \right)^{\, 2}\, , 
                                                                                            \eqno (5.38)
$$
$$
P_{\, norm}\, =
\, \frac{3-2\, \beta^{\, 2}}{3}\, \, \frac{e^{\, 2}}{m^{\, 2}\, c^{\, 3}}\, \, 
\gamma^{\, 4}\, \left(\, \frac{d\vec{p}}{dt}\, \right)^{\, 2}\, , 
                                                                                            \eqno (5.39)
$$
$$
P_{\, coll}\, =
\, (\, 1+\beta^{\, 2})\, \, \frac{e^{\, 2}}{m^{\, 2}\, c^{\, 3}}\, \, \gamma^{\, 2}\, 
\left(\, \frac{d\vec{p}}{dt}\, \right)^{\, 2}\, . 
                                                                                            \eqno (5.40)
$$

For a given magnitude of the instantaneous force applied to a particle with a given speed, these expressions demonstrate, in particular, that
$$
P_{\mathcal{G}^{\, \ell}\, norm}\, =\, 
\frac{1}{1+5\, \beta^{\, 2}}\, \, \gamma^{\, 2}\, P_{\mathcal{G}^{\, \ell}\, coll}\, ,             \eqno (5.41)
$$
$$
P_{\mathcal{F}\mathcal{G}_{d}^{\, t}\, norm}\, =\, 
\gamma^{\, 2}\, P_{\mathcal{F}\mathcal{G}_{d}^{\, t}\, coll}\, ,                               \eqno (5.42)
$$
$$
P_{\, norm}\, =\, \frac{3-2\, \beta^{\, 2}}{\, 3\, (\, 1+\beta^{\, 2}\, )}\, \, 
\gamma^{\, 2}\, P_{\, coll}\, ,                                                             \eqno (5.43)
$$

showing that each of these instantaneous radiation powers of a relativistic particle under transverse acceleration exceeds the corresponding instantaneous radiation power under longitudinal acceleration by a factor of approximately $\gamma^2$.

Relation (5.42) coincides with the well-known relation between the instantaneous powers of electromagnetic radiation from an electrically charged particle under transverse and longitudinal acceleration, respectively, for a given magnitude of the instantaneous force applied to a particle moving with a given instantaneous speed \cite{Jack}.

Following \cite{La}, we next consider the elementary instantaneous powers defined by relation (4.1), presenting the corresponding radiation patterns of these powers for different values of $\beta$, while still restricting the analysis to the two special cases in which, at the instant of radiation, the vector $\dot{\vec{\beta}}$ is collinear with the vector $\vec{\beta}$ and in which, at the same instant, the vector $\dot{\vec{\beta}}$ is normal to the vector $\vec{\beta}$.

As before, for comparison, we simultaneously consider such radiation patterns for both longitudinal electrojeitonic and transverse electromagnetic-jeitonic radiations.

\subsection*{5.1. Radiation Patterns of the Elementary Instantaneous Powers of Radiation from a Point Charge for $\dot{\vec\beta}\, coll\, \vec\beta$.}

\addcontentsline{toc}{subsection}{5.1. Radiation Patterns of the Elementary Instantaneous Powers of Radiation from a Point Charge for $\dot{\vec\beta}\, coll\, \vec\beta$.}

In this case, successive use of relations (3.33), (3.43), and (3.53) in (4.1) yields the following relations determining the angular distributions of the elementary instantaneous powers of the radiation under consideration, as recorded at the observation point at time $t$:
$$
dP_{\mathcal{G}^{\ell}} = \frac{e^2\dot{\vec{\beta}}^{\,2}}{4\pi c}
\frac{\cos^2\theta}{(1-\beta\cos\theta)^6}\,d\Omega,
\eqno (5.44)
$$
$$
dP_{\mathcal{F}\mathcal{G}_{d}^{t}} = \frac{e^2\dot{\vec{\beta}}^{\,2}}{4\pi c}
\frac{\sin^2\theta}{(1-\beta\cos\theta)^6}\,d\Omega,
\eqno (5.45)
$$
$$
dP = \frac{e^2\dot{\vec{\beta}}^{\,2}}{4\pi c}
\frac{1}{(1-\beta\cos\theta)^6}\,d\Omega.
\eqno (5.46)
$$
$$
dP_{\mathcal{F}} = \frac{1}{2}dP_{\mathcal{F}\mathcal{G}_{d}^{t}} =
dP_{\mathcal{G}_{d}^{t}}.
\eqno (5.47)
$$

The right-hand side of relation (5.45) again coincides with that of relation \cite[(73.13)]{La}, which gives the elementary instantaneous intensity of electromagnetic radiation from a point electrically charged particle in this particular case.

Relations (5.44)--(5.46), in turn, define the following radiation-pattern functions for the elementary instantaneous powers of the radiation under consideration:
$$
 g(\theta, \, \beta)\, :=\, \frac{\cos^{\, 2}\theta}{(\, 1-\beta\, \cos\theta\, )^{\, 6}},    \eqno (5.48)
 $$
 $$
 f(\theta, \, \beta)\, :=\, \frac{\sin^{\, 2}\theta}{(\, 1-\beta\, \cos\theta\, )^{\, 6}},    \eqno (5.49)
 $$
 $$
 t(\theta, \, \beta)\, :=\, \frac{1}{(\, 1-\beta\, \cos\theta\, )^{\, 6}},    \eqno (5.50)
 $$
 which are plotted for various fixed values of $\beta$ in Figures 5.5--5.11.
\newpage

\begin{figure}[!h]
\begin{center}
\vspace{-6pt}
\includegraphics[width=100mm]{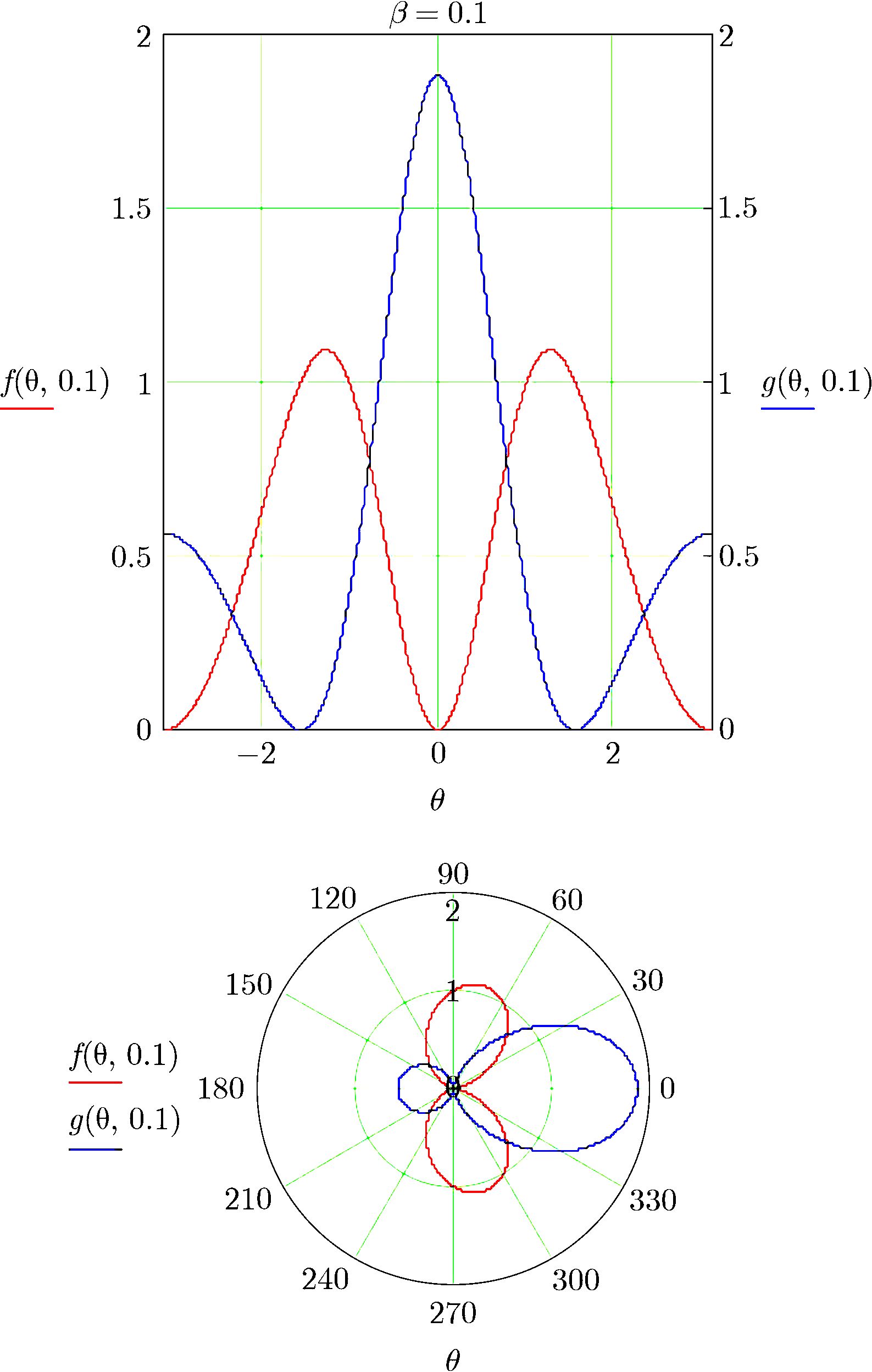}
\vspace{+30pt}
\caption{}
\end{center}
\vspace{-1mm}
\end{figure}
\newpage

\begin{figure}[!h]
\begin{center}
\vspace{-6pt}
\includegraphics[width=100mm]{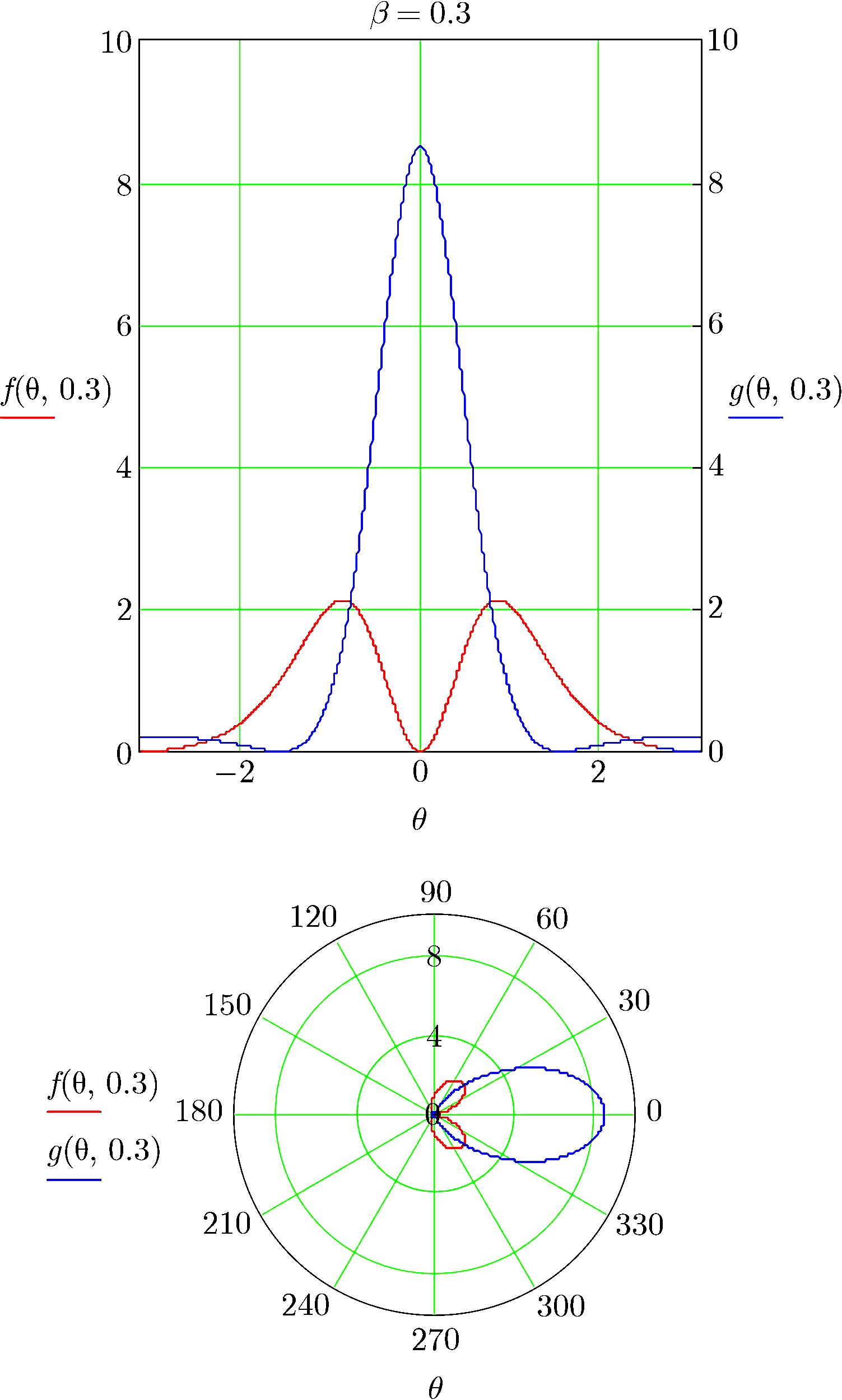}
\vspace{+30pt}
\caption{}
\end{center}
\vspace{-1mm}
\end{figure}
\newpage

\begin{figure}[!h]
\begin{center}
\vspace{-6pt}
\includegraphics[width=110mm]{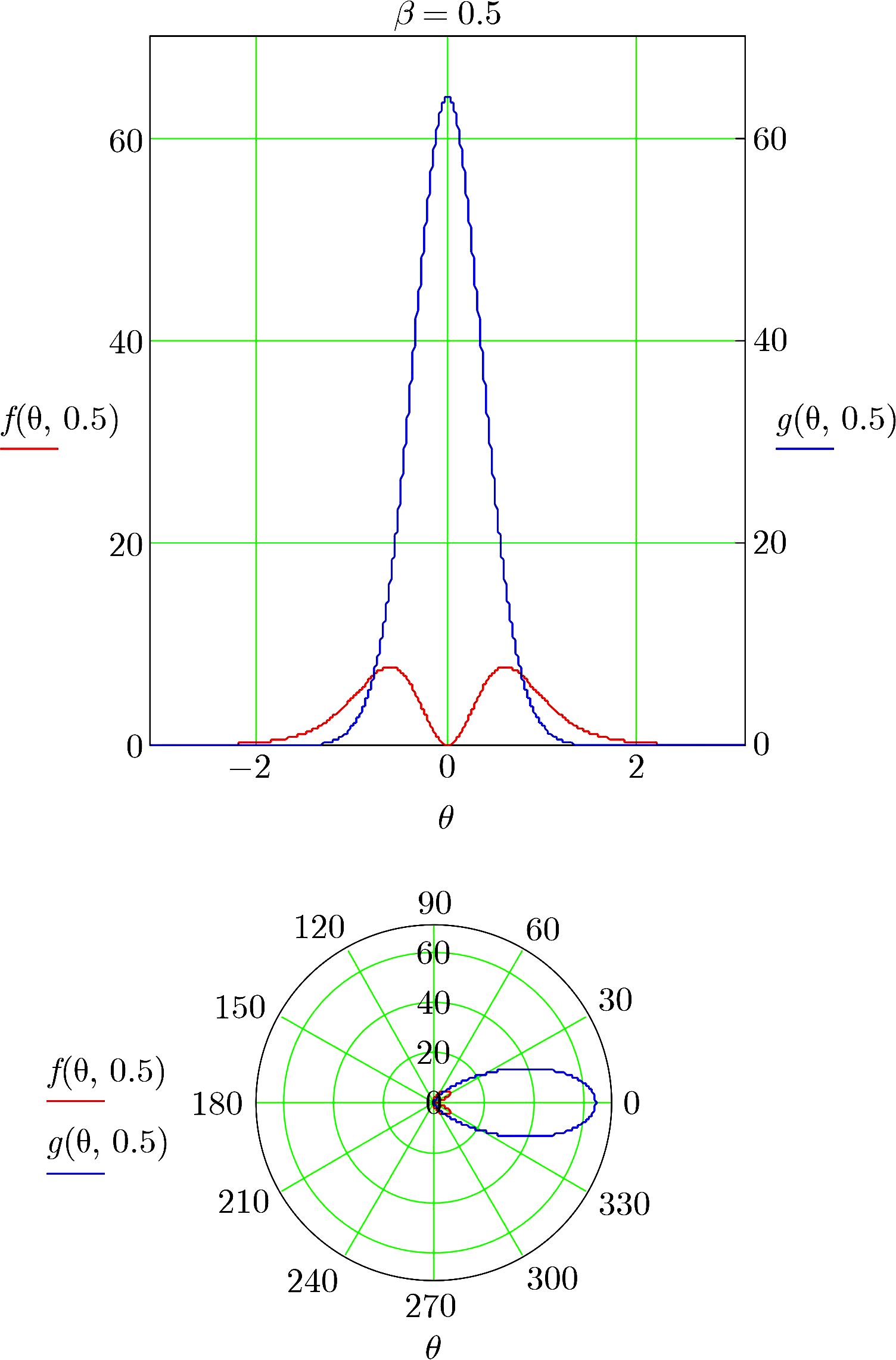}
\vspace{+30pt}
\caption{}
\end{center}
\vspace{-1mm}
\end{figure}
\newpage

\begin{figure}[!h]
\begin{center}
\vspace{-6pt}
\includegraphics[width=120mm]{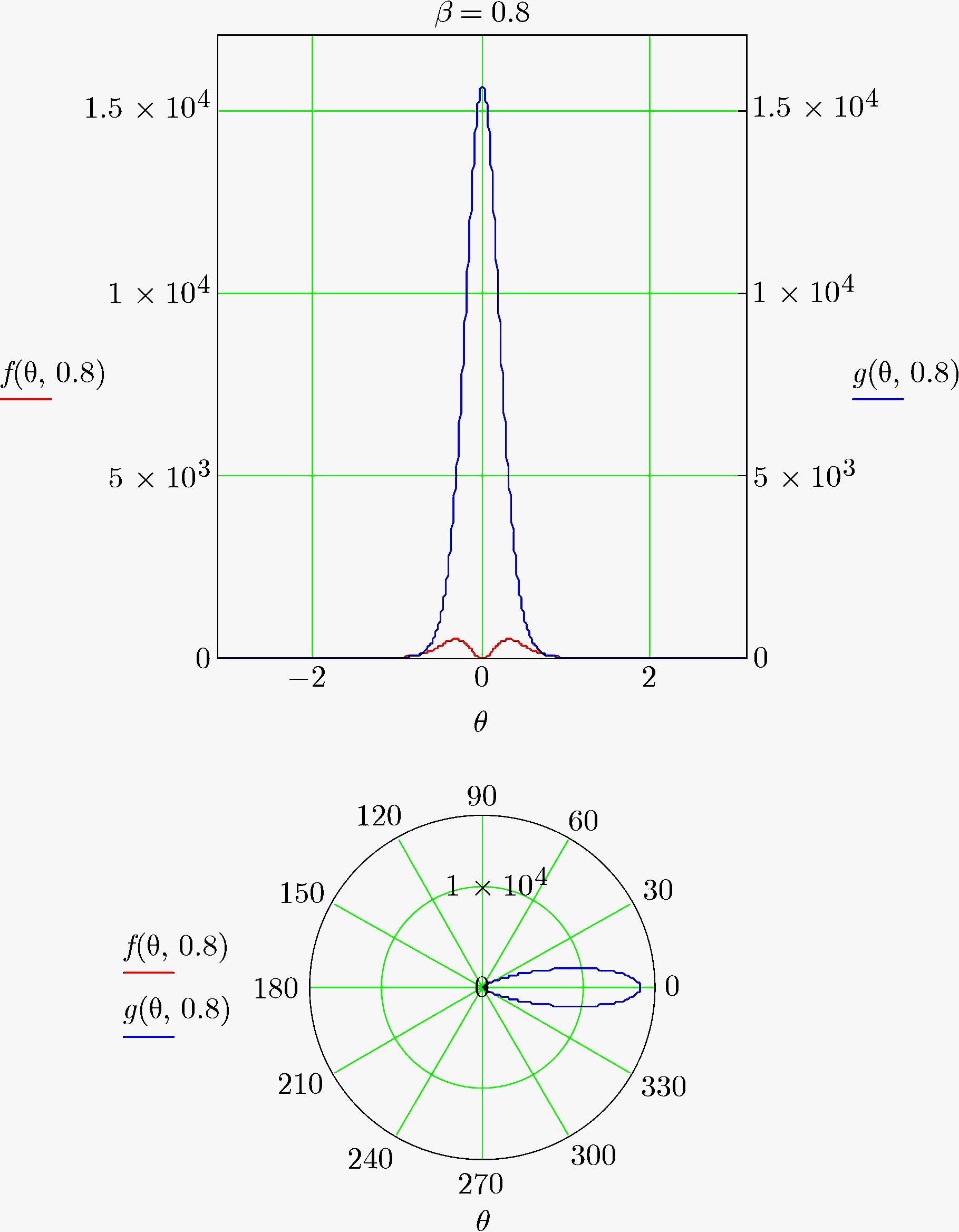}
\vspace{+20pt}
\caption{}
\end{center}
\vspace{-1mm}
\end{figure}
\newpage

\begin{figure}[!h]
\begin{center}
\vspace{-6pt}
\includegraphics[width=90mm]{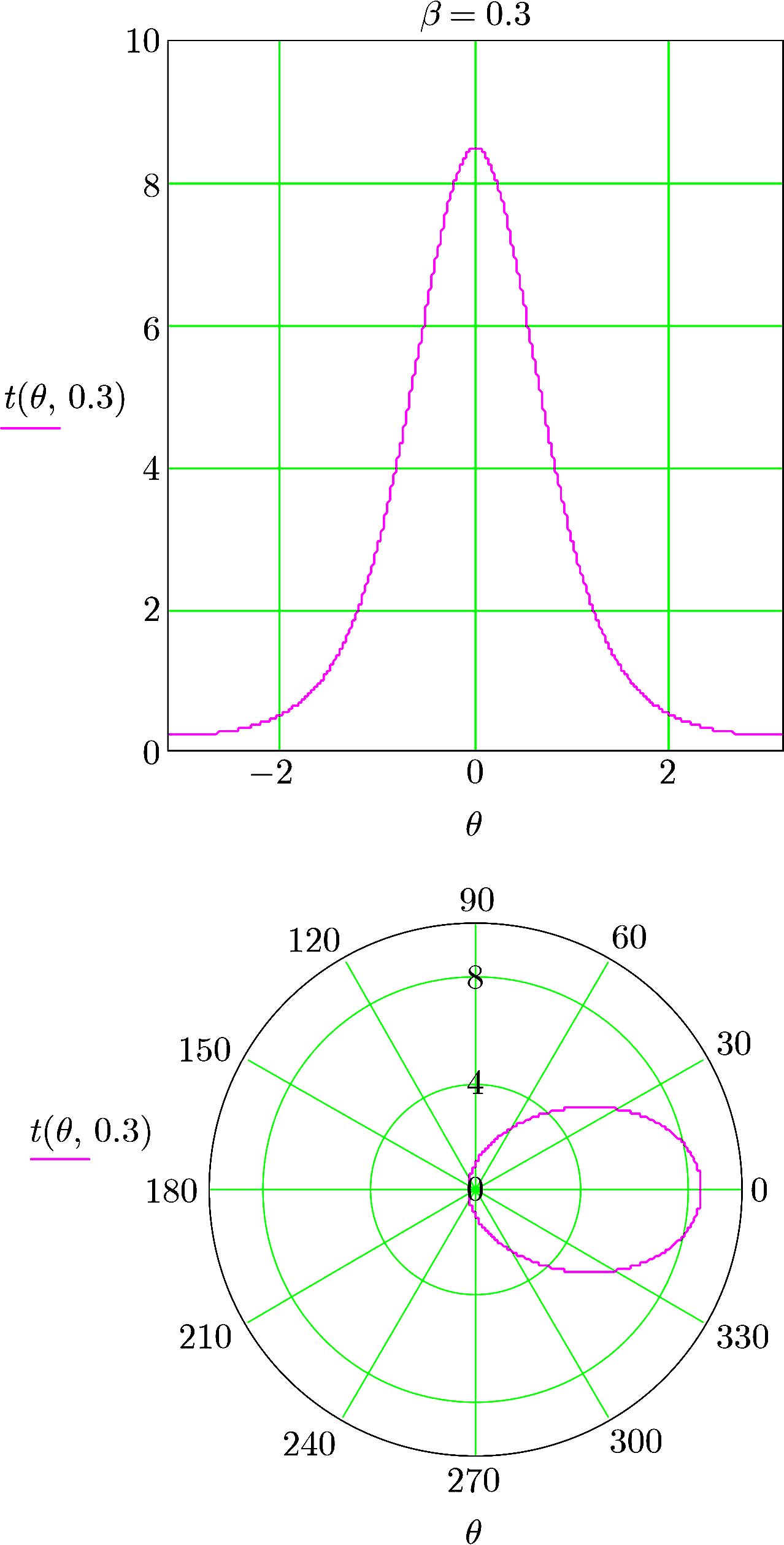}
\vspace{+20pt}
\caption{}
\end{center}
\vspace{-1mm}
\end{figure}
\newpage

\begin{figure}[!h]
\begin{center}
\vspace{-6pt}
\includegraphics[width=90mm]{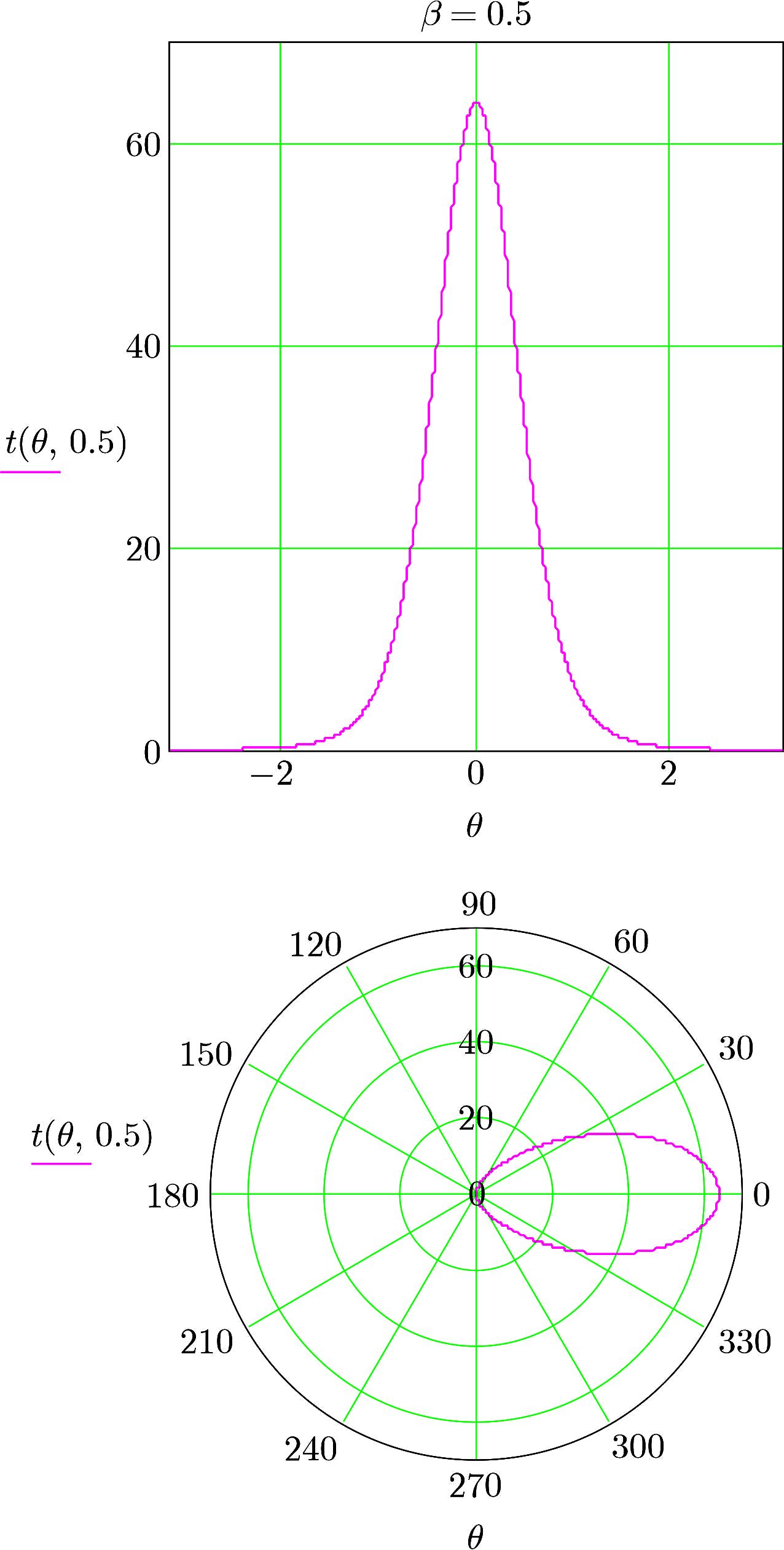}
\vspace{+20pt}
\caption{}
\end{center}
\vspace{-1mm}
\end{figure}
\newpage

\begin{figure}[!h]
\begin{center}
\vspace{-6pt}
\includegraphics[width=100mm]{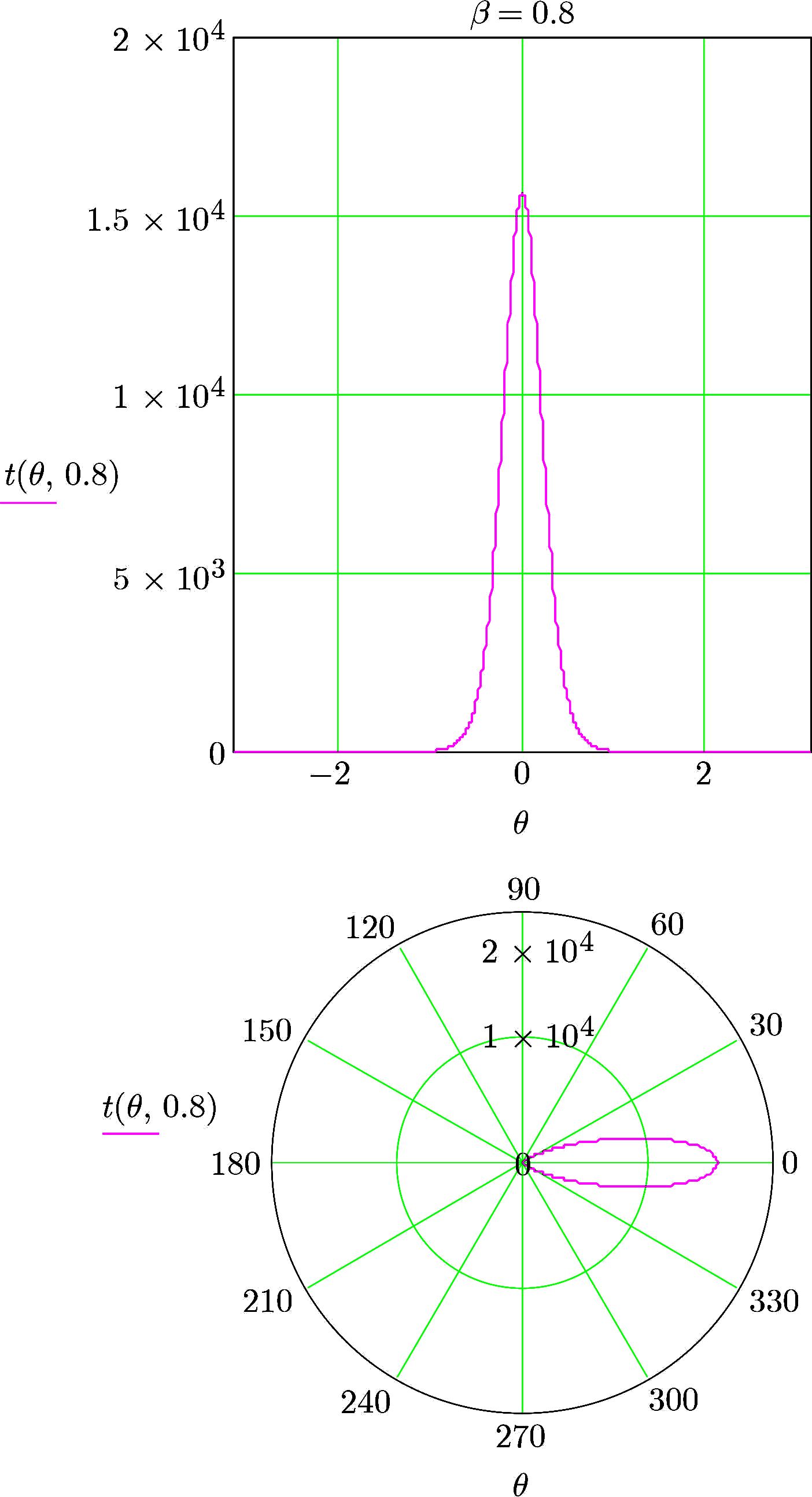}
\vspace{+20pt}
\caption{}
\end{center}
\vspace{-1mm}
\end{figure}
\newpage

\subsection*{5.2. Radiation Patterns of the Elementary Instantaneous Powers of Radiation from a Point charge for $\dot{\vec\beta}\, norm\, \vec\beta$.}

\addcontentsline{toc}{subsection}{5.2. Radiation Patterns of the Elementary Instantaneous Powers of Radiation from a Point charge for $\dot{\vec\beta}\, norm\, \vec\beta$.}

In this case, successive use of relations (3.33), (3.43), and (3.53) in (4.1) yields the following relations determining the angular distributions of the elementary instantaneous powers of the radiation under consideration, as recorded at the observation point at time $t$:
$$
 dP_{\mathcal{G}^{\, \ell}}\, =\, \frac{e^{\, 2}\, \dot{\vec{\beta}}^{\, \, 2}}{4\, \pi\, c}\, \, 
 \frac{\sin^{\, 2}\theta\, \cos^{\, 2}\varphi}{(\, 1-\beta\, \cos\theta\, )^{\, 6}}\, \, d\, \Omega\, , 
                                                                                         \eqno (5.51)
$$
$$
 dP_{\mathcal{F}\mathcal{G}_{d}^{\, t}}\, =\, \frac{e^{\, 2}\, \dot{\vec{\beta}}^{\, \, 2}}{4\, \pi\, c}\, \, 
 \left(\frac{1}{(\, 1-\beta\, \cos\theta\, )^{\, 4}}-\frac{(1-\beta^{\, 2})\sin^{\, 2}\theta\, \cos^{\, 2}\varphi}{(\, 1-
 \beta\, \cos\theta\, )^{\, 6}}\right)\, d\, \Omega\, , 
                                                                                             \eqno (5.52)
$$
$$
 dP\, =\, \frac{e^{\, 2}\, \dot{\vec{\beta}}^{\, \, 2}}{4\, \pi\, c}\, \, 
 \left(\frac{1}{(\, 1-\beta\, \cos\theta\, )^{\, 4}}+\frac{\beta^{\, 2}\, \sin^{\, 2}\theta\, \cos^{\, 2}\varphi}{(\, 1-
 \beta\, \cos\theta\, )^{\, 6}}\right)\, d\, \Omega\, , 
                                                                                             \eqno (5.53)
$$
$$
dP_{\, \mathcal{F}}\, =\, \frac{1}{2}\, dP_{\, \mathcal{F}\mathcal{G}_{d}^{\, t}}\, =
\, dP_{\, \mathcal{G}_{d}^{\, t}}\, .                                                           \eqno (5.54)
$$

The right-hand side of relation (5.52) again coincides with that of relation \cite[(73.14)]{La}, which gives the elementary instantaneous intensity of electromagnetic radiation from a point electrically charged particle when, at the instant of radiation, the velocity and acceleration of the particle are mutually perpendicular.

Relations (5.51)--(5.53), in turn, define the following radiation-pattern functions for the elementary instantaneous powers of the radiation under consideration
$$
g(\theta, \varphi, \beta):=\, \frac{\sin^{\, 2}\theta\, \cos^{\, 2}\varphi}{(\, 1-
\beta\, \cos\theta\, )^{\, 6}}\, ,                                                     \eqno (5.55)
$$
$$
f(\theta, \varphi, \beta):=\, \frac{1}{(\, 1-\beta\, \cos\theta\, )^{\, 4}}-
\frac{(1-\beta^{\, 2})\sin^{\, 2}\theta\, \cos^{\, 2}\varphi}{(\, 1-
 \beta\, \cos\theta\, )^{\, 6}}\, ,                                              \eqno (5.56)
$$
$$
t(\theta, \varphi, \beta):=\, \frac{1}{(\, 1-\beta\, \cos\theta\, )^{\, 4}}+
\frac{\beta^{\, 2}\, \sin^{\, 2}\theta\, \cos^{\, 2}\varphi}{(\, 1-
 \beta\, \cos\theta\, )^{\, 6}}\, .                                               \eqno (5.57)
$$

Relations (5.55)--(5.57), in turn, yield the following relations determining the angular distributions of the elementary instantaneous powers of the radiation under consideration in the osculating plane $\mathcal{Q}(t')$ $(\varphi=0)$, which are plotted for various fixed values of $\beta$ in Figures 5.12--5.22.
 $$
g(\theta, \beta)\, :=\, g(\theta, 0, \beta)=\, \frac{\sin^{\, 2}\theta}{(\, 1-
\beta\, \cos\theta\, )^{\, 6}}\, ,                                                     \eqno (5.58)
$$
$$
f(\theta, \beta)\, :=\, f(\theta, 0, \beta)=\, \frac{(\, \cos\theta-\beta\, )^{\, 2}}{(\, 1-\beta\, \cos\theta\, )^{\, 6}}\, , 
                                                                                      \eqno (5.59)
$$
$$
t(\theta, \beta)\, :=\, t(\theta, 0, \beta)=\, \frac{(\, \cos\theta-\beta\, )^{\, 2}+\sin^{\, 2}\theta}{(\, 1-\beta\, \cos\theta\, )^{\, 6}}\, .
                                                                                          \eqno (5.60)
$$

\newpage
\begin{figure}[!h]
\begin{center}
\vspace{-6pt}
\includegraphics[width=110mm]{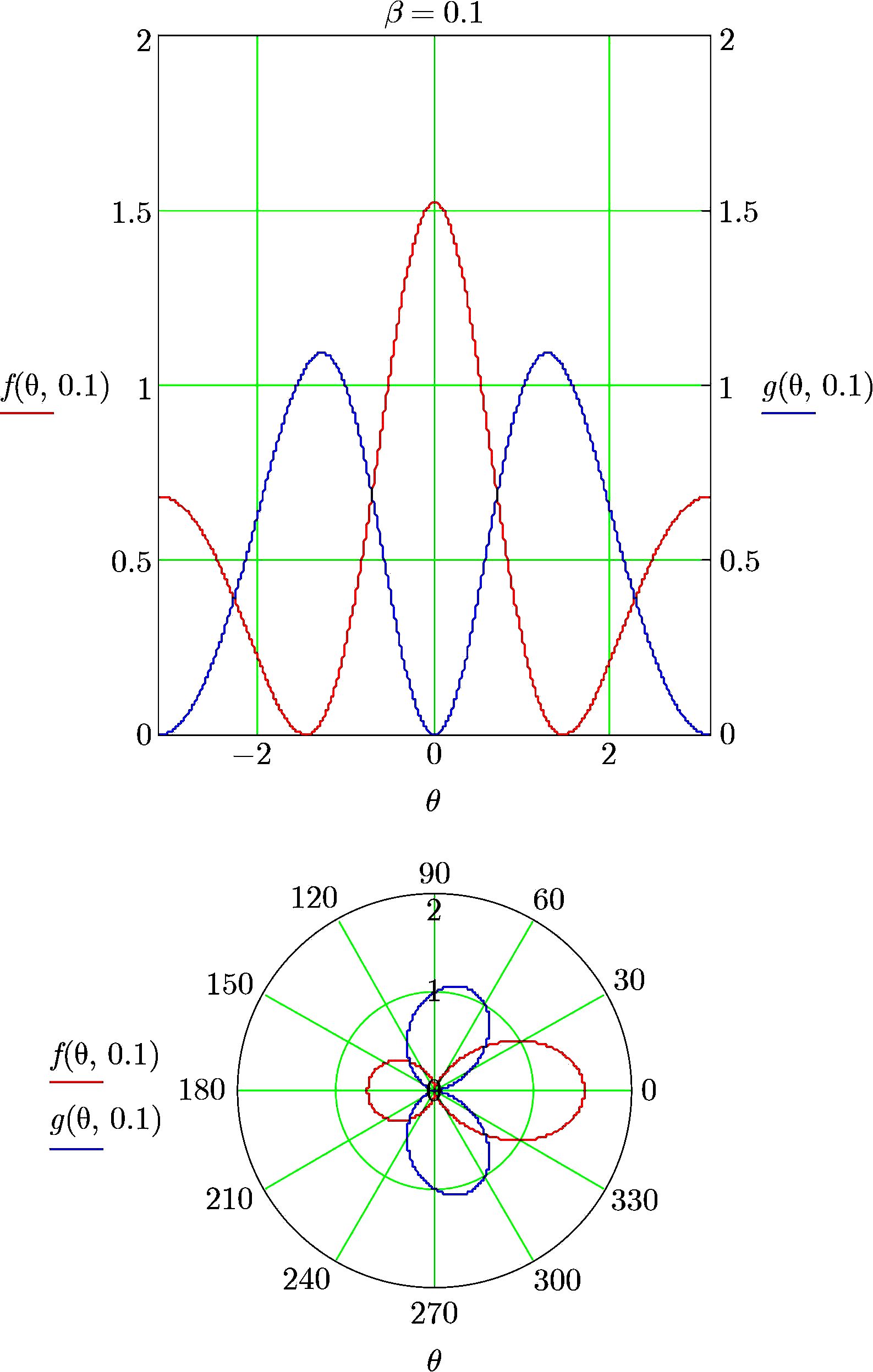}
\vspace{+20pt}
\caption{}
\end{center}
\vspace{-1mm}
\end{figure}
\newpage

\begin{figure}[!h]
\begin{center}
\vspace{-6pt}
\includegraphics[width=110mm]{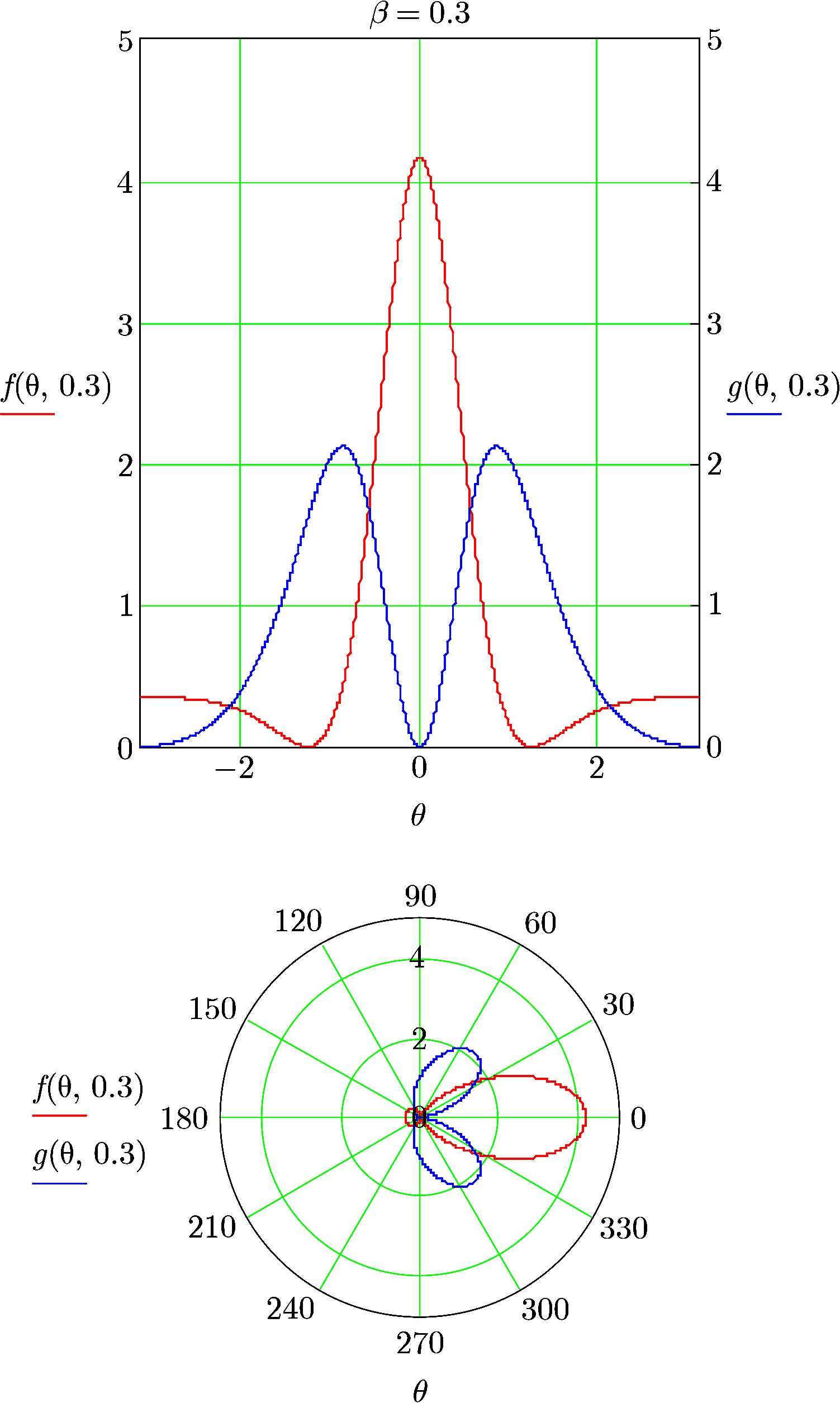}
\vspace{+20pt}
\caption{}
\end{center}
\vspace{-1mm}
\end{figure}
\newpage

\begin{figure}[!h]
\begin{center}
\vspace{-6pt}
\includegraphics[width=110mm]{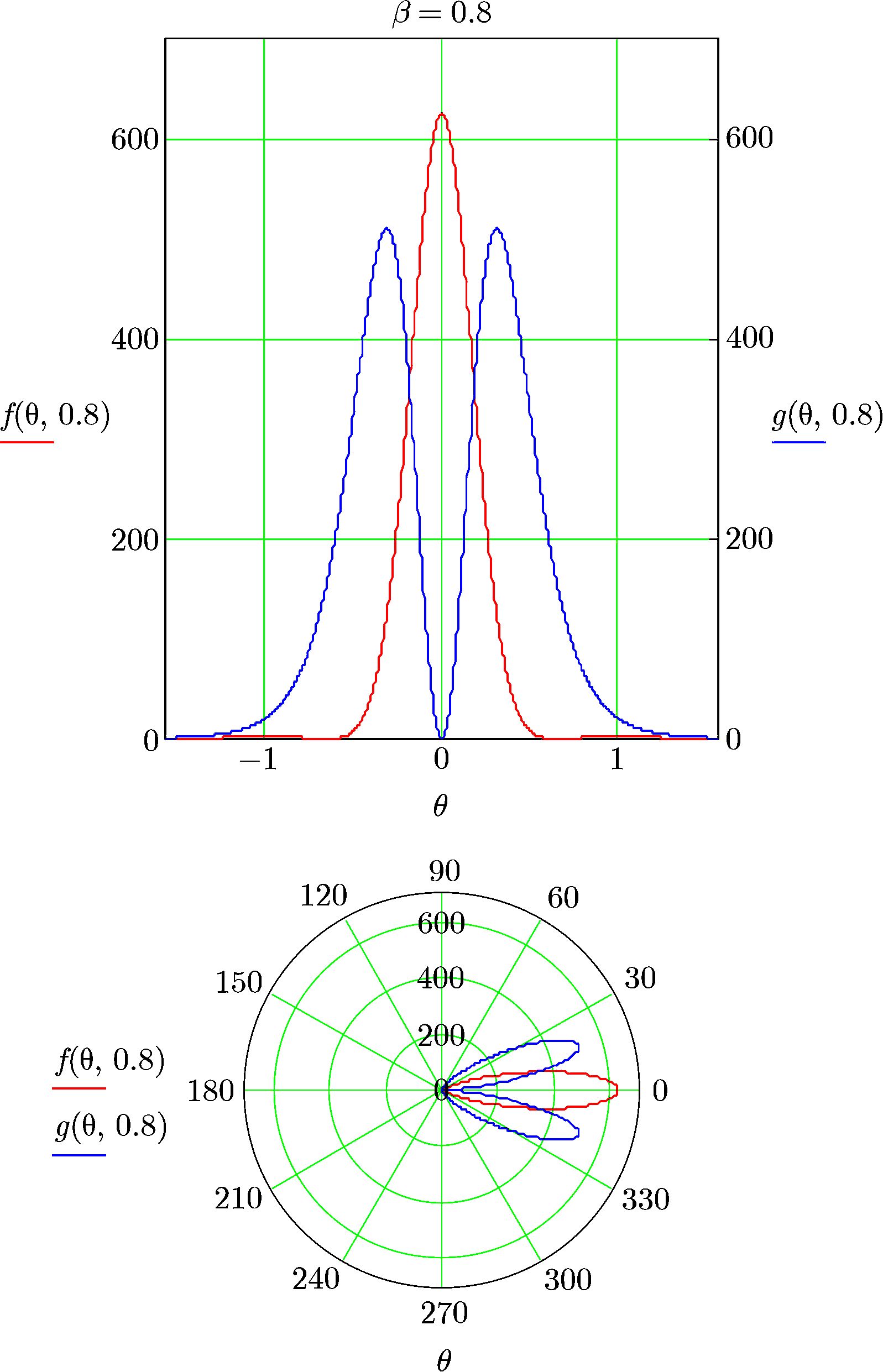}
\vspace{+20pt}
\caption{}
\end{center}
\vspace{-1mm}
\end{figure}
\newpage

\begin{figure}[!h]
\begin{center}
\vspace{-6pt}
\includegraphics[width=120mm]{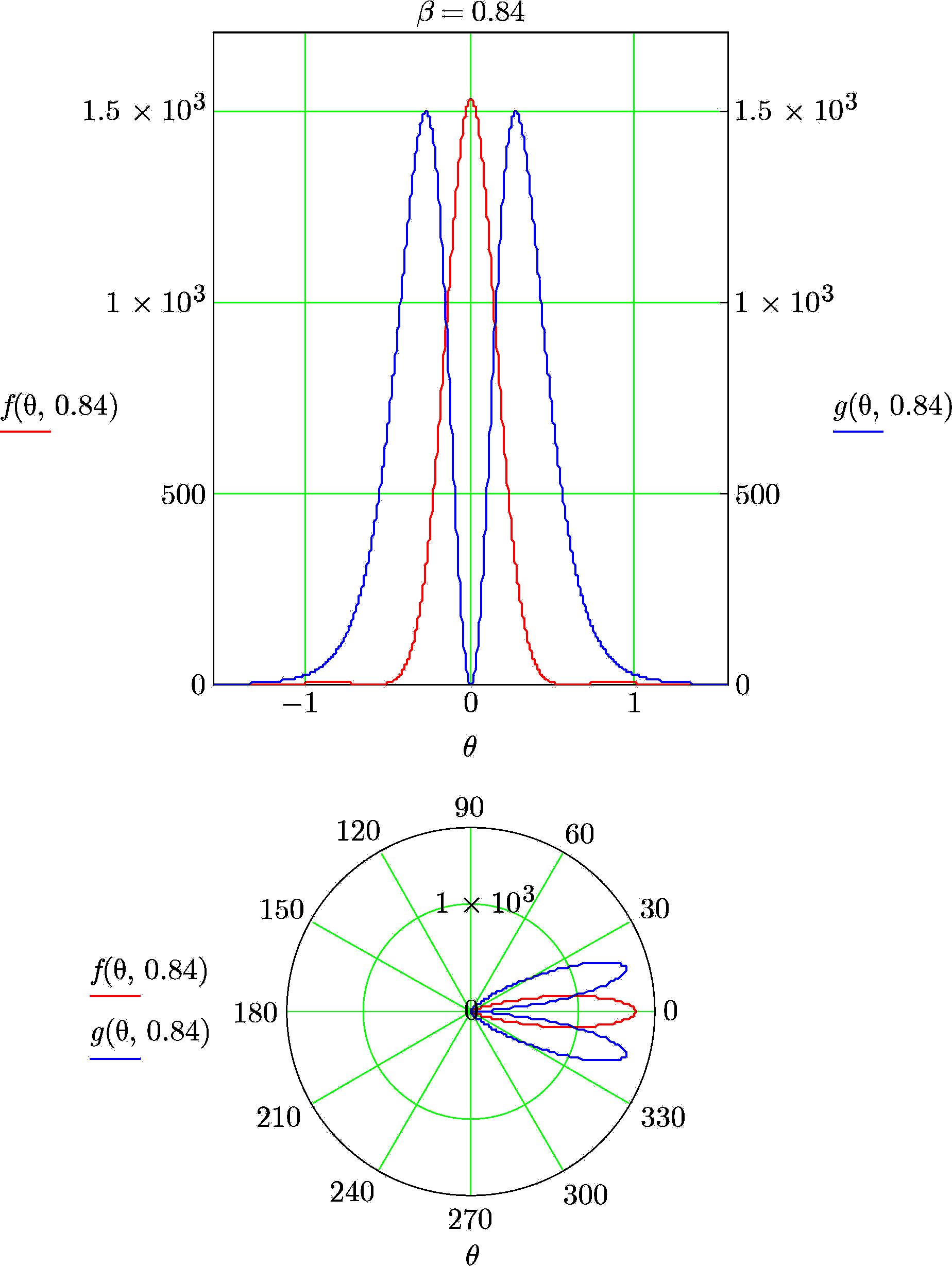}
\vspace{+20pt}
\caption{}
\end{center}
\vspace{-1mm}
\end{figure}
\newpage

\begin{figure}[!h]
\begin{center}
\vspace{-6pt}
\includegraphics[width=110mm]{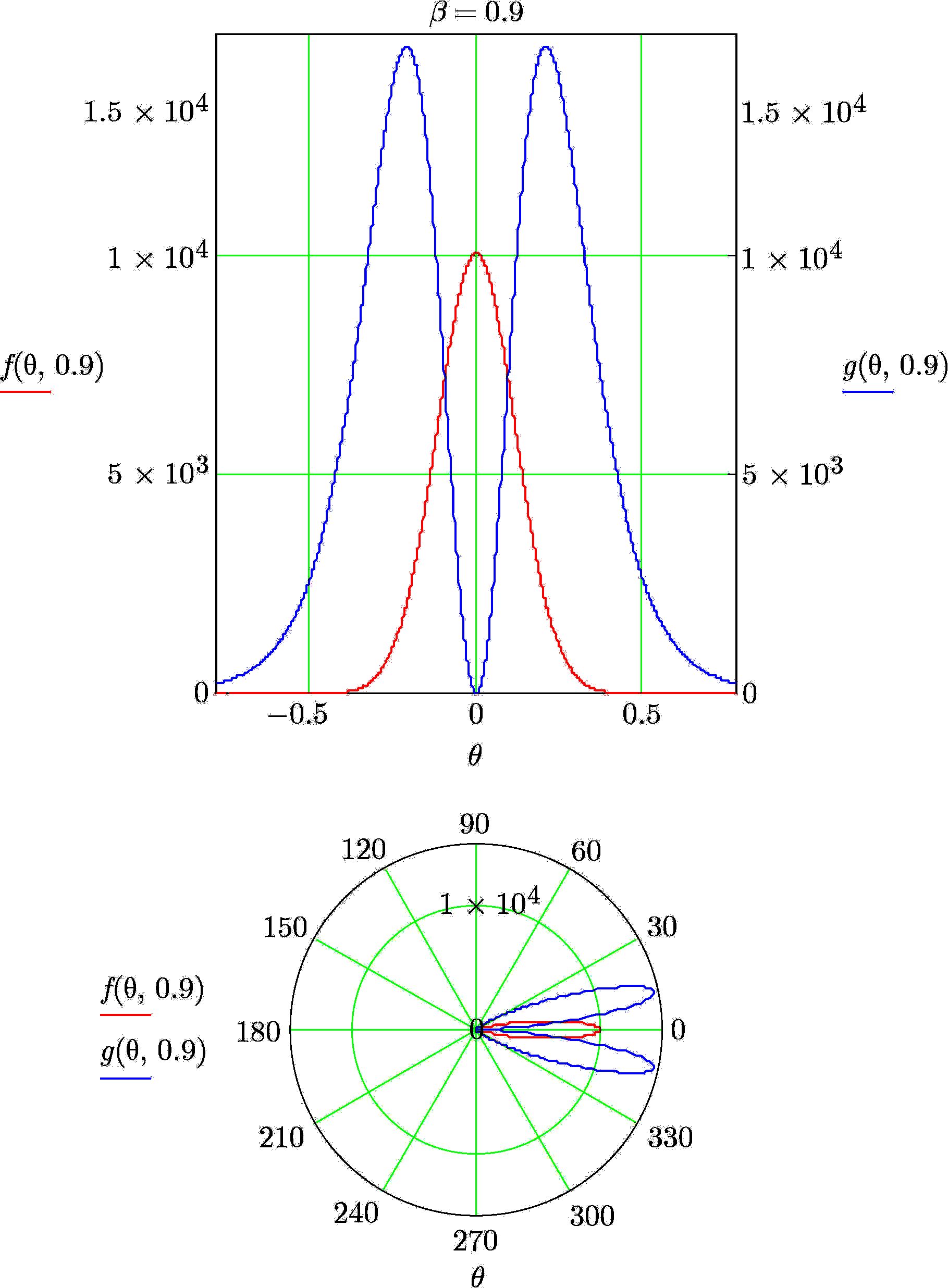}
\vspace{+20pt}
\caption{}
\end{center}
\vspace{-1mm}
\end{figure}
\newpage

\begin{figure}[!h]
\begin{center}
\vspace{-6pt}
\includegraphics[width=120mm]{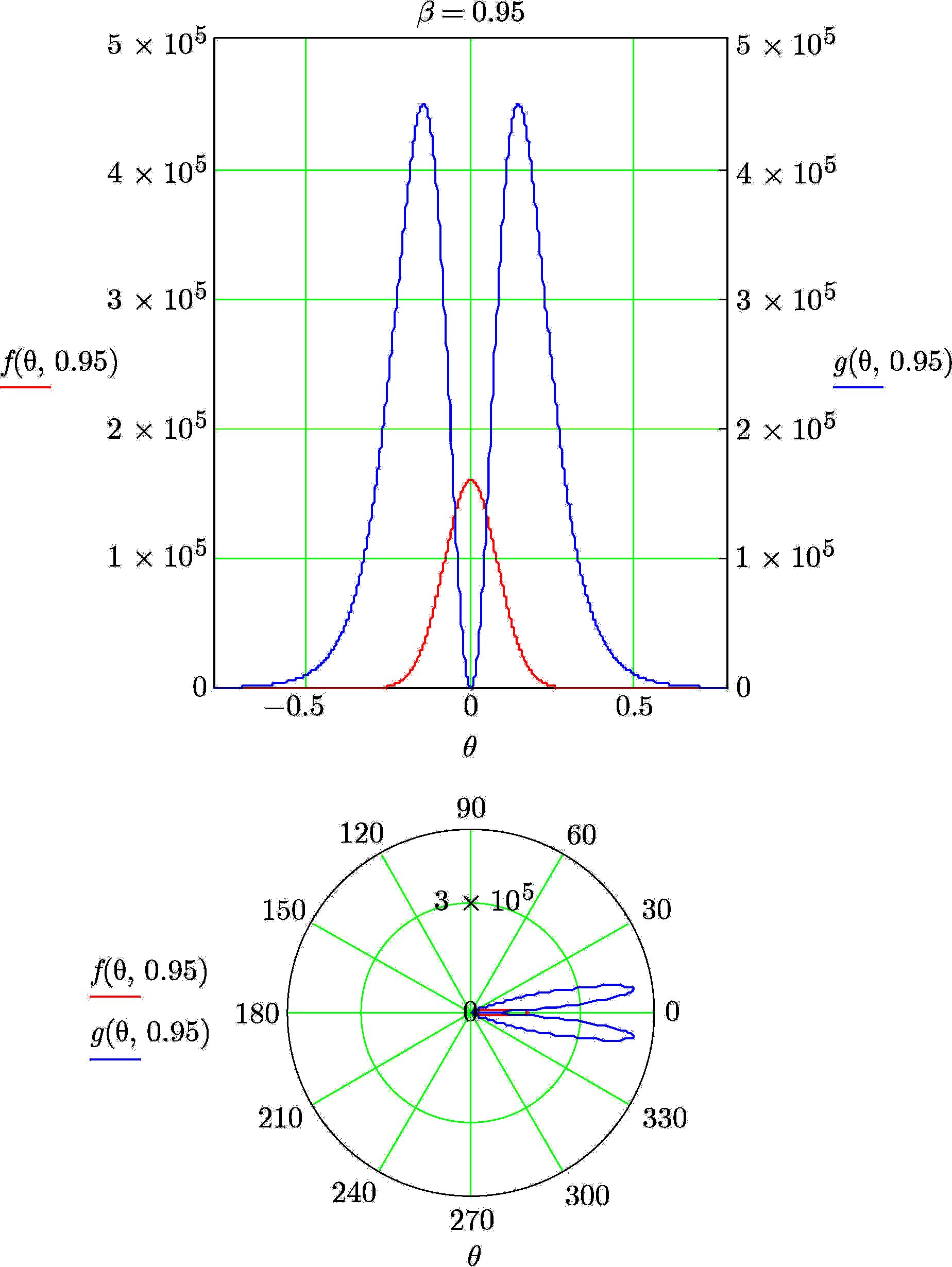}
\vspace{+20pt}
\caption{}
\end{center}
\vspace{-1mm}
\end{figure}
\newpage

\begin{figure}[!h]
\begin{center}
\vspace{-6pt}
\includegraphics[width=120mm]{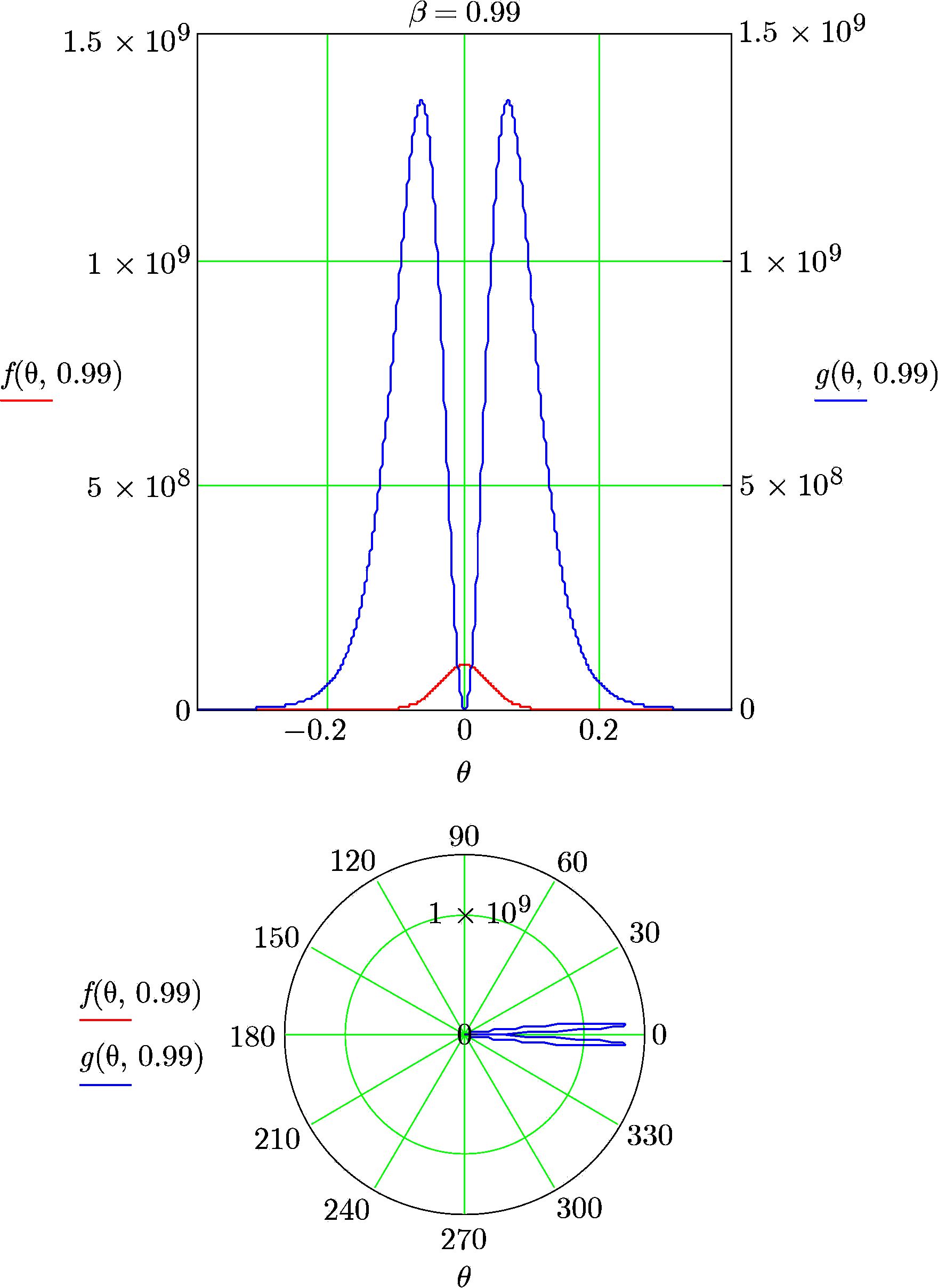}
\vspace{+20pt}
\caption{}
\end{center}
\vspace{-1mm}
\end{figure}
\newpage

\begin{figure}[!h]
\begin{center}
\vspace{-6pt}
\includegraphics[width=90mm]{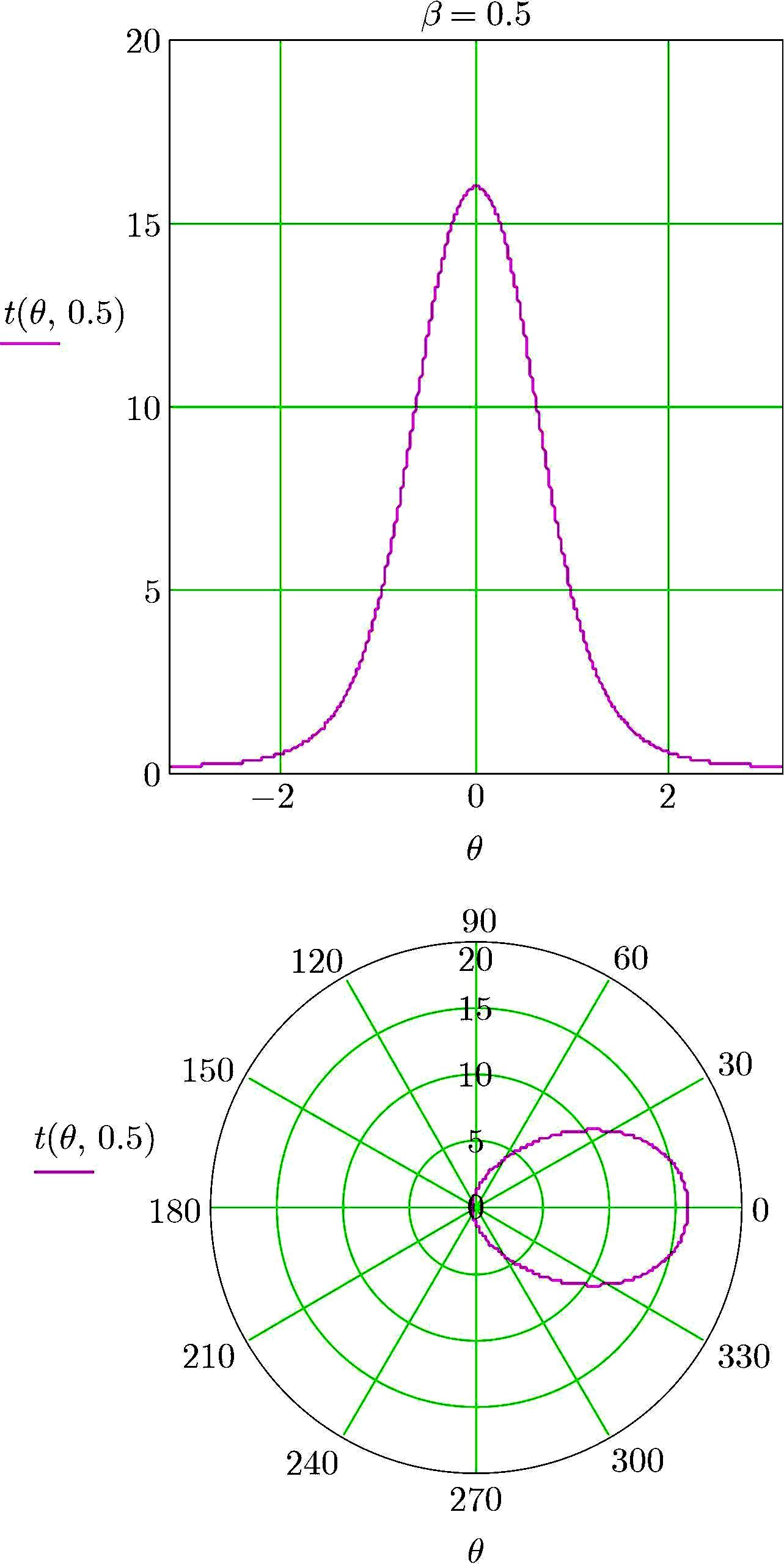}
\vspace{+20pt}
\caption{}
\end{center}
\vspace{-1mm}
\end{figure}
\newpage

\begin{figure}[!h]
\begin{center}
\vspace{-6pt}
\includegraphics[width=90mm]{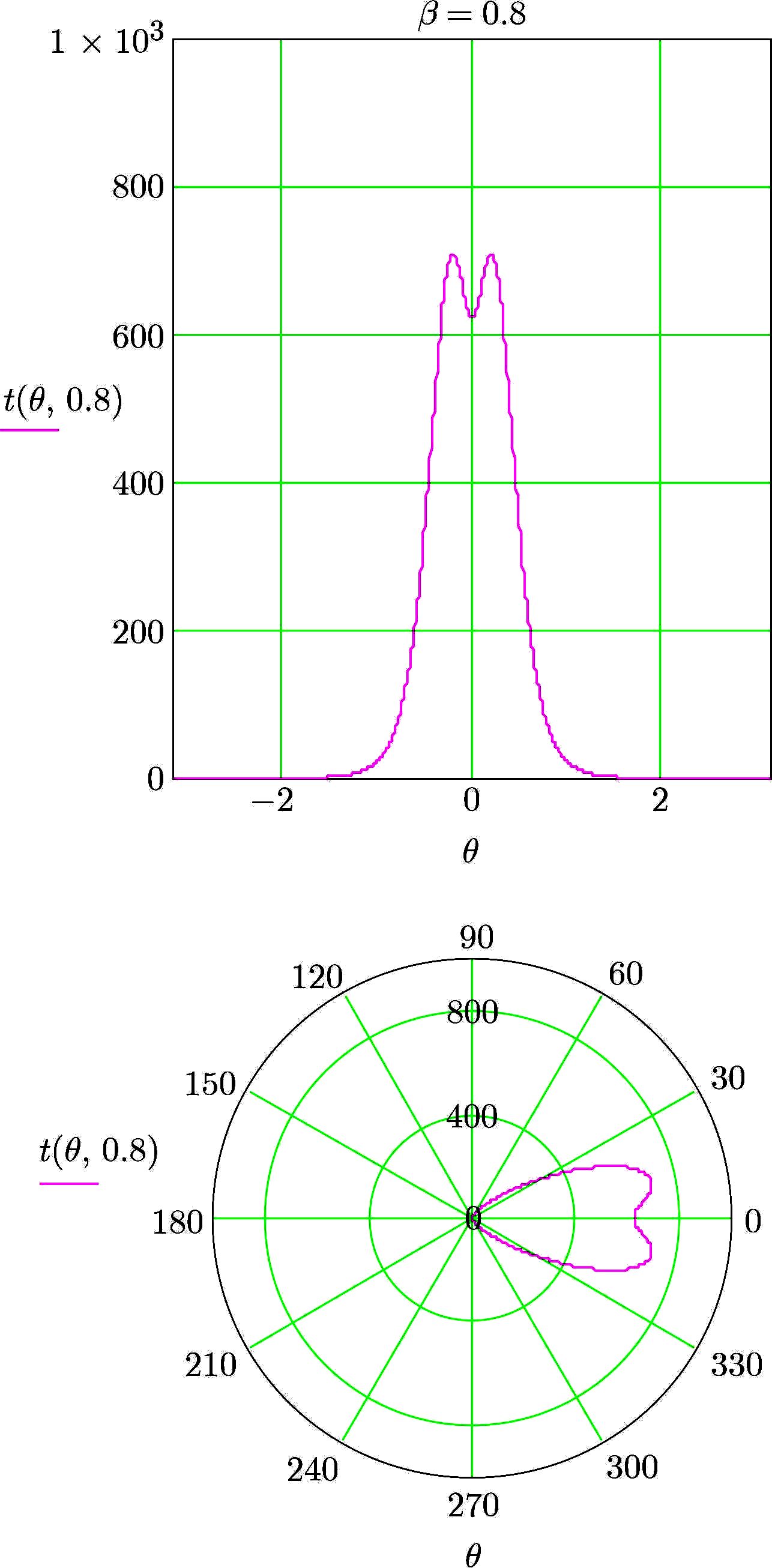}
\vspace{+20pt}
\caption{}
\end{center}
\vspace{-1mm}
\end{figure}
\newpage

\begin{figure}[!h]
\begin{center}
\vspace{-6pt}
\includegraphics[width=90mm]{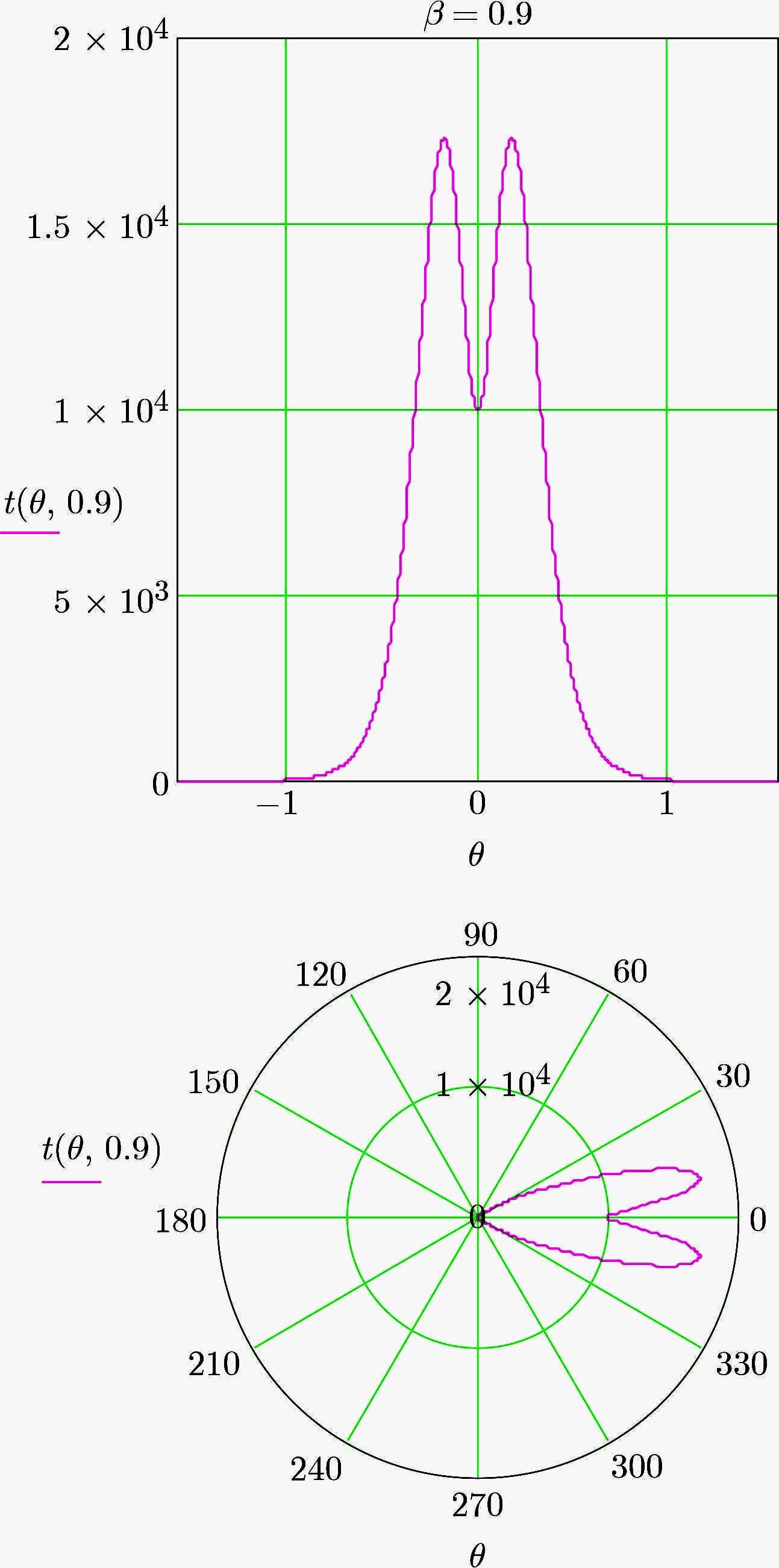}
\vspace{+20pt}
\caption{}
\end{center}
\vspace{-1mm}
\end{figure}
\newpage

\begin{figure}[!h]
\begin{center}
\vspace{-6pt}
\includegraphics[width=90mm]{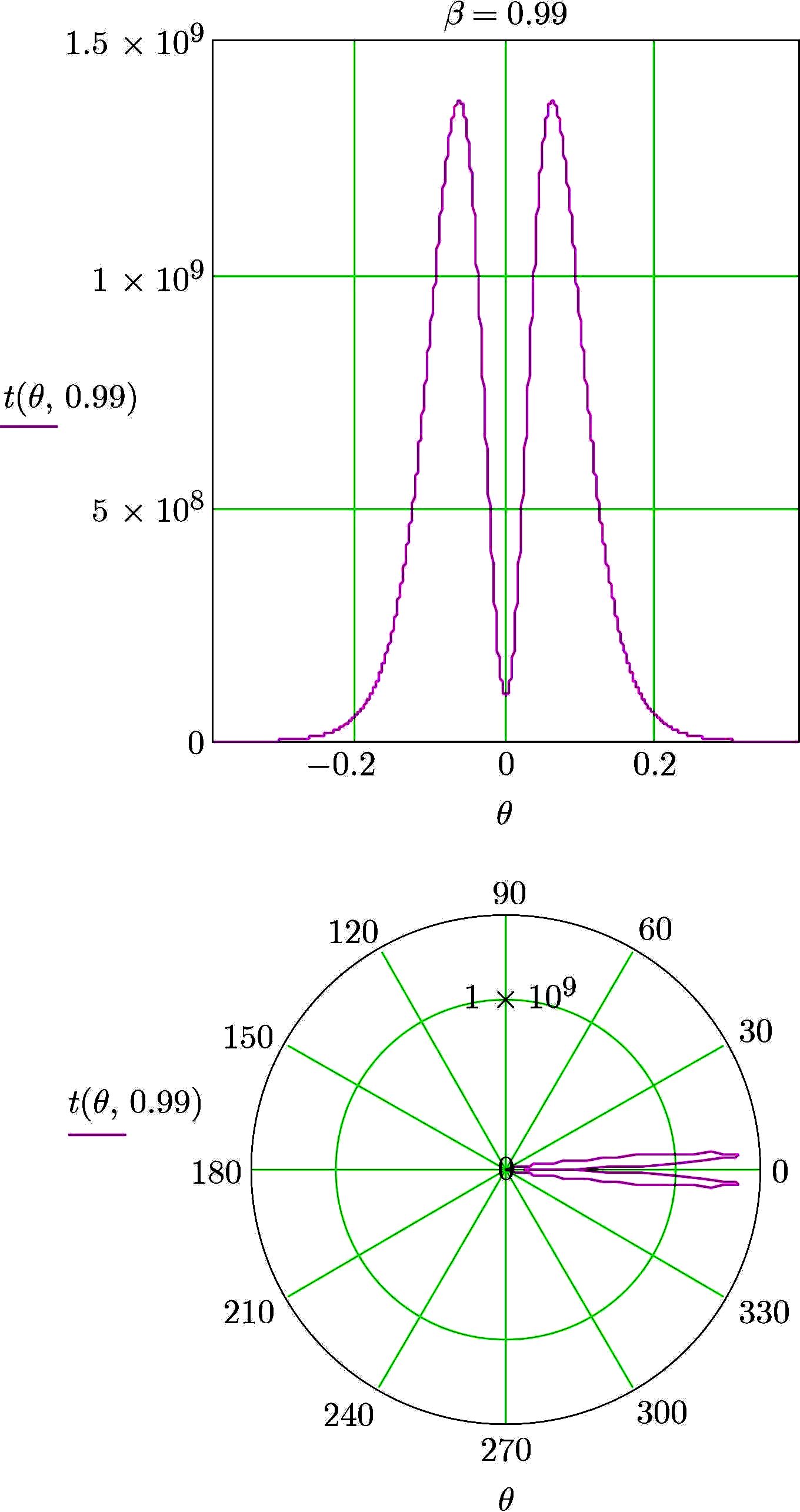}
\vspace{+20pt}
\caption{}
\end{center}
\vspace{-1mm}
\end{figure}
\newpage

The radiation patterns shown in Figures 5.5--5.8, together with relation (5.14), demonstrate that the ``number of unusual and interesting effects for relativistic particles'' \cite{Jack} is supplemented by the \emph{effect (phenomenon) of dominance of the longitudinal electrojeitonic radiation} of such particles over the well-known transverse electromagnetic-jeitonic radiation of these particles.

An even more unusual and interesting effect of the dominance of longitudinal electrojeitonic radiation from ultrarelativistic particles is demonstrated, together with relation (5.31), by the radiation patterns shown in Figures 5.12--5.18.

As a consequence of this dominance, we also obtain an ``unusual'' topology of the radiation patterns shown in Figures 5.20--5.22.

Thus, both for $\dot{\vec{\beta}}\, coll \, \vec{\beta}$ and for $\dot{\vec{\beta}}\, norm \, \vec{\beta}$, the radiation properties of an ultrarelativistic point electrically charged particle are determined predominantly by the properties of the dominant component, represented by the longitudinal electrojeitonic radiation of the particle.

Since, at any instant, the radiation from a point electrically charged particle undergoing arbitrary curvilinear motion can also be regarded as a coherent superposition of the radiations associated with the normal and tangential components of the particle's instantaneous acceleration~\cite{Jack}, the dominance of longitudinal electrojeitonic radiation under consideration extends to the case of arbitrary curvilinear motion of such a particle.

\section{Simplest Radiating Systems of Electric Charges and Currents}

\subsection*{\S \, 6. Simplest Radiating Systems of Electric Charges and Currents}
\addcontentsline{toc}{subsection}{\S \, 6. Simplest Radiating Systems of Electric Charges and Currents}
\setcounter{section}{6}
\setcounter{figure}{0}

As in \cite[Ch.~9]{Jack}, we consider examples of the simplest radiating systems with a simple geometry of current distribution, such that the integral defining the vector potential can be evaluated in a relatively simple closed form \cite[Ch.~9]{Jack}. As an example of such a radiating system, we consider a linear traveling-wave electric antenna \cite[Ch.~X,~\S1]{Eis} formed by a continuous straight cylindrical conductor, which is long compared with $\lambda$, and is located in free space (the main radiating part of the Beverage antenna and the OBE-2 Kharchenko antenna).
A distinctive feature of such an antenna is that the phase shift between the radiation-field strengths created by its elementary sections in the direction of the OZ axis \cite[Fig.~1.1.X]{Eis}, at $V\rightarrow C$, tends to zero \cite[Ch.~X,~\S1]{Eis}, creating favorable conditions for the generating radiation in the direction of this axis.
Let the conductor be located along the OZ axis, directed in the same direction as the propagation of the traveling current wave in it, and let the beginning of the conductor be located at the origin of the OZ axis (Fig.~6.1; see also \cite[Fig.~1.2.X]{Eis}).

\begin{figure}[H]
\begin{center}
\vspace{0pt}
\includegraphics[width=80mm]{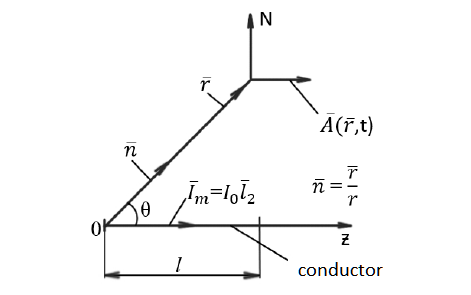}
\vspace{-2pt}
\caption{Geometry used to derive the radiation-pattern functions of a linear traveling-wave electric antenna.}
\end{center}
\end{figure}

This antenna differs from the well-known linear electric vibrators \cite[Ch. VII, \S1]{Eis} in that, for $v\rightarrow C$, where $v$ is the phase velocity of the traveling current wave in the conductor, the radiation generated by any elementary section of the antenna in the direction of the OZ axis is in phase \cite[Ch. X, \S1]{Eis}. This provides favorable conditions for the generating longitudinal electrojeitonic waves by the antenna along this axis.
The symmetric surface wave \cite[Fig. 1.10. IV]{Eis} propagating along the conductor, like any other surface wave, is also characterized by a significant concentration of its field energy in the vicinity of the conductor, thereby favoring a higher concentration of radiation energy along this axis. \cite[Ch. IV, \S10]{Eis}
The asymptotic expression for the retarded vector potential, $\vec{A}(\vec{r},t)$ \cite[Ch. II.\S1]{Med}, generated by this antenna, for zero attenuation coefficient $\beta=0$, no reflections at the ends of the antenna, and $V\rightarrow C$, is given by relation \cite[Ch. X, \S2]{Eis}

\begin{equation}\label{bla1}
  \vec{A}(\vec{r}, t) = \frac{I_0 \vec{e_z} e^{-i\omega t}}{cr} \int_{0}^{l} e^{-ikz(\cos\theta -1)} dz :=B_0 \vec{e_z} e^{-i\omega t'} \int_{0}^{l} e^{-ikz(\cos\theta -1)} dz
  \tag{1}
\end{equation}

\begin{equation}\label{bla2}
  t'=t-r/c
  \tag{1.1}
\end{equation}

\begin{equation}\label{bla3}
  B_0:=\frac{I_0}{cr}
  \tag{1.2}
\end{equation}

After evaluating the integral on the right-hand side, this relation takes the form:

\begin{equation}\label{bla4}
  \vec{A}(\vec{r}, t) =B_0 \vec{e_z} e^{-i\omega t'} i \frac{e^{-ikl(\cos \theta -1)}-1}{k(\cos \theta -1)},
  \tag{2}
\end{equation}

which leads to the following expression for the electric-field strength accompanying the electrojeitonic field of the radiation from the antenna at the given instant:

\begin{equation}\label{bla5}
   \vec{E_A}(\vec{r}, t) =-ik\vec{A}(\vec{r}, t) =B_0 \vec{e_z} e^{-i\omega t'} \frac{e^{-ikl(\cos \theta -1)}-1}{\cos \theta -1}.
  \tag{3}
\end{equation}

Relation (3) gives

\begin{equation}\label{bla6}
   \vec{E_A^l}(\vec{r}, t) = \vec{n}(\vec{E_A}\cdot \vec{n}) = \vec{n} \cos \theta B_0 e^{-i\omega t'} \frac{e^{-ikl(\cos \theta -1)}-1}{\cos \theta -1}, 
  \tag{4}
\end{equation}

the use of which leads to the expression for the time-averaged elementary power flux of the longitudinal electrojeitonic radiation from the antenna at time $t$:

\begin{equation}\label{bla7}
  dP_{G^l} = \vec{S_{G^l}}\cdot \vec(n).
  \tag{4.1}
\end{equation}

According to (3.7) and (4.1):

\begin{equation}\label{bla8}
  \vec{S_{G^l}} = \frac{c}{4\pi} \overline{(E^l_A)^2} \cdot \vec{n} \Rightarrow \vec{dP_{G^l}} = \frac{c}{8\pi} R_l(\vec{E^l_A} \cdot \vec{E^l_A}^*)r^2 d\Omega .
  \tag{5}
\end{equation}

Substitution of (4) into (5) gives

\begin{equation}\label{bla9}
  \vec{dP_{G^l}} = \frac{I_0^2}{2\pi c} \frac{\cos^2 \theta \cdot \sin^2 (\frac{kl}{2}(\cos \theta -1))}{(cos \theta -1)^2} d\Omega , 
  \tag{6}
\end{equation}

which, for $l=n\lambda$, becomes

\begin{equation}\label{bla10}
  \vec{dP_{G^l}} = \frac{I_0^2}{2\pi c} \frac{\cos^2 \theta \cdot \sin^2 (\pi n(\cos \theta -1))}{(cos \theta -1)^2} d\Omega . 
  \tag{7}
\end{equation}

This relation defines the radiation-pattern function for the time-averaged power of the longitudinal electrojeitonic radiation from the antenna:

\begin{equation}\label{bla11}
  g(\theta , n) = \left( \frac{\cos \theta \cdot \sin (\pi n(\cos \theta -1))}{(cos \theta -1)} \right) ^2 .
  \tag{8}
\end{equation}

The radiation patterns defined by this relation are shown in Figure 6.2:

\begin{figure}[H]
\begin{center}
\vspace{0pt}
\includegraphics[width=140mm]{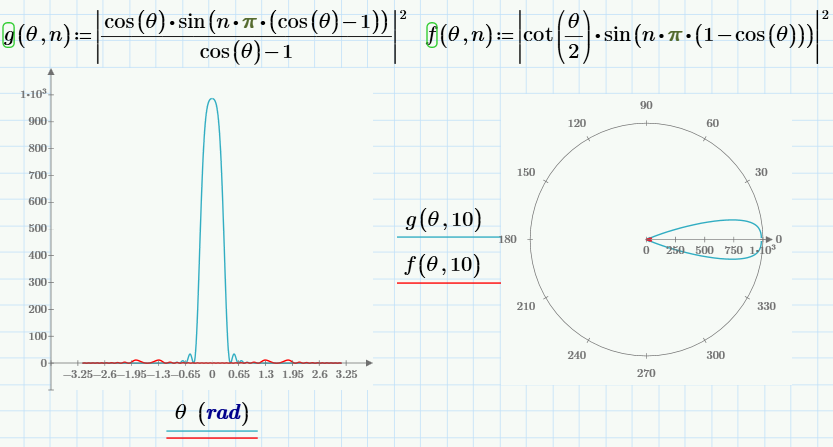}
\vspace{-2pt}
\caption{Radiation patterns.}
\end{center}
\end{figure}

\newpage

\section*{Acknowledgements}
\addcontentsline{toc}{section}{Acknowledgements}

In conclusion, the author again expresses his deep gratitude to N. N. Bogolyubov, Sr., D. V. Shirkov, and D. I. Blokhintsev for their stimulating discussions concerning the fundamental idea underlying the theory under consideration. He also expresses his gratitude to the Directorate of the Laboratory of Theoretical Physics of the Joint Institute for Nuclear Research for having provided the opportunity to undertake a course project (1968) and subsequently a diploma thesis (1970) ~\cite{Al7} at the Laboratory. These works provided a mathematical justification of the above-mentioned fundamental idea concerning the \emph{physical} existence of a classical field defined by the \emph{symmetric} derivative 4\nobreakdash-tensor of the classical electrodynamic 4\nobreakdash-potential. They also demonstrated a nonzero contribution of this field to the energy--momentum tensor of an arbitrary classical electrodynamic field system, thereby establishing the material nature of this field.

The author also expresses his deep gratitude to A. B. Arbuzov, Professor of the Russian Academy of Sciences and Doctor of Physical and Mathematical Sciences, Laboratory of Theoretical Physics, Joint Institute for Nuclear Research (Dubna); B. Zh. Zalikhanov, Doctor of Physical and Mathematical Sciences; and V. A. Baranov, Candidate of Physical and Mathematical Sciences, Laboratory of Nuclear Problems, Joint Institute for Nuclear Research (Dubna), for their interest in and attention to the works discussed.

The author expresses his deep gratitude to D. V. Podlesny, Candidate of Pedagogical Sciences, for his long-standing and sincere interest in the author's work.

The author also expresses his deep gratitude to V. A. Nesterin, Full Member of the Academy of Electrical Engineering Sciences of the Chuvash Republic and Doctor of Technical Sciences, for submitting the author's works \cite{Al2}, \cite{Al3}, and \cite{Al5} for publication in the ``Proceedings of the Academy of Electrical Engineering Sciences of the Chuvash Republic''.

\addcontentsline{toc}{section}{References}
\renewcommand{\bibname}{References}

\end{document}